\documentclass[runningheads]{llncs}

\usepackage[utf8]{inputenc}
\usepackage{amsmath}
\usepackage{amssymb}

\let\subparagraph\paragraph
\usepackage[compact]{titlesec}
\usepackage{thmtools}
\usepackage[english]{babel}
\usepackage{pdfpages}
\usepackage{paralist}
\usepackage{graphicx}
\usepackage{subcaption}
\usepackage[labelfont=bf]{caption}
\usepackage{longtable}
\usepackage{booktabs}
\usepackage{tabu}
\usepackage[referable]{threeparttablex}
\usepackage{varwidth}
\usepackage{multirow}
\usepackage{wrapfig}

\usepackage{float}

\usepackage{longtable}

\usepackage{pifont}

\usepackage[noend, ruled, linesnumbered]{algorithm2e}
\usepackage{parskip}

\usepackage{hyperref}
\usepackage{thm-restate}
\usepackage[capitalise]{cleveref}

\crefname{figure}{Figure}{Figure}
\crefformat{footnote}{#2\footnotemark[#1]#3}
\usepackage{tikz}
\usetikzlibrary{calc}
\usepackage{comment}

\usepackage{todonotes}
\usepackage{tikz}
\usetikzlibrary{arrows,automata,shapes,decorations,decorations.markings,calc, matrix,decorations.pathmorphing, patterns,backgrounds}

\usepackage{bm}
\usepackage{pgfplots}
\usepackage{pbox}

\usepackage{appendix}
\usepackage{calc}
\usepackage{fp}
\usepackage{multirow}
\usepackage{pbox}
\usepackage[clock]{ifsym}

\usepackage{booktabs}   
\usepackage{subcaption} 

\usepackage{enumerate,enumitem}

\setlist{nosep}
\renewcommand{\smallskip}{}
\DeclareMathAlphabet{\mathpzc}{OT1}{pzc}{m}{it}

\usepackage{graphicx}

\newcommand{\Paragraph}[1]{{\smallskip\noindent\bf #1}}

\newcommand{\bad}[1]{\mkern 1.5mu\overline{\mkern-1.5mu#1\mkern-1.5mu}\mkern 1.5mu}

\newcommand{\initev}{\mathsf{init\_event}}
\newcommand{\true}{\mathsf{true}}
\newcommand{\false}{\mathsf{false}}
\newcommand{\RF}[1]{\mathsf{RF}_{#1}}
\newcommand{\RVF}{\mathsf{RVF}}
\newcommand{\RVFE}{\sim_{\RVF}}
\newcommand{\mutate}{\mathsf{mutate}}
\newcommand{\backtrack}{\mathsf{backtrack}}
\newcommand{\ancestors}{\mathsf{ancestors}}
\newcommand{\forbids}{\mathsf{forbids}}

\newcommand{\VSC}{\operatorname{VSC}}

\newcommand{\AlgoSC}{\operatorname{VerifySC}}

\newcommand{\MemoryMap}{\operatorname{MMap}}

\newcommand{\Threads}{\operatorname{Threads}}
\newcommand{\RVFSMC}{\operatorname{RVF-SMC}}

\mathchardef\mhyphen="2D 
\newcommand{\Nats}{\mathbb{N}}

\newcommand{\Worklist}{\mathcal{S}}
\newcommand{\DoneSet}{\mathsf{Done}}

\newcommand{\Source}{\mathsf{Nidhugg/source}}

\newcommand{\ReadsFrom}{\mathsf{Nidhugg/rfsc}}

\newcommand{\VCDPOR}{\operatorname{VC-DPOR}}
\newcommand{\DCDPOR}{\operatorname{DC-DPOR}}

\newcommand{\SC}{\operatorname{SC}}

\newcommand{\NegativeAnnotation}{\mathcal{C}}

\newcommand{\Seq}{\tau}
\newcommand{\Trace}{\sigma}

\newcommand{\SysReads}{\mathcal{R}}
\newcommand{\SysWrites}{\mathcal{W}}

\newcommand{\RMW}{\mathit{rmw}}
\newcommand{\CAS}{\mathit{cas}}
\newcommand{\Read}{r}

\newcommand{\TO}{\mathsf{PO}}
\newcommand{\Write}{w}

\newcommand{\Event}{e}

\newcommand{\CHB}[3]{#1\mathsf{\mapsto}_{#2}#3}
\newcommand{\NCHB}[3]{#1\mathsf{\not \mapsto}_{#2}#3}

\newcommand{\Value}{\mathsf{val}}
\newcommand{\CandidateSet}{\operatorname{F}}

\newcommand{\Unordered}[3]{#1\parallel_{#2} #3}
\usepackage{centernot}
\newcommand{\Ordered}[3]{#1 \centernot \parallel_{\hspace{-1.4mm}#2}\hspace{0.8mm} #3}
\newcommand{\Refines}{\sqsubseteq}

\newcommand{\Confl}[2]{#1 \Join #2}

\newcommand{\Enabled}{\mathsf{enabled}}

\newcommand{\Concat}{\circ}
\newcommand{\Events}[1]{\SysEvents(#1)}
\newcommand{\Reads}[1]{\SysReads(#1)}
\newcommand{\Writes}[1]{\SysWrites(#1)}

\newcommand{\Project}{|}
\newcommand{\Domain}{\mathsf{dom}}
\newcommand{\ValueDomain}{\mathcal{D}}

\newcommand{\System}{\mathcal{H}}

\newcommand{\Process}{\mathsf{thr}}
\newcommand{\Proc}[1]{\Process(#1)}

\newcommand{\Globals}{\mathcal{G}}

\newcommand{\Aquire}{\mathsf{acquire}}
\newcommand{\Release}{\mathsf{release}}
\newcommand{\Location}[1]{\mathsf{var}(#1)}

\newcommand{\SysEvents}{\mathcal{E}}

\newcommand{\GoodWrites}{\mathsf{GoodW}}

\newcommand{\VisibleWrites}{\mathsf{VisibleW}}

\newcommand{\myblue}{blue!80!black}
\newcommand{\myred}{red!80!black}

\usetikzlibrary{arrows.meta,calc,decorations.markings,math,arrows.meta}

\makeatletter
\g@addto@macro\bfseries{\boldmath}
\makeatother

\pgfdeclarelayer{bg}    
\pgfsetlayers{bg,main}  

\def \darkred {black!20!red}
\def \darkgreen {black!20!green}
\SetKw{Continue}{continue}

\SetCommentSty{mycommfont}

\begin{document}

\title{Stateless Model Checking under a Reads-Value-From Equivalence}

\author{
Pratyush Agarwal\inst{1} \and
Krishnendu Chatterjee\inst{2} \and
Shreya Pathak\inst{1} \and
Andreas Pavlogiannis\inst{3} \and
Viktor Toman\inst{2}
}
\authorrunning{P. Agarwal \and K. Chatterjee \and A. Pavlogiannis \and S. Pathak \and V. Toman}

\institute{
IIT Bombay, Mumbai, India
\and
IST Austria, Klosterneuburg, Austria
\and
Aarhus University, Aarhus, Denmark
}

\maketitle

\setcounter{footnote}{0}
\begin{abstract}
Stateless model checking (SMC) is one of the standard approaches to the verification of concurrent programs.
As scheduling non-determinism creates exponentially large spaces of thread interleavings,
SMC attempts to partition this space into equivalence classes and explore only a few representatives from each class.
The efficiency of this approach depends on two factors:
(a)~the coarseness of the partitioning, and
(b)~the time to generate representatives in each class.
For this reason, the search for coarse partitionings that are efficiently explorable is an active research challenge.

In this work we present $\RVFSMC$, a new SMC algorithm that uses a novel \emph{reads-value-from (RVF)} partitioning.
Intuitively, two interleavings are deemed equivalent if they
agree on the value obtained
in each read event, and read events induce consistent causal orderings between them.
The RVF partitioning is provably coarser than recent approaches based on Mazurkiewicz and ``reads-from'' partitionings.
Our experimental evaluation reveals that RVF is quite often a very effective
equivalence, as the underlying partitioning is exponentially coarser than other approaches.
Moreover, $\RVFSMC$ generates representatives very efficiently, as
the reduction in the partitioning is often met with significant speed-ups in the model checking task.

\end{abstract}

\section{Introduction}\label{sec:intro}

The verification of concurrent programs is one of the key challenges in formal methods.
Interprocess communication adds a new dimension of non-determinism in program behavior, which is resolved by a scheduler.
As the programmer has no control over the scheduler, program correctness has to be guaranteed under all possible schedulers, i.e., the scheduler is adversarial to the program and can generate erroneous behavior if one can arise out of scheduling decisions.
On the other hand, during program testing, the adversarial nature of the scheduler is to hide erroneous runs, making bugs extremely difficult to reproduce by testing alone~(aka Heisenbugs\cite{Musuvathi08}).
Consequently, the verification of concurrent programs rests on rigorous model checking techniques~\cite{Clarke00} that cover all possible program behaviors that can arise out of scheduling non-determinism, leading to early tools such as VeriSoft~\cite{Godefroid97,Godefroid05} and {\sc CHESS}~\cite{Musuvathi07b}.

To battle with the state-space explosion problem, effective model checking for concurrency is stateless.
A stateless model checker (SMC) explores the behavior of the concurrent program by manipulating traces instead of states,
where each (concurrent) trace is an interleaving of event sequences of the corresponding threads~\cite{G96}.
To further improve performance, various techniques try to reduce the number of explored traces, such as context bounded techniques~\cite{Musuvathi07,Lal09,Chini2017,Baumann2021}
As many interleavings induce the same program behavior, SMC partitions the interleaving space into equivalence classes and attempts to sample a few representative traces from each class.
The most popular approach in this domain is partial-order reduction techniques~\cite{Clarke99,G96,Peled93}, which deems interleavings as equivalent based on the way that conflicting memory accesses are ordered, also known as the Mazurkiewicz equivalence~\cite{Mazurkiewicz87}.
Dynamic partial order reduction~\cite{Flanagan05} constructs this equivalence dynamically, when all memory accesses are known, and thus does not suffer from the imprecision of earlier approaches based on static information.
Subsequent works managed to explore the Mazurkiewicz partitioning optimally~\cite{Abdulla14,Nguyen18}, while spending only polynomial time per class.

The performance of an SMC algorithm is generally a product of two factors:
(a)~the size of the underlying partitioning that is explored, and
(b)~the total time spent in exploring each class of the partitioning.
Typically, the task of visiting a class requires solving a consistency-checking problem,
where the algorithm checks whether a semantic abstraction, used to represent the class, has a consistent concrete interleaving that witnesses the class.
For this reason, the search for effective SMC is reduced to the search of coarse partitionings for which the consistency problem is tractable, and has become a very active research direction in recent years.
In~\cite{Aronis18}, the Mazurkiewicz partitioning was further reduced by ignoring the order of conflicting write events that are not observed, while retaining polynomial-time consistency checking.
Various other works refine the notion of dependencies between events, yielding coarser abstractions~\cite{Godefroid93,Elvira17,Kokologiannakis19b}.
The work of~\cite{Chalupa17} used a reads-from abstraction and showed that the consistency problem admits a fully polynomial solution in acyclic communication topologies.
Recently, this approach was generalized to arbitrary topologies, with an algorithm that remains polynomial for a bounded number of threads~\cite{Abdulla19}.
Finally, recent approaches define value-centric partitionings~\cite{Chatterjee19}, as well as partitionings based on maximal causal models~ \cite{HUANG15}.
These partitionings are very coarse, as they attempt to distinguish only between traces which differ in the values read by their corresponding read events.
We illustrate the benefits of value-based partitionings with a motivating example.

\subsection{Motivating Example}\label{subsec:motivating}

Consider a simple concurrent program shown in~\cref{fig:motivating}.
The program has 98 different orderings of the conflicting memory
accesses, and each ordering corresponds to a separate class of the
Mazurkiewicz partitioning. Utilizing the reads-from abstraction
reduces the number of partitioning classes to 9.
However, when taking into consideration the values that the events
can read and write, the number of cases to consider can be reduced even
further. In this specific example, there is only a single behaviour
the program may exhibit, in which both read events read
the only observable value.

\begin{figure}[h]
\vspace{-1mm}
  \small
  \centering
  \begin{minipage}{0.15\textwidth}
  \begin{align*}
    \text{Th}&\text{read}_{1}\\
    \hline\\[-1em]
    1.~& \textcolor{\myblue}{\Write(x, 1)}\\
    \\[-1.6em]
    2.~& \textcolor{\myred}{\Write(y, 1)}\\
    \\[-0.1em]
  \end{align*}
  \end{minipage}
  \begin{minipage}{0.15\textwidth}
  \begin{align*}
    \text{Th}&\text{read}_{2}\\
    \hline\\[-1em]
    1.~& \textcolor{\myblue}{\Write(x, 1)}\\
    \\[-1.6em]
    2.~& \textcolor{\myred}{\Write(y, 1)}\\
    \\[-1.6em]
    3.~& \textcolor{\myblue}{\Read(x)}\\
  \end{align*}
  \end{minipage}
  \begin{minipage}{0.15\textwidth}
  \begin{align*}
    \text{Th}&\text{read}_{3}\\
    \hline\\[-1em]
    1.~& \textcolor{\myblue}{\Write(x, 1)}\\
    \\[-1.6em]
    2.~& \textcolor{\myred}{\Write(y, 1)}\\
    \\[-1.6em]
    3.~& \textcolor{\myred}{\Read(y)}\\
  \end{align*}
  \end{minipage}
  \qquad\qquad
  \begin{minipage}{0.15\textwidth}
  \begin{align*}
    \text{Equivalence classe}&\text{s:}\\
    \hline\\[-1em]
    \text{Mazurkiewicz~\cite{Abdulla14}}     ~~~~& 98\\
    \\[-1.6em]
    \text{reads-from~\cite{Abdulla19}}       ~~~~& \;\:9\\
    \\[-1.6em]
    \text{value-centric~\cite{Chatterjee19}}    ~~~~& \;\:7\\
    \\[-1.6em]
    \text{this work}    ~~~~& \;\:1\\
  \end{align*}
  \end{minipage}
  \vspace{-2mm}
  \caption{
    Concurrent program and its underlying partitioning classes.
  }
  \label{fig:motivating}
\end{figure}

The above benefits have led to recent attempts in performing SMC using a value-based equivalence~\cite{HUANG15,Chatterjee19}.
However, as the realizability problem is NP-hard in general~\cite{Gibbons97}, both approaches suffer significant drawbacks.
In particular, the work of~\cite{Chatterjee19} combines the value-centric approach with the Mazurkiewicz partitioning, which creates a refinement with exponentially many more classes than potentially necessary.
The example program in~\cref{fig:motivating} illustrates this, where
while both read events can only observe one possible value, the work
of~\cite{Chatterjee19} further enumerates all Mazurkiewicz orderings
of all-but-one threads, resulting in 7 partitioning classes.
Separately,
the work of~\cite{HUANG15} relies on SMT solvers, thus spending exponential time to solve the realizability problem.
Hence, each approach suffers an exponential blow-up a-priori, which motivates the following question: is there an efficient \emph{parameterized} algorithm for the consistency problem?
That is, we are interested in an algorithm that is exponential-time in the worst case (as the problem is NP-hard in general), but efficient when certain natural parameters of the input are small, and thus only becomes slow in extreme cases.

Another disadvantage of these works is that each of the exploration algorithms can end up to the same class of the partitioning many times, further hindering performance.
To see an example, consider the program in~\cref{fig:motivating} again.
The work of~\cite{Chatterjee19} assigns values to reads one by one,
and in this example, it needs to consider as
separate cases both permutations of the two reads as the orders
for assigning the values. This is to ensure completeness in cases
where there are write events causally dependent on some read events
(e.g., a write event appearing only if its thread-predecessor reads
a certain value). However, no causally dependent write events are
present in this program, and our work uses a principled approach
to detect this and avoid the redundant exploration.
While an example to demonstrate \cite{HUANG15} revisiting
partitioning classes is a bit more involved one, this property
follows from the lack of information sharing between spawned
subroutines, enabling the approach to be massively parallelized,
which has been discussed already in prior works~\cite{Chalupa17,AbdullaAJN18,Chatterjee19}.

\subsection{Our Contributions}\label{subsec:contributions}

In this work we tackle the two challenges illustrated in the motivating example in a principled, algorithmic way.
In particular, our contributions are as follows.
\begin{compactenum}
\item\label{item:contr1} We study the problem of verifying the sequentially consistent executions.
The problem is known to be NP-hard~\cite{Gibbons97} in general, already for 3 threads.
We show that the problem can be solved in $O(k^{d+1}\cdot n^{k+1})$ time for an input of $n$ events, $k$ threads and $d$ variables.
Thus, although the problem NP-hard in general, it can be solved in polynomial time when the number of threads and number of variables is bounded.
Moreover, our bound reduces to $O(n^{k+1})$ in the class of programs where every variable is written by only one thread (while read by many threads).
Hence, in this case the bound is polynomial for a fixed number of threads and without any dependence on the number of variables.

\item\label{item:contr2} We define a new equivalence between concurrent traces, called the \emph{reads-value-from (RVF)} equivalence.
Intuitively, two traces are RVF-equivalent if they agree on the value obtained in each read event, and read events induce consistent causal orderings between them.
We show that RVF induces a coarser partitioning than the partitionings explored by recent well-studied SMC algorithms ~\cite{Abdulla14,Chalupa17,Chatterjee19}, and thus reduces the search space of the model checker.

\item\label{item:contr3} We develop a novel SMC algorithm called $\RVFSMC$, and show that it is sound and complete for local safety properties such as assertion violations.
Moreover, $\RVFSMC$ has complexity $k^d\cdot n^{O(k)}\cdot \beta$, where $\beta$ is the size of the underlying RVF partitioning.
Under the hood, $\RVFSMC$ uses our consistency-checking algorithm of \cref{item:contr1} to visit each RVF class during the exploration.
Moreover, $\RVFSMC$ uses a novel heuristic to significantly reduce the number of revisits in any given RVF class, compared to the value-based explorations of~\cite{HUANG15,Chatterjee19}.

\item\label{item:contr4} We implement $\RVFSMC$ in the stateless model checker Nidhugg~\cite{Abdulla2015}.
Our experimental evaluation reveals that RVF is
quite often a very effective
equivalence, as the underlying partitioning is exponentially coarser than other approaches.
Moreover, $\RVFSMC$ generates representatives very efficiently, as
the reduction in the partitioning is often met with significant speed-ups in the model checking task.
\end{compactenum}

\section{Preliminaries}\label{sec:prel}

\Paragraph{General notation.}
Given a natural number $i\geq 1$, we let $[i]$ be the set $\{ 1,2,\dots, i \}$.
Given a map $f\colon X\to Y$, we let $\Domain(f)=X$ denote the domain of $f$.
We represent maps $f$ as sets of tuples $\{ (x, f(x))\}_x$.
Given two maps $f_1, f_2$ over the same domain $X$, we write $f_1=f_2$ if
for every $x\in X$ we have $f_1(x)=f_2(x)$.
Given a set $X'\subset X$, we denote by $f\Project X'$ the restriction of $f$ to $X'$.
A binary relation $\sim$ on a set $X$ is an {\em equivalence} iff $\sim$ is reflexive, symmetric and transitive.

\subsection{Concurrent Model}\label{subsec:prel_model}

Here we describe the computational model of concurrent programs with
shared memory under the Sequential Consistency ($\SC$) memory model.
We follow a standard exposition of stateless model checking,
similarly to~\cite{Flanagan05,Abdulla14,Chalupa17,Abdulla19,Kokologiannakis19,Chatterjee19},

\Paragraph{Concurrent program.}
We consider a concurrent program $\System=\{ \Process_i \}_{i=1}^k$ of $k$ deterministic threads.
The threads communicate over a shared memory 
$\Globals$ of global variables with a finite value domain $\ValueDomain$.
Threads execute \emph{events} of the following types.
\begin{enumerate}[noitemsep,topsep=0pt,partopsep=0px]
\item A \emph{write event} $\Write$ writes a value $v\in\ValueDomain$ to a global variable $x\in \Globals$.
\item A \emph{read event} $\Read$ reads the value $v\in\ValueDomain$ of a global variable $x\in \Globals$.
\end{enumerate}
Additionally, threads can execute local events which do not access
global variables and thus are not modeled explicitly.

Given an event $\Event$, we denote by $\Proc{\Event}$ its thread and
by $\Location{\Event}$ its global variable.
We denote by $\SysEvents$ the set of all events,
and by $\SysReads$ ($\SysWrites$) the set of read (write) events.
Given two events $\Event_1, \Event_2\in \SysEvents$,
we say that they \emph{conflict}, denoted $\Confl{\Event_1}{\Event_2}$,
if they access the same global variable and at least one of them is a write event.

\Paragraph{Concurrent program semantics.}
The semantics of $\System$ are defined by means of a transition system
over a state space of global states.
A global state consists of
(i) a memory function that maps every global variable to a value, and
(ii) a local state for each thread, which contains the values
of the local variables and the program counter of the thread.
We consider the standard setting of Sequential Consistency ($\SC$),
and refer to~\cite{Flanagan05} for formal details. As usual,
$\System$ is execution-bounded, which means that the state space is finite and acyclic.

\Paragraph{Event sets.}
Given a set of events $X\subseteq \SysEvents$,
we write $\Reads{X}=X\cap \SysReads$ for the set of read events of $X$,
and $\Writes{X}=X\cap \SysWrites$ for the set of write events of $X$.
Given a set of events $X\subseteq \SysEvents$ and a thread $\Process$,
we denote by $X_{\Process}$ and $X_{\neq\Process}$ the events
of $\Process$, and the events of all other threads in $X$, respectively.

\Paragraph{Sequences and Traces.}
Given a sequence of events $\Seq=\Event_1,\dots,\Event_j$, we denote by
$\Events{\Seq}$ the set of events that appear in $\Seq$.
We further denote $\Reads{\Seq} = \Reads{\Events{\Seq}}$ and
$\Writes{\Seq} = \Writes{\Events{\Seq}}$.

Given a sequence $\Seq$ and two events
$\Event_1, \Event_2 \in \Events{\Seq}$, we write
$\Event_1 <_\Seq \Event_2$ when $\Event_1$ appears before $\Event_2$
in $\Seq$, and $\Event_1 \leq_\Seq \Event_2$ to denote that
$\Event_1 <_\Seq \Event_2$ or $\Event_1 = \Event_2$.
Given a sequence $\Seq$ and a set of events $A$, we denote by
$\Seq \Project A$ the \emph{projection} of $\Seq$ on $A$, which is
the unique subsequence of $\Seq$ that contains all events of
$A \cap \Events{\Seq}$, and only those events.
Given a sequence $\Seq$ and a thread $\Process$,
let $\Seq_{\Process}$ be the subsequence of $\Seq$ with events of $\Process$,
i.e., $\Seq \Project \Events{\Seq}_{\Process}$.
Given two sequences $\Seq_1$ and $\Seq_2$, we denote by
$\Seq_1 \Concat \Seq_2$ the sequence that results in appending
$\Seq_2$ after $\Seq_1$.

A (concrete, concurrent) \emph{trace} is a sequence of events $\Trace$ that
corresponds to a concrete valid execution of $\System$.
We let $\Enabled(\Trace)$ be the set of enabled
events after $\Trace$ is executed, and call $\Trace$ \emph{maximal} if
$\Enabled(\Trace)=\emptyset$.
As $\System$ is bounded,
all executions of $\System$ are finite and
the length of the longest execution in $\System$ is a parameter of the input.

\Paragraph{Reads-from and Value functions.}
Given a sequence of events $\Seq$, we define the
\emph{reads-from function} of $\Seq$, denoted
$\RF{\Seq}\colon\Reads{\Seq}\to \Writes{\Seq}$, as follows.
Given a read event $\Read \in \Reads{\Seq}$, we have that
$\RF{\Seq}(\Read)$ is the latest write (of any thread)
conflicting with $\Read$ and occurring before $\Read$ in $\Seq$, i.e.,
(i) $\Confl{\RF{\Seq}(\Read)}{\Read}$,
(ii) $\RF{\Seq}(\Read) <_\Seq \Read$, and
(iii) for each $\bad{\Write} \in \Writes{\Seq}$ such that
$\Confl{\bad{\Write}}{\Read}$ and $\bad{\Write} <_\Seq \Read$,
we have $\bad{\Write} \leq_\Seq \RF{\Seq}(\Read)$.
We say that $\Read$ reads-from $\RF{\Seq}(\Read)$ in $\Seq$.
For simplicity, we assume that $\System$ has an initial salient
write event on each variable.

Further, given a trace $\Trace$, we define the
\emph{value function} of $\Trace$, denoted
$\Value_{\Trace}\colon\Events{\Trace}\to \ValueDomain$, such that
$\Value_{\Trace}(\Event)$ is the value of the global variable
$\Location{\Event}$ after the prefix of $\Trace$ up to and including
$\Event$ has been executed. Intuitively, $\Value_{\Trace}(\Event)$
captures the value that a read (resp. write) event $\Event$
shall read (resp. write) in $\Trace$.
The value function $\Value_{\Trace}$ is well-defined as
$\Trace$ is a valid trace and the
threads of $\System$ are deterministic.

\subsection{Partial Orders}\label{subsec:prel_partialorders}

In this section we present relevant notation around partial orders,
which are a central object in this work.

\Paragraph{Partial orders.}
Given a set of events $X\subseteq \SysEvents$, a \emph{(strict) partial order} $P$ over $X$ is an irreflexive,
antisymmetric and transitive relation over $X$ (i.e., $<_{P}\,\subseteq X\times X$).
Given two events $\Event_1,\Event_2\in X$, we write $\Event_1\leq_P \Event_2$ to denote that $\Event_1<_P\Event_2$ or $\Event_1=\Event_2$.
Two distinct events $\Event_1,\Event_2\in X$ are \emph{unordered} by $P$, denoted $\Unordered{\Event_1}{P}{\Event_2}$,
if neither $\Event_1<_{P}\Event_2$ nor $\Event_2<_{P} \Event_1$, and \emph{ordered} (denoted $\Ordered{\Event_1}{P}{\Event_2}$) otherwise.
Given a set $Y\subseteq X$, we denote by $P\Project Y$ the \emph{projection} of $P$ on the set $Y$,
where for every pair of events $\Event_1, \Event_2\in Y$,
we have that $\Event_1<_{P\Project Y} \Event_2$ iff $\Event_1<_{P} \Event_2$.
Given two partial orders $P$ and $Q$ over a common set $X$, we say that $Q$ \emph{refines} $P$, denoted by $Q\Refines P$, if
for every pair of events $\Event_1, \Event_2\in X$, if $\Event_1<_{P}\Event_2$ then $\Event_1<_{Q}\Event_2$.
A \emph{linearization} of $P$ is a total order that refines $P$.

\Paragraph{Lower  sets.}
Given a pair $(X,P)$, where $X$ is a set of events and $P$ is a partial order over $X$,
a \emph{lower set} of $(X,P)$ is a set $Y\subseteq X$ such that
for every event $\Event_1\in Y$ and event $\Event_2\in X$ with
$\Event_2\leq_{P} \Event_1$, we have $\Event_2\in Y$.

\Paragraph{Visible writes.}
Given a partial order $P$ over a set $X$,
and a read event $\Read\in \Reads{X}$,
the set of \emph{visible writes} of $\Read$ is defined as
\vspace{-1mm}
\begin{align*}
\VisibleWrites_{P}(\Read)=&\{
\;\Write\in \Writes{X}:~\text{(i) } \Confl{\Read}{\Write}
\text{ and (ii) }   \Read\not <_{P}\Write
\text{ and (iii) for each }\\
&   \;\;\, \Write'\in \Writes{X}
\text{ with }  \Confl{\Read}{\Write'}\text{,}
\text{ if }  \Write<_{P} \Write'   \text{ then }  \Write'\not  <_{P} \Read\;
\}
\end{align*}
i.e., the set of write events
$\Write$ conflicting with $\Read$ that are not ``hidden'' to
$\Read$ by $P$.

\Paragraph{The program order $\TO$.}
The \emph{program order} $\TO$ of $\System$ is a partial order
$<_{\TO}\subseteq \SysEvents\times \SysEvents$ that defines
a fixed order between some pairs of events of the same thread,
reflecting the semantics of $\System$.

A set of events $X\subseteq \SysEvents$ is \emph{proper} if
(i) it is a lower set of $(\SysEvents, \TO)$, and
(ii) for each thread $\Process$,
the events $X_{\Process}$ are totally ordered in $\TO$
(i.e., for each distinct $\Event_1,\Event_2 \in X_{\Process}$ we have $\Ordered{\Event_1}{\TO}{\Event_2}$).
A sequence $\Seq$ is \emph{well-formed} if
(i) its set of events $\Events{\Seq}$ is proper, and
(ii) $\Seq$ respects the program order
(formally, $\Seq\Refines \TO\Project \Events{\Seq}$).
Every trace $\Trace$ of $\System$
is well-formed, as it corresponds to a concrete valid execution of $\System$.
Each event of $\System$ is then uniquely identified by its $\TO$ predecessors,
and by the values its $\TO$ predecessor reads have read.

\begin{figure}[h]
\vspace{-1mm}
  \begin{minipage}{0.40\textwidth}
    \centering
    \scalebox{0.95}{
    \begin{tikzpicture}[thick,
    pre/.style={<-,shorten >= 2pt, shorten <=2pt, very thick},
    post/.style={->,shorten >= 2pt, shorten <=2pt,  very thick},
    seqtrace/.style={->, line width=1},
    aux_seqtrace/.style={->, line width=1, draw=gray},
    und/.style={very thick, draw=gray},
    event/.style={rectangle, minimum height=3.5mm, draw=black, fill=white, minimum width=11.5mm,   line width=1pt, inner sep=2, font={\footnotesize}},
    aux_event/.style={event, draw=gray},
    virt/.style={circle,draw=black!50,fill=black!20, opacity=0},
    bad/.style={preaction={fill, white}, pattern color=red!20, pattern=north east lines},
    good/.style={preaction={fill, white}, pattern color=green!40, pattern=north west lines},
    enabled/.style={event, draw=gray, dashed, fill=black!05},
    isLabel/.style={rectangle, fill opacity=0.5, fill=white, text opacity=1},
    recnode/.style={circle, draw=black},
    ]

    \newcommand{\xstep}{3.4}
    \newcommand{\ystep}{0.78}


    \newcommand{\xmove}{0.}
    \newcommand{\ymove}{0.}
    \node[]                  (t1_0)   at (\xmove + 0*\xstep, \ymove + -0.2*\ystep) {\small$\Process_1$};
    \node[]                  (t1_end) at (\xmove + 0*\xstep, \ymove + -3.8*\ystep) {};
    \node[]                  (t2_0)   at (\xmove + 0.5*\xstep, \ymove + -0.2*\ystep) {\small$\Process_2$};
    \node[]                  (t2_end) at (\xmove + 0.5*\xstep, \ymove + -4.3*\ystep) {};
    \node[]                  (t3_0)   at (\xmove + 1*\xstep, \ymove + -0.2*\ystep) {\small$\Process_3$};
    \node[]                  (t3_end) at (\xmove + 1*\xstep, \ymove + -4.8*\ystep) {};
    \draw[seqtrace]          (t1_0) to (t1_end);
    \draw[seqtrace]          (t2_0) to (t2_end);
    \draw[seqtrace]          (t3_0) to (t3_end);
    \node[event]             (t1_w1)  at (\xmove + 0*\xstep, \ymove + -1*\ystep) {\textcolor{\myblue}{$\Write(x, 1)$}};
    \node[event]             (t1_r1)  at (\xmove + 0*\xstep, \ymove + -3*\ystep) {\textcolor{\myblue}{$\Read(x)$}};
    \node[event]             (t2_w2)  at (\xmove + 0.5*\xstep, \ymove + -1.5*\ystep) {\textcolor{\myblue}{$\Write(x, 1)$}};
    \node[event]             (t2_r2)  at (\xmove + 0.5*\xstep, \ymove + -2.5*\ystep) {\textcolor{\myblue}{$\Read(x)$}};
    \node[event]             (t2_w3)  at (\xmove + 0.5*\xstep, \ymove + -3.5*\ystep) {\textcolor{\myred}{$\Write(y, 2)$}};
    \node[event]             (t3_w4)  at (\xmove + 1*\xstep, \ymove + -2*\ystep) {\textcolor{\myred}{$\Write(y, 1)$}};
    \node[event]             (t3_r3)  at (\xmove + 1*\xstep, \ymove + -4*\ystep) {\textcolor{\myred}{$\Read(y)$}};
    \draw[post, \darkred, dashed]   (t2_w2) to[out=190, in=60, distance=0.6cm] (t1_r1)
                             node [midway, xshift=15pt, yshift=-38pt, rotate=30, isLabel] {\small read by};
    \draw[post, \darkred, dashed]   (t2_w2) to[out=0, in=0, distance=0.5cm] (t2_r2);
    \draw[post, \darkred, dashed]   (t2_w3) to[out=0, in=160] (t3_r3);

    \end{tikzpicture}
    }
  \end{minipage}
  \qquad\quad
  \begin{minipage}{0.50\textwidth}
    \caption{
      A trace $\Trace$, the displayed events $\Events{\Trace}$
      are vertically ordered as they appear in
      $\Trace$. The solid black edges represent the program order $\TO$.
      The dashed red edges represent the reads-from function
      $\RF{\Trace}$. The transitive closure of all the edges then
      gives us the causally-happens-before partial order
      $\CHB{}{\Trace}{}$.
    }
  \label{fig:causallyhb}
  \end{minipage}
\end{figure}
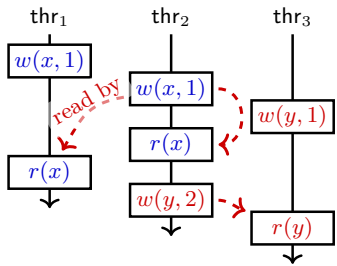

\Paragraph{Causally-happens-before partial orders.}
A trace $\Trace$ induces a \emph{causally-happens-before} partial order
$\CHB{}{\Trace}{} \subseteq \Events{\Trace}\times \Events{\Trace}$,
which is the weakest partial order such that
(i)~it refines the program order (i.e., $\CHB{}{\Trace}{}\Refines \TO\Project\Events{\Trace}$),
and (ii)~for every read event $\Read\in \Reads{\Trace}$, its
reads-from $\RF{\Trace}(\Read)$ is ordered before it
(i.e., $\RF{\Trace}(\Read)\;\, \CHB{}{\Trace}{} \;\,\Read$).
Intuitively, $\CHB{}{\Trace}{}$ contains the causal orderings in $\Trace$,
i.e., it captures the flow of write events into
read events in $\Trace$ together with the program order.
\cref{fig:causallyhb} presents an example of a trace and its causal orderings.

\section{Reads-Value-From Equivalence}\label{sec:equiv}

In this section we present our new equivalence on traces,
called the \emph{reads-value-from} equivalence
($\RVF$ equivalence, or $\RVFE$, for short).
Then we illustrate that $\RVFE$ has some desirable properties for stateless model checking.

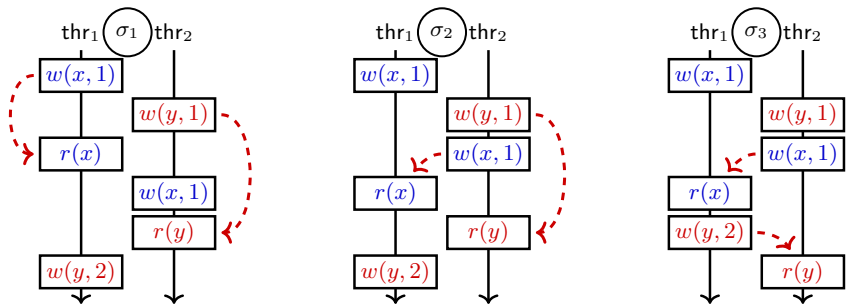
\begin{figure}[h]
  \centering
  \scalebox{0.95}{
  \begin{tikzpicture}[thick,
  pre/.style={<-,shorten >= 2pt, shorten <=2pt, very thick},
  post/.style={->,shorten >= 2pt, shorten <=2pt,  very thick},
  seqtrace/.style={->, line width=1},
  aux_seqtrace/.style={->, line width=1, draw=gray},
  und/.style={very thick, draw=gray},
  event/.style={rectangle, minimum height=3.5mm, draw=black, fill=white, minimum width=11.5mm,   line width=1pt, inner sep=2, font={\footnotesize}},
  aux_event/.style={event, draw=gray},
  virt/.style={circle,draw=black!50,fill=black!20, opacity=0},
  bad/.style={preaction={fill, white}, pattern color=red!20, pattern=north east lines},
  good/.style={preaction={fill, white}, pattern color=green!40, pattern=north west lines},
  enabled/.style={event, draw=gray, dashed, fill=black!05},
  isLabel/.style={rectangle, fill opacity=0.5, fill=white, text opacity=1},
  recnode/.style={circle, draw=black},
  ]

  \newcommand{\xstep}{2.6}
  \newcommand{\ystep}{1.1}


  \newcommand{\xmove}{0.}
  \newcommand{\ymove}{0.}
  \node[recnode]           (1)        at (\xmove + 0.25*\xstep, \ymove + -0.45*\ystep) {$\Trace_1$};
  \node[]                  (1_t1_0)   at (\xmove + 0*\xstep, \ymove + -0.45*\ystep) {\small$\Process_1$};
  \node[]                  (1_t1_end) at (\xmove + 0*\xstep, \ymove + -4.02*\ystep) {};
  \node[]                  (1_t2_0)   at (\xmove + 0.5*\xstep, \ymove + -0.45*\ystep) {\small$\Process_2$};
  \node[]                  (1_t2_end) at (\xmove + 0.5*\xstep, \ymove + -4.02*\ystep) {};
  \draw[seqtrace]          (1_t1_0) to (1_t1_end);
  \draw[seqtrace]          (1_t2_0) to (1_t2_end);
  \node[event]             (1_t1_w1)  at (\xmove + 0*\xstep, \ymove + -1*\ystep) {\textcolor{\myblue}{$\Write(x, 1)$}};
  \node[event]             (1_t1_r1)  at (\xmove + 0*\xstep, \ymove + -2*\ystep) {\textcolor{\myblue}{$\Read(x)$}};
  \node[event]             (1_t1_w2)  at (\xmove + 0*\xstep, \ymove + -3.5*\ystep) {\textcolor{\myred}{$\Write(y, 2)$}};
  \node[event]             (1_t2_w3)  at (\xmove + 0.5*\xstep, \ymove + -1.5*\ystep) {\textcolor{\myred}{$\Write(y, 1)$}};
  \node[event]             (1_t2_w4)  at (\xmove + 0.5*\xstep, \ymove + -2.5*\ystep) {\textcolor{\myblue}{$\Write(x, 1)$}};
  \node[event]             (1_t2_r2)  at (\xmove + 0.5*\xstep, \ymove + -3*\ystep) {\textcolor{\myred}{$\Read(y)$}};
  \draw[post, \darkred, dashed]   (1_t1_w1) to[out=180, in=180, distance=0.5cm] (1_t1_r1);
  \draw[post, \darkred, dashed]   (1_t2_w3) to[out=0, in=0, distance=0.6cm] (1_t2_r2);

  \renewcommand{\xmove}{4.4}
  \renewcommand{\ymove}{0.}
  \node[recnode]           (1)        at (\xmove + 0.25*\xstep, \ymove + -0.45*\ystep) {$\Trace_2$};
  \node[]                  (1_t1_0)   at (\xmove + 0*\xstep, \ymove + -0.45*\ystep) {\small$\Process_1$};
  \node[]                  (1_t1_end) at (\xmove + 0*\xstep, \ymove + -4.02*\ystep) {};
  \node[]                  (1_t2_0)   at (\xmove + 0.5*\xstep, \ymove + -0.45*\ystep) {\small$\Process_2$};
  \node[]                  (1_t2_end) at (\xmove + 0.5*\xstep, \ymove + -4.02*\ystep) {};
  \draw[seqtrace]          (1_t1_0) to (1_t1_end);
  \draw[seqtrace]          (1_t2_0) to (1_t2_end);
  \node[event]             (1_t1_w1)  at (\xmove + 0*\xstep, \ymove + -1*\ystep) {\textcolor{\myblue}{$\Write(x, 1)$}};
  \node[event]             (1_t1_r1)  at (\xmove + 0*\xstep, \ymove + -2.5*\ystep) {\textcolor{\myblue}{$\Read(x)$}};
  \node[event]             (1_t1_w2)  at (\xmove + 0*\xstep, \ymove + -3.5*\ystep) {\textcolor{\myred}{$\Write(y, 2)$}};
  \node[event]             (1_t2_w3)  at (\xmove + 0.5*\xstep, \ymove + -1.5*\ystep) {\textcolor{\myred}{$\Write(y, 1)$}};
  \node[event]             (1_t2_w4)  at (\xmove + 0.5*\xstep, \ymove + -2.*\ystep) {\textcolor{\myblue}{$\Write(x, 1)$}};
  \node[event]             (1_t2_r2)  at (\xmove + 0.5*\xstep, \ymove + -3*\ystep) {\textcolor{\myred}{$\Read(y)$}};
  \draw[post, \darkred, dashed]   (1_t2_w4) to[out=180, in=50] (1_t1_r1);
  \draw[post, \darkred, dashed]   (1_t2_w3) to[out=0, in=0, distance=0.6cm] (1_t2_r2);

  \renewcommand{\xmove}{8.8}
  \renewcommand{\ymove}{0.}
  \node[recnode]           (1)        at (\xmove + 0.25*\xstep, \ymove + -0.45*\ystep) {$\Trace_3$};
  \node[]                  (1_t1_0)   at (\xmove + 0*\xstep, \ymove + -0.45*\ystep) {\small$\Process_1$};
  \node[]                  (1_t1_end) at (\xmove + 0*\xstep, \ymove + -4.02*\ystep) {};
  \node[]                  (1_t2_0)   at (\xmove + 0.5*\xstep, \ymove + -0.45*\ystep) {\small$\Process_2$};
  \node[]                  (1_t2_end) at (\xmove + 0.5*\xstep, \ymove + -4.02*\ystep) {};
  \draw[seqtrace]          (1_t1_0) to (1_t1_end);
  \draw[seqtrace]          (1_t2_0) to (1_t2_end);
  \node[event]             (1_t1_w1)  at (\xmove + 0*\xstep, \ymove + -1*\ystep) {\textcolor{\myblue}{$\Write(x, 1)$}};
  \node[event]             (1_t1_r1)  at (\xmove + 0*\xstep, \ymove + -2.5*\ystep) {\textcolor{\myblue}{$\Read(x)$}};
  \node[event]             (1_t1_w2)  at (\xmove + 0*\xstep, \ymove + -3*\ystep) {\textcolor{\myred}{$\Write(y, 2)$}};
  \node[event]             (1_t2_w3)  at (\xmove + 0.5*\xstep, \ymove + -1.5*\ystep) {\textcolor{\myred}{$\Write(y, 1)$}};
  \node[event]             (1_t2_w4)  at (\xmove + 0.5*\xstep, \ymove + -2.*\ystep) {\textcolor{\myblue}{$\Write(x, 1)$}};
  \node[event]             (1_t2_r2)  at (\xmove + 0.5*\xstep, \ymove + -3.5*\ystep) {\textcolor{\myred}{$\Read(y)$}};
  \draw[post, \darkred, dashed]   (1_t2_w4) to[out=180, in=50] (1_t1_r1);
  \draw[post, \darkred, dashed]   (1_t1_w2) to[out=0, in=120] (1_t2_r2);

  \end{tikzpicture}
  }
  \vspace{-2mm}
  \caption{
    Three traces $\Trace_1$, $\Trace_2$, $\Trace_3$, events of each
    trace are vertically ordered as they appear in the trace.
    Traces $\Trace_1$ and $\Trace_2$ are $\RVF$-equivalent
    ($\Trace_1$ $\RVFE$ $\Trace_2$), as they have the same events,
    same value function, and the two read events are causally unordered
    in both. Trace $\Trace_3$ is not $\RVF$-equivalent with either of
    $\Trace_1$ and $\Trace_2$. Compared to $\Trace_1$ resp. $\Trace_2$,
    the value function of $\Trace_3$ differs
    (\textcolor{\myred}{$\Read(y)$} reads a different value),
    and the causal orderings of the reads differ
    ($\CHB{\textcolor{\myblue}{\Read(x)}}{\Trace_3}{\textcolor{\myred}{\Read(y)}}$).
  }
  \label{fig:rvfequivalence}
\end{figure}

\Paragraph{Reads-Value-From equivalence.}
Given two traces $\Trace_1$ and $\Trace_2$, we say that they are
\emph{reads-value-from-equivalent},
written $\Trace_1 \RVFE \Trace_2$, if the following hold.
\begin{enumerate}[noitemsep,topsep=0pt,partopsep=0px]
\item $\Events{\Trace_1} = \Events{\Trace_2}$, i.e., they consist of the same set of events.
\item $\Value_{\Trace_1} = \Value_{\Trace_2}$, i.e., each event reads resp. writes the same value in both.
\item $\CHB{}{\Trace_1}{} \Project \SysReads = \CHB{}{\Trace_2}{} \Project \SysReads$,
i.e., their causal orderings agree on the read events.
\end{enumerate}
\cref{fig:rvfequivalence} presents an intuitive example of
$\RVF$-(in)equivalent traces.

\Paragraph{Soundness.}
The $\RVF$ equivalence induces a partitioning on
the maximal traces of $\System$.
Any algorithm that explores each class of this partitioning provably
discovers every reachable local state of every thread, and thus
$\RVF$ is a sound equivalence for local safety properties,
such as assertion violations,
in the same spirit as in other recent works~\cite{Abdulla19,Chalupa17,Chatterjee19,HUANG15}.
This follows from the fact that for any
two traces $\Trace_1$ and $\Trace_2$
with $\Events{\Trace_1} = \Events{\Trace_2}$ and
$\Value_{\Trace_1} = \Value_{\Trace_2}$, the local states of each thread
are equal after executing $\Trace_1$ and $\Trace_2$.

\begin{figure}[h]
  \centering
  \begin{tikzpicture}[thick,
  pre/.style={<-,shorten >= 2pt, shorten <=2pt, very thick},
  post/.style={->,shorten >= 2pt, shorten <=2pt,  very thick},
  seqtrace/.style={->, line width=1},
  aux_seqtrace/.style={->, line width=1, draw=gray},
  und/.style={very thick, draw=gray},
  event/.style={rectangle, minimum height=3.5mm, draw=black, fill=white, minimum width=11.5mm,   line width=1pt, inner sep=2, font={\footnotesize}},
  aux_event/.style={event, draw=gray},
  virt/.style={circle,draw=black!50,fill=black!20, opacity=0},
  bad/.style={preaction={fill, white}, pattern color=red!20, pattern=north east lines},
  good/.style={preaction={fill, white}, pattern color=green!40, pattern=north west lines},
  enabled/.style={event, draw=gray, dashed, fill=black!05},
  isLabel/.style={rectangle, fill opacity=0.5, fill=white, text opacity=1},
  recnode/.style={circle, draw=black},
  ]

  \newcommand{\xstep}{2.8}
  \newcommand{\ystep}{0.6}

  \newcommand{\xmove}{0.}
  \newcommand{\ymove}{0.}

  \node[]                  (rvf)  at (\xmove + 0*\xstep, \ymove + 0*\ystep) {{\bf reads-value-from}};
  \node[]                  (rf)   at (\xmove + 1*\xstep, \ymove + 1*\ystep) {{\bf reads-from\cite{Abdulla19,Kokologiannakis19}}};
  \node[]                  (vc)   at (\xmove + 1*\xstep, \ymove + -1*\ystep) {{\bf value-centric\cite{Chatterjee19}}};
  \node[]                  (dc)   at (\xmove + 2*\xstep, \ymove + 0*\ystep) {{\bf data-centric\cite{Chalupa17}}};
  \node[]                  (maz)  at (\xmove + 3.3*\xstep, \ymove + 0*\ystep) {{\bf Mazurkiewicz\cite{Flanagan05,Abdulla14,Kokologiannakis17}}};

  \draw[post, dashed]   (rf) to[out=180, in=40] (rvf);
  \draw[post, dashed]   (vc) to[out=180, in=-40] (rvf);
  \draw[post, dashed]   (dc) to[out=140, in=0] (rf);
  \draw[post, dashed]   (dc) to[out=220, in=0] (vc);
  \draw[post, dashed]   (maz) to[] (dc);

  \end{tikzpicture}
\caption{
  SMC trace equivalences. An edge from X to Y signifies that Y is always
  at least as coarse, and sometimes coarser, than X.
}
\label{fig:coarseness}
\end{figure}
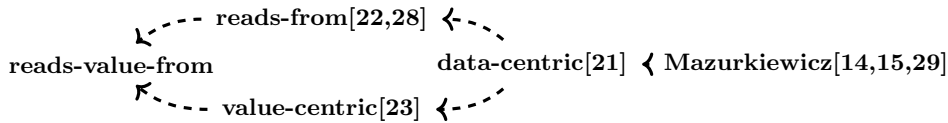

\Paragraph{Coarseness.}
Here we describe the coarseness properties of the $\RVF$ equivalence,
as compared to other equivalences used by state-of-the-art approaches
in stateless model checking. \cref{fig:coarseness} summarizes the comparison.

The SMC algorithms of~\cite{Abdulla19} and~\cite{Kokologiannakis19}
operate on a \emph{reads-from equivalence}, which deems two traces
$\Trace_1$ and $\Trace_2$ equivalent if
\begin{enumerate}[noitemsep,topsep=0pt,partopsep=0px]
\item they consist of the same events ($\Events{\Trace_1} = \Events{\Trace_2}$), and
\item their reads-from functions coincide ($\RF{\Trace_1} = \RF{\Trace_2}$).
\end{enumerate}
The above two conditions imply that the induced causally-happens-before
partial orders are equal, i.e., $\CHB{}{\Trace_1}{} = \CHB{}{\Trace_2}{}$,
and thus trivially also $\CHB{}{\Trace_1}{} \Project \SysReads = \CHB{}{\Trace_2}{} \Project \SysReads$.
Further, by a simple inductive argument
the value functions of the two traces are also equal,
i.e., $\Value_{\Trace_1} = \Value_{\Trace_2}$. Hence any two
reads-from-equivalent traces are also $\RVF$-equivalent, which makes the
$\RVF$ equivalence always at least as coarse as the reads-from equivalence.

The work of~\cite{Chatterjee19} utilizes a \emph{value-centric equivalence},
which deems two traces equivalent if they satisfy all the conditions
of our $\RVF$ equivalence, and also some further conditions (note that
these conditions are necessary for correctness of the
SMC algorithm in~\cite{Chatterjee19}).
Thus the $\RVF$ equivalence is trivially always at least as coarse.
The value-centric equivalence preselects a single thread $\Process$,
and then requires two extra conditions for the traces to be
equivalent, namely:
\begin{enumerate}[noitemsep,topsep=0pt,partopsep=0px]
\item For each read of $\Process$, either the read reads-from a write of
$\Process$ in both traces,
or it does not read-from a write of $\Process$ in either of the two traces.
\item For each conflicting pair of events not belonging to $\Process$,
the ordering of the pair is equal in the two traces.
\end{enumerate}

Both the reads-from equivalence and the value-centric equivalence are
in turn as coarse as the \emph{data-centric equivalence} of~\cite{Chalupa17}.
Given two traces, the data-centric equivalence has the equivalence
conditions of the reads-from equivalence, and additionally,
it preselects a single thread $\Process$
(just like the value-centric equivalence)
and requires the second extra condition of the value-centric equivalence,
i.e., equality of orderings for each conflicting pair of events outside of
$\Process$.

Finally, the data-centric equivalence is as coarse as the classical
\emph{Mazurkiewicz equivalence}~\cite{Mazurkiewicz87}, the baseline
equivalence for stateless model
checking~\cite{Flanagan05,Abdulla14,Kokologiannakis17}.
Mazurkiewicz equivalence deems two traces equivalent if they consist
of the same set of events and they agree on their ordering of conflicting events.

While $\RVF$ is always at least as coarse,
it can be (even exponentially) coarser,
than each of the other above-mentioned equivalences.
We illustrate this in~\cref{sec:app_equiv}.
We summarize these observations in the following proposition.

\begin{proposition}
$\RVF$ is at least as coarse as each of
the Mazurkiewicz equivalence (\cite{Abdulla14}),
the data-centric equivalence (\cite{Chalupa17}),
the reads-from equivalence (\cite{Abdulla19}),
and the value-centric equivalence (\cite{Chatterjee19}).
Moreover, $\RVF$ can be exponentially coarser than each of these equivalences.
\end{proposition}

In this work we develop our SMC algorithm $\RVFSMC$ around the $\RVF$ equivalence, with the guarantee that the algorithm explores
at most one maximal trace per class of the $\RVF$ partitioning, and thus can perform significantly fewer steps than  algorithms based on the above equivalences.
To utilize $\RVF$, the algorithm in each step solves an instance of the verification of sequential consistency problem,
which we tackle in the next section.
Afterwards, we present $\RVFSMC$.

\section{Verifying Sequential Consistency}\label{sec:verification}

In this section we present our contributions towards
the problem of verifying sequential consistency ($\VSC$).
We present an algorithm $\AlgoSC$ for $\VSC$, and we
show how it can be efficiently used in stateless model checking.

\Paragraph{The $\VSC$ problem.}
Consider an input pair $(X, \GoodWrites)$ where
\begin{enumerate}[noitemsep,topsep=0pt,partopsep=0px]
\item $X \subseteq \SysEvents$ is a proper set of events, and
\item $\GoodWrites \colon \Reads{X} \to 2^{\Writes{X}}$ is a good-writes
function such that $\Write \in \GoodWrites(\Read)$
only if $\Confl{\Read}{\Write}$.
\end{enumerate}
A \emph{witness} of $(X, \GoodWrites)$ is a linearization $\Seq$
of $X$ (i.e., $\Events{\Seq} = X$) respecting the program order
(i.e., $\Seq \Refines \TO \Project X$), such that each read
$\Read \in \Reads{\Seq}$ reads-from one of its good-writes in $\Seq$,
formally $\RF{\Seq}(\Read) \in \GoodWrites(\Read)$
(we then say that $\Seq$ \emph{satisfies} the good-writes
function $\GoodWrites$).
The task is to decide whether $(X, \GoodWrites)$ has a witness,
and to construct one in case it exists.

\Paragraph{$\VSC$ in Stateless Model Checking.}
The $\VSC$ problem naturally ties in with our SMC approach
enumerating the equivalence classes of the $\RVF$ trace partitioning.
In our approach, we shall generate instances $(X, \GoodWrites)$
such that
(i) each witness $\Trace$ of $(X, \GoodWrites)$ is a valid
program trace, and
(ii) all witnesses $\Trace_1, \Trace_2$
of $(X, \GoodWrites)$ are pairwise $\RVF$-equivalent
($\Trace_1 \RVFE \Trace_2$).

\Paragraph{Hardness of $\VSC$.}
Given an input $(X, \GoodWrites)$ to the $\VSC$ problem,
let $n=|X|$, let $k$ be the number of threads appearing in $X$, and
let $d$ be the number of variables accessed in $X$.
The classic work of~\cite{Gibbons97} establishes
two important lower bounds on the complexity of $\VSC$:
\begin{enumerate}[noitemsep,topsep=0pt,partopsep=0px]
\item $\VSC$ is NP-hard even when restricted only to inputs with $k=3$.
\item $\VSC$ is NP-hard even when restricted only to inputs with $d=2$.
\end{enumerate}
The first bound eliminates the possibility of any algorithm with time
complexity $O( n^{ f(k) })$, where $f$ is an arbitrary computable
function. Similarly, the second bound eliminates algorithms
with complexity $O( n^{ f(d) })$ for any computable $f$.

In this work we show that the problem is parameterizable in $k+d$, and thus admits efficient (polynomial-time) solutions when both variables are bounded.


\subsection{Algorithm for $\VSC$}\label{sec:verification_algo}

In this section we present our algorithm $\AlgoSC$ for the
problem $\VSC$. First we define some relevant notation.
In our definitions we consider a fixed input
pair $(X, \GoodWrites)$ to the $\VSC$ problem,
and a fixed sequence $\Seq$ with $\Events{\Seq} \subseteq X$.

\Paragraph{Active writes.}
A write $\Write \in \Writes{\Seq}$ is \emph{active} in $\Seq$
if it is the last write of its variable in $\Seq$. Formally,
for each $\Write' \in \Writes{\Seq}$ with
$\Location{\Write'} = \Location{\Write}$ we have
$\Write' \leq_{\Seq} \Write$. We can then say that
$\Write$ is the active write of the variable $\Location{\Write}$
in $\Seq$.

\Paragraph{Held variables.}
A variable $x \in \Globals$ is \emph{held} in $\Seq$
if there exists a read $\Read \in \Reads{X} \setminus \Events{\Seq}$
with $\Location{\Read} = x$ such that for each its good-write
$\Write \in \GoodWrites(\Read)$ we have $\Write \in \Seq$.
In such a case we say that $\Read$ \emph{holds} $x$ in $\Seq$.
Note that several distinct reads may hold a single variable in $\Seq$.

\Paragraph{Executable events.}
An event $\Event \in \Events{X} \setminus \Events{\Seq}$ is
\emph{executable} in $\Seq$ if $\Events{\Seq} \cup \{\Event\}$
is a lower set of $(X,\TO)$ and the following hold.
\begin{enumerate}[noitemsep,topsep=0pt,partopsep=0px]
\item If $\Event$ is a read, it has an active good-write
$\Write \in \GoodWrites(\Event)$ in $\Seq$.
\item If $\Event$ is a write, its variable $\Location{\Event}$
is not held in $\Seq$.
\end{enumerate}

\Paragraph{Memory maps.}
A \emph{memory map} of $\Seq$ is a function
from global variables to thread indices
$\MemoryMap_{\Seq}\colon \Globals \to [k]$ where for each
variable $x \in \Globals$,
the map $\MemoryMap_{\Seq}(x)$
captures the thread of the active write of $x$ in $\Seq$.

\Paragraph{Witness states.}
The sequence $\Seq$ is a \emph{witness prefix} if the following hold.
\begin{enumerate}[noitemsep,topsep=0pt,partopsep=0px]
\item $\Seq$ is a witness of $(\Events{\Seq}, \,\GoodWrites\Project \Reads{\Seq})$.
\item For each $\Read \in X \setminus \Reads{\Seq}$ that holds its
variable $\Location{\Read}$ in $\Seq$, one of its good-writes
$\Write \in \GoodWrites(\Read)$ is active in $\Seq$.
\end{enumerate}
Intuitively, $\Seq$ is a witness prefix if it satisfies all $\VSC$
requirements modulo its events, and if each read not in
$\Seq$ has at least one good-write still available to
read-from in potential extensions of $\Seq$.
For a witness prefix $\Seq$ we call its corresponding
event set and memory map a \emph{witness state}.

\begin{figure}[h]
  \raggedright
  \begin{minipage}{0.42\textwidth}
    \begin{tikzpicture}[thick,
    pre/.style={<-,shorten >= 2pt, shorten <=2pt, very thick},
    post/.style={->,shorten >= 2pt, shorten <=2pt,  very thick},
    seqtrace/.style={->, line width=1},
    aux_seqtrace/.style={->, line width=1, draw=gray},
    und/.style={very thick, draw=gray},
    event/.style={rectangle, minimum height=5mm, draw=black, fill=white, minimum width=8mm,   line width=1pt, inner sep=2, font={\footnotesize}},
    aux_event/.style={event, draw=gray},
    virt/.style={circle,draw=black!50,fill=black!20, opacity=0},
    bad/.style={preaction={fill, white}, pattern color=red!20, pattern=north east lines},
    good/.style={preaction={fill, white}, pattern color=green!40, pattern=north west lines},
    enabled/.style={event, draw=gray, dashed, fill=black!05},
    isLabel/.style={rectangle, fill opacity=0.5, fill=white, text opacity=1},
    recnode/.style={circle, draw=black},
    ]

    \newcommand{\xstep}{2.6}
    \newcommand{\ystep}{0.78}


    \newcommand{\xmove}{0.}
    \newcommand{\ymove}{0.}
    \node[]                  (t1_0)   at (\xmove + 0*\xstep, \ymove + -0.2*\ystep) {\small$\Process_1$};
    \node[]                  (t1_end) at (\xmove + 0*\xstep, \ymove + -5.5*\ystep) {};
    \node[]                  (t2_0)   at (\xmove + 0.5*\xstep, \ymove + -0.2*\ystep) {\small$\Process_2$};
    \node[]                  (t2_end) at (\xmove + 0.5*\xstep, \ymove + -5.5*\ystep) {};
    \node[]                  (t3_0)   at (\xmove + 1*\xstep, \ymove + -0.2*\ystep) {\small$\Process_3$};
    \node[]                  (t3_end) at (\xmove + 1*\xstep, \ymove + -5.5*\ystep) {};
    \node[]                  (t4_0)   at (\xmove + 1.5*\xstep, \ymove + -0.2*\ystep) {\small$\Process_4$};
    \node[]                  (t4_end) at (\xmove + 1.5*\xstep, \ymove + -5.5*\ystep) {};
    \draw[seqtrace]          (t1_0) to (t1_end);
    \draw[seqtrace]          (t2_0) to (t2_end);
    \draw[seqtrace]          (t3_0) to (t3_end);
    \draw[seqtrace]          (t4_0) to (t4_end);
    \node[event]             (t1_e1)  at (\xmove + 0*\xstep, \ymove + -2.5*\ystep) {\textcolor{\myblue}{$\Write_x$}};
    \node[event, enabled]    (t1_e2)  at (\xmove + 0*\xstep, \ymove + -3.5*\ystep) {\textcolor{\myblue}{$\Read_x$}};
    \node[event, enabled]    (t2_e1)  at (\xmove + 0.5*\xstep, \ymove + -3.5*\ystep) {\textcolor{\myblue}{$\Write'_x$}};
    \node[event, enabled]    (t2_e2)  at (\xmove + 0.5*\xstep, \ymove + -4.5*\ystep) {\textcolor{\myred}{$\Read_y$}};
    \node[event]             (t3_e1)  at (\xmove + 1*\xstep, \ymove + -1.5*\ystep) {\textcolor{\myred}{$\Write_y$}};
    \node[event, enabled]    (t3_e2)  at (\xmove + 1*\xstep, \ymove + -3.5*\ystep) {\textcolor{\myred}{$\bad{\Read_y}$}};
    \node[event]             (t4_e1)  at (\xmove + 1.5*\xstep, \ymove + -1.0*\ystep) {\textcolor{\myblue}{$\bad{\Write_x}$}};
    \node[event]             (t4_e2)  at (\xmove + 1.5*\xstep, \ymove + -2.0*\ystep) {\textcolor{\myblue}{$\bad{\Read_x}$}};
    \node[event, enabled]    (t4_e3)  at (\xmove + 1.5*\xstep, \ymove + -3.5*\ystep) {\textcolor{\myred}{$\bad{\Write_y}$}};
    \draw[post, \darkred, dashed]   (t4_e1) to[out=180, in=160, distance=0.4cm] (t4_e2);
    \draw[post, dotted, \darkgreen]   (t1_e1) to[out=0, in=30, distance=0.4cm] (t1_e2);
    \draw[post, dotted, \darkgreen]   (t2_e1) to[out=180, in=0] (t1_e2);
    \draw[post, dotted, \darkgreen]   (t3_e1) to[out=220, in=30] (t2_e2);
    \draw[post, dotted, \darkgreen]   (t4_e3) to[out=180, in=0] (t3_e2);

    \end{tikzpicture}
  \end{minipage}
  \begin{minipage}{0.57\textwidth}
    \caption{
      Event set $X$, and the good-writes function $\GoodWrites$ denoted by the green
      dotted edges. The solid nodes are ordered vertically as they
      appear in $\Seq$. The grey dashed nodes are in $X \setminus \Events{\Seq}$.
      Events $\textcolor{\myblue}{\Read_x}$ and
      $\textcolor{\myblue}{\Write'_x}$ are executable in $\Seq$.
      Event \textcolor{\myred}{$\bad{\Read_y}$} is not, its good-write
      is not active in $\Seq$. Event \textcolor{\myred}{$\bad{\Write_y}$}
      is also not executable, as its variable \textcolor{\myred}{$y$}
      is held by \textcolor{\myred}{$\Read_y$}. The memory map of $\Seq$
      is $\MemoryMap_{\Seq}(\textcolor{\myblue}{x}) = 1$ and
      $\MemoryMap_{\Seq}(\textcolor{\myred}{y}) = 3$. $\Seq$ is
      a witness prefix, and $\Events{\Seq}$ with $\MemoryMap_{\Seq}$
      together form its witness state.
    }
    \label{fig:vscnotation}
  \end{minipage}
\end{figure}
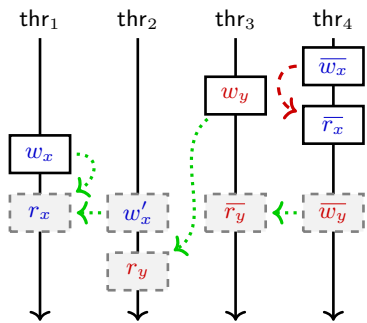

\cref{fig:vscnotation} provides an example illustrating
the above concepts, where for brevity of presentation,
the variables are subscripted and the values are not displayed.

\begin{algorithm}[h]
\small
\SetInd{0.4em}{0.4em}
\DontPrintSemicolon
\caption{$\AlgoSC$}\label{algo:vsc}
\KwIn{
Proper event set $X$ and good-writes function $\GoodWrites\colon \Reads{X}\to 2^{\Writes{X}}$
}
\KwOut{
A witness $\Seq$ of $(X, \GoodWrites)$ if $(X, \GoodWrites)$ has a witness, else $\Seq=\bot$
}
\BlankLine
$\Worklist\gets \{\epsilon\}$; $\DoneSet\gets \{\epsilon\}$\label{algo:vsc_init}\\
\While{$\Worklist\neq \emptyset$}{\label{algo:vsc_main_while}
Extract a sequence $\Seq$ from $\Worklist$\label{algo:vsc_extract_worklist}\\
\lIf(\tcp*[f]{All events executed, witness found}){$\Events{\Seq}=X$}{\label{algo:vsc_test_done}
\Return{$\Seq$}
}
\ForEach{event $\Event$ executable in $\Seq$}{\label{algo:vsc_ifextend}
Let $\Seq_{\Event}\gets \Seq \Concat \Event$\tcp*[f]{Execute $\Event$}\label{algo:vsc_execute_event}\\
\uIf{$\not \exists\Seq'\in \DoneSet$ s.t. $\Events{\Seq_{\Event}}=\Events{\Seq'}$
and $\MemoryMap_{\Seq_{\Event}}=\MemoryMap_{\Seq'}$}{\label{algo:vsc_if_new}
Insert $\Seq_{\Event}$ in $\Worklist$ and in $\DoneSet$\tcp*[f]{New witness state reached}\label{algo:vsc_insert_worklist}\\
}
}
}
\Return{$\bot$}\tcp*[f]{No witness exists}\label{algo:vsc_no_witness}
\end{algorithm}

\Paragraph{Algorithm.}
We are now ready to describe our algorithm $\AlgoSC$, in
\cref{algo:vsc} we present the pseudocode.
We attempt to construct a witness of $(X, \GoodWrites)$ by
enumerating the witness states reachable by the following process.
We start (\cref{algo:vsc_init}) with an empty sequence $\epsilon$ as the
first witness prefix (and state). We maintain a worklist $\Worklist$
of so-far unprocessed witness prefixes, and a set $\DoneSet$ of
reached witness states. Then we iteratively obtain new witness prefixes
(and states) by considering an already obtained
prefix (\cref{algo:vsc_extract_worklist})
and extending it with each possible executable event
(\cref{algo:vsc_execute_event}).
Crucially, when we arrive at a sequence $\Seq_{\Event}$, we include it
only if no sequence $\Seq'$ with equal corresponding
witness state has been reached yet (\cref{algo:vsc_if_new}).
We stop when we successfully create a witness (\cref{algo:vsc_test_done})
or when we process all reachable witness
states (\cref{algo:vsc_no_witness}).

\Paragraph{Correctness and Complexity.}
We now highlight the correctness and complexity properties of $\AlgoSC$, while we refer to \cref{sec:app_verification} for the proofs.
The soundness follows straightforwardly by the fact that each sequence in $\Worklist$
is a witness prefix.
This follows from a simple inductive argument
that extending a witness prefix with an executable event
yields another witness prefix.
The completeness follows from the fact that given two witness prefixes
$\Seq_1$ and $\Seq_2$ with equal induced witness state, these prefixes
are ``equi-extendable'' to a witness. Indeed, if a suffix $\Seq^*$ exists
such that $\Seq_1 \Concat \Seq^*$ is a witness of $(X, \GoodWrites)$,
then $\Seq_2 \Concat \Seq^*$ is also a witness of $(X, \GoodWrites)$.
The time complexity of $\AlgoSC$ is bounded by $O(n^{k + 1} \cdot k^{d + 1})$,
for $n$ events, $k$ threads and $d$ variables.
The bound follows from the fact that there are
at most $n^{k} \cdot k^{d}$ pairwise distinct witness states.
We thus have the following theorem.

\begin{restatable}{theorem}{themvsc}\label{them:vsc}
$\VSC$ for $n$ events, $k$ threads and $d$ variables is solvable in \mbox{$O(n^{k + 1} \cdot k^{d + 1})$} time.
Moreover, if each variable is written by only one thread, $\VSC$ is solvable in \mbox{$O(n^{k + 1} )$} time.
\end{restatable}

\Paragraph{Implications.}
We now highlight some important implications of \cref{them:vsc}.
Although $\VSC$ is NP-hard~\cite{Gibbons97}, the theorem shows that the problem is parameterizable in $k+d$,
and thus in polynomial time when both parameters are bounded.
Moreover, even when only $k$ is bounded, the problem is fixed-parameter tractable in $d$, meaning that $d$ only exponentiates a constant as opposed to $n$ (e.g., we have a polynomial bound even when $d=\log n$ ).
Finally, the algorithm is polynomial for a fixed number of threads regardless of $d$,
when every memory location is written by only
one thread (e.g., in producer-consumer settings, or in the
concurrent-read-exclusive-write (CREW)
concurrency model).
These important facts brought forward by \cref{them:vsc} indicate that $\VSC$ is likely to be efficiently solvable in many practical settings, which in turn makes $\RVF$ a good equivalence for SMC.

\subsection{Practical heuristics for $\AlgoSC$ in SMC}\label{sec:verification_efficiency}

We now turn our attention to some practical heuristics that are expected to further improve the performance of $\AlgoSC$
in the context of SMC.

\Paragraph{1. Limiting the Search Space.}
We employ two straightforward improvements to $\AlgoSC$ that significantly
reduce the search space in practice.
Consider the for-loop in
\cref{algo:vsc_ifextend} of~\cref{algo:vsc}
enumerating the possible extensions of $\Seq$.
This enumeration can be sidestepped by the following
two greedy approaches.
\begin{enumerate}[noitemsep,topsep=0pt,partopsep=0px]
\item If there is a read $\Read$ executable in $\Seq$, then
extend $\Seq$ with $\Read$ and do not enumerate other options.
\item Let $\bad{\Write}$ be an active write in $\Seq$ such that
$\bad{\Write}$ is not a good-write of any
$\Read \in \Reads{X} \setminus \Events{\Seq}$.
Let $\Write \in \Writes{X} \setminus \Events{\Seq}$ be a write of the
same variable ($\Location{\Write} = \Location{\bad{\Write}}$),
note that $\Write$ is executable in $\Seq$.
If $\Write$ is also not a good-write of any
$\Read \in \Reads{X} \setminus \Events{\Seq}$,
then extend $\Seq$ with $\Write$ and do not enumerate other options.
\end{enumerate}
The enumeration of \cref{algo:vsc_ifextend} then proceeds only
if neither of the above two techniques can be applied for $\Seq$.
This extension of $\AlgoSC$ preserves completeness
(not only when used during SMC, but in general),
and it can be significantly faster in practice.
For clarity of presentation we do not fully formalize this
extended version, as its worst-case complexity remains the same.

\Paragraph{2. Closure.}
We introduce \emph{closure}, a low-cost filter for early
detection of $\VSC$ instances $(X, \GoodWrites)$ with no witness.
The notion of closure, its beneficial properties and construction
algorithms are well-studied for the
\emph{reads-from consistency verification}
problems~\cite{Chalupa17,Abdulla19,Pavlogiannis20}, i.e., problems
where a desired reads-from function is provided as input instead of
a desired good-writes function $\GoodWrites$.
Further, the work of~\cite{Chatterjee19} studies closure with respect
to a good-writes function, but only for partial orders of
Mazurkiewicz width 2 (i.e., for partial orders with no
triplet of pairwise conflicting and pairwise unordered events).
Here we define closure for all good-writes instances
$(X, \GoodWrites)$, with the underlying partial order
(in our case, the program order $\TO$) of arbitrary
Mazurkiewicz width.

Given a $\VSC$ instance $(X, \GoodWrites)$, its closure
$P(X)$ is the weakest partial order that refines the program
order ($P\Refines \TO\Project X$) and further satisfies the
following conditions.
Given a read $\Read \in \Reads{X}$,
let $Cl(\Read) = \GoodWrites(\Read) \cap \VisibleWrites_{P}(\Read)$.
The following must hold.
\begin{enumerate}[noitemsep,topsep=0pt,partopsep=0px]
\item $Cl(\Read) \neq \emptyset$.
\item If $(Cl(\Read), \,P\Project Cl(\Read))$ has a least element
$\Write$, then $\Write <_{P} \Read$.
\item If $(Cl(\Read), \,P\Project Cl(\Read))$ has a greatest element
$\Write$, then for each $\bad{\Write} \in \Writes{X} \setminus \GoodWrites(\Read)$
with $\Confl{\Read}{\bad{\Write}}$, if $\bad{\Write} <_{P} \Read$ then
$\bad{\Write} <_{P} \Write$.
\item For each $\bad{\Write} \in \Writes{X} \setminus \GoodWrites(\Read)$
with $\Confl{\Read}{\bad{\Write}}$, if each
$\Write \in Cl(\Read)$ satisfies $\Write <_{P} \bad{\Write}$,
then we have $\Read <_{P} \bad{\Write}$.
\end{enumerate}

If $(X, \GoodWrites)$ has no closure
(i.e., there is no $P$ with the above conditions), then
$(X, \GoodWrites)$ provably has no witness.
If $(X, \GoodWrites)$ has closure $P$, then
each witness $\Seq$ of $\VSC(X, \GoodWrites)$
provably refines $P$ (i.e., $\Seq\Refines P$).

Finally, we explain how closure can be used by $\AlgoSC$.
Given an input $(X, \GoodWrites)$,
the closure procedure is carried out before $\AlgoSC$ is called.
Once the closure $P$ of $(X, \GoodWrites)$ is constructed,
since each solution of $\VSC(X, \GoodWrites)$ has to refine $P$,
we restrict $\AlgoSC$ to only consider sequences refining $P$.
This is ensured by an extra condition in~\cref{algo:vsc_ifextend}
of~\cref{algo:vsc}, where we proceed with an event $\Event$
only if it is minimal in $P$ restricted to events not yet
in the sequence.
This preserves completeness, while further reducing
the search space to consider for $\AlgoSC$.

\Paragraph{3. $\AlgoSC$ guided by auxiliary trace.}
In our SMC approach, each time we generate a $\VSC$ instance
$(X, \GoodWrites)$, we further have available an auxiliary
trace $\widetilde{\Trace}$. In $\widetilde{\Trace}$,
either all-but-one, or all, good-writes conditions of $\GoodWrites$
are satisfied.
If all good writes in $\GoodWrites$ are satisfied, we already have $\widetilde{\Trace}$
as a witness of $(X, \GoodWrites)$ and hence we do not need to run
$\AlgoSC$ at all.
On the other hand, if case all-but-one are satisfied,
we use $\widetilde{\Trace}$ to guide the search of $\AlgoSC$,
as described below.

We guide the search by deciding the order in which we process the sequences of
the worklist $\Worklist$ in~\cref{algo:vsc}.
We use the auxiliary trace $\widetilde{\Trace}$
with $\Events{\widetilde{\Trace}} = X$.
We use $\Worklist$ as a last-in-first-out
stack, that way we search for a witness in a depth-first fashion.
Then, in~\cref{algo:vsc_ifextend} of~\cref{algo:vsc}
we enumerate the extension events in the reverse order of how
they appear in $\widetilde{\Trace}$. We enumerate in reverse order, as
each resulting extension is pushed into our worklist $\mathcal{S}$,
which is a stack (last-in-first-out). As a result,
in~\cref{algo:vsc_extract_worklist} of the
subsequent iterations of the main while loop, we pop extensions
from $\mathcal{S}$ in order induced by $\widetilde{\sigma}$.

\section{Stateless Model Checking}\label{sec:smc}

We are now ready to present our SMC algorithm
$\RVFSMC$ that uses $\RVF$ to model check a concurrent program.
$\RVFSMC$ is a sound and complete algorithm for local safety properties,
i.e., it is guaranteed to discover all local states that each thread visits.

$\RVFSMC$ is a recursive algorithm. Each recursive call of $\RVFSMC$
is argumented by a tuple
$(X, \GoodWrites, \Trace, \NegativeAnnotation)$ where:
\begin{enumerate}[noitemsep,topsep=0pt,partopsep=0px]
\item $X$ is a proper set of events.
\item $\GoodWrites \colon \Reads{X} \to 2^{\Writes{X}}$
is a desired good-writes function.
\item $\Trace$ is a valid trace that is a witness of $(X, \GoodWrites)$.
\item $\NegativeAnnotation \colon \SysReads \to \Threads \to \Nats$
is a partial function called \emph{causal map} that tracks implicitly,
for each read $\Read$, the writes that have already been considered
as reads-from sources of $\Read$.
\end{enumerate}
Further, we maintain a function
$\ancestors \colon \Reads{X} \to \{ \true, \false \}$,
where for each read $\Read \in \Reads{X}$, $\ancestors(\Read)$ stores
a boolean \emph{backtrack signal} for $\Read$.
We now provide details on the notions of causal maps and backtrack signals.

\Paragraph{Causal maps.}
The causal map $\NegativeAnnotation$ serves to ensure that no more than
one maximal trace is explored per $\RVF$ partitioning class.
Given a read $\Read \in \Enabled(\Trace)$ enabled in a trace $\Trace$,
we define $\forbids_{\Trace}^{\NegativeAnnotation}(\Read)$ as
the set of
writes in $\Trace$ such that
$\NegativeAnnotation$ forbids $\Read$ to read-from them. Formally,
$\forbids_{\Trace}^{\NegativeAnnotation}(\Read) = \emptyset$
if $\Read \not\in \Domain(\NegativeAnnotation)$, otherwise
$\forbids_{\Trace}^{\NegativeAnnotation}(\Read) =
\{
\Write \in \Writes{\Trace} \;|\;
\Write \textrm{ is within first }\NegativeAnnotation(\Read)(\Proc{\Write}) \textrm{ events of }\Trace_{\Process}
\}$.
We say that a trace $\Trace$ \emph{satisfies} $\NegativeAnnotation$
if for each $\Read \in \Reads{\Trace}$ we have
$\RF{\Trace}(\Read) \not\in \forbids_{\Trace}^{\NegativeAnnotation}(\Read)$.

\Paragraph{Backtrack signals.}
Each call of $\RVFSMC$ (with its $\GoodWrites$)
operates with a trace $\widetilde{\Trace}$ satisfying $\GoodWrites$
that has only reads as enabled events. Consider one of those enabled
reads $\Read \in \Enabled(\widetilde{\Trace})$. Each
maximal trace satisfying $\GoodWrites$ shall contain $\Read$, and
further, one of the following two cases is true:
\begin{enumerate}[noitemsep,topsep=0pt,partopsep=0px]
\item In all maximal traces $\Trace'$ satisfying $\GoodWrites$,
we have that $\Read$ reads-from some write of $\Writes{\widetilde{\Trace}}$ in $\Trace'$.
\item There exists a maximal trace $\Trace'$ satisfying $\GoodWrites$,
such that $\Read$ reads-from a write not in $\Writes{\widetilde{\Trace}}$ in $\Trace'$.
\end{enumerate}
Whenever we can prove that the first above case is true for $\Read$,
we can use this fact to prune away some recursive calls of $\RVFSMC$
while maintaining completeness.
Specifically, we leverage the following crucial lemma.

\begin{restatable}{lemma}{lembacktrack}\label{lem:lembacktrack}
Consider a call $\RVFSMC(X, \GoodWrites, \Trace, \NegativeAnnotation)$
and a trace $\widetilde{\Trace}$ extending $\Trace$ maximally such
that no event of the extension is a read.
Let $\Read \in \Enabled(\widetilde{\Trace})$
such that $\Read \not\in \Domain(\NegativeAnnotation)$.
If there exists a trace $\Trace'$ that
(i) satisfies $\GoodWrites$ and $\NegativeAnnotation$, and
(ii) contains $\Read$ with $\RF{\Trace'}(\Read) \not\in \Writes{\widetilde{\Trace}}$,
then there exists a trace $\bad{\Trace}$ that
(i) satisfies $\GoodWrites$ and $\NegativeAnnotation$,
(ii) contains $\Read$ with $\RF{\bad{\Trace}}(\Read) \in \Writes{\widetilde{\Trace}}$, and
(iii) contains a write $\Write \not\in \Writes{\widetilde{\Trace}}$
with $\Confl{\Read}{\Write}$ and $\Process(\Read) \neq \Process(\Write)$.
\end{restatable}

We then compute a boolean \emph{backtrack signal} for a given
$\RVFSMC$ call and read $\Read \in \Enabled(\widetilde{\Trace})$
to capture satisfaction of the consequent of \cref{lem:lembacktrack}.
If the computed backtrack signal is false, we can safely stop the
$\RVFSMC$ exploration of this specific call and backtrack to its
recursion parent.

\begin{algorithm}[h]
\small
\SetInd{0.4em}{0.4em}
\DontPrintSemicolon
\caption{$\RVFSMC(X, \GoodWrites, \Trace, \NegativeAnnotation)$}\label{algo:exploration}
\KwIn{Proper set of events $X$,
good-writes function $\GoodWrites$,
valid trace $\Trace$ that is a witness of $(X, \GoodWrites)$,
causal map $\NegativeAnnotation$.
}
\BlankLine
$\widetilde{\Trace} \gets \Trace \Concat \widehat{\Trace}$
where $\widehat{\Trace}$ extends $\Trace$ maximally such that no event of $\widehat{\Trace}$ is a read\label{line:exploration_extendtrace}\\
\ForEach(\tcp*[f]{All extension events are writes})
{$\Write\in \Events{\widehat{\Trace}}$}
{\label{line:exploration_foreachextwrite}
  \ForEach(\tcp*[f]{All ancestor mutations are reads})
  {$\Read\in \Domain(\ancestors)$}
  {\label{line:exploration_foreachancestor}
    \If(\tcp*[f]{Potential new source for $\Read$ to read-from})
    {$\Confl{\Read}{\Write}$ and $\Process(\Read) \neq \Process(\Write)$}
    {\label{line:exploration_ifnewsource}
      $\ancestors(\Read) \gets \true$\label{line:exploration_setbacktrack}\tcp*[f]{Set backtrack signal to true}\\
    }
  }
}
$\mutate \gets \epsilon$\tcp*[f]{Construct a sequence of enabled reads}\label{line:exploration_initmutate}\\
\ForEach(\tcp*[f]{Enabled events in $\widetilde{\Trace}$ are reads})
{$\Read \in \Enabled(\widetilde{\Trace})$}
{\label{line:exploration_foreachenabled}
  \If(\tcp*[f]{Causal map $\NegativeAnnotation$ is defined for $\Read$})
  {$\Read \in \Domain(\NegativeAnnotation)$}
  {\label{line:exploration_ifcausaldef}
    $\mutate \gets \mutate \Concat \Read$\tcp*[f]{Insert $\Read$ to the end of $\mutate$}\label{line:exploration_lin}\\
  }
  \Else(\tcp*[f]{Causal map $\NegativeAnnotation$ is undefined for $\Read$})
  {\label{line:exploration_ifcausalndef}
    $\mutate \gets \Read \Concat \mutate$\tcp*[f]{Insert $\Read$ to the beginning of $\mutate$}\label{line:exploration_lin}\\
  }
}
$\backtrack \gets \true$\label{line:exploration_backinit}\\
\While
{$\backtrack = \true$ and $\mutate \neq \epsilon$}
{\label{line:exploration_whileback}
  $\Read \gets$ pop front of $\mutate$\label{line:exploration_popread}\tcp*[f]{Process next read of $\mutate$}\\
  \If
  {$\Read \not\in \Domain(\NegativeAnnotation)$}
  {\label{line:exploration_ifcausalnotdef}
    $\backtrack \gets \false$\label{line:exploration_backfalse}\\
  }
  $\CandidateSet_{\Read} \gets \VisibleWrites_{\TO\Project \Events{\widetilde{\Trace}}}(\Read) \setminus
                               \forbids_{\widetilde{\Trace}}^{\NegativeAnnotation}(\Read)$
                               \tcp*[f]{Visible writes not forbidden by $\NegativeAnnotation$}\label{line:exploration_mutwrites}\\
  $\ValueDomain_{\Read}\gets \{ \Value_{\widetilde{\Trace}}(\Write):~\Write\in \CandidateSet_{\Read} \}$
  \tcp*[f]{The set of values that $\Read$ may read}\label{line:exploration_mutvalues}\\
  \ForEach(\tcp*[f]{Process each value})
  {$v \in \ValueDomain_{\Read}$}
  {\label{line:exploration_foreachvalue}
    $X' \gets X \cup \Events{\widetilde{\Trace}} \cup \{\Read\}$\tcp*[f]{New event set}\label{line:exploration_newevents}\\
    $\GoodWrites' \gets \GoodWrites \cup \{  (\Read, \{\; \Write \in \CandidateSet_{\Read} \,|\, \Value_{\widetilde{\Trace}}(\Write) = v \;\}) \}$
    \tcp*[f]{New good-writes}\label{line:exploration_newgw}\\
    $\Trace' \gets \AlgoSC(X', \GoodWrites')$\tcp*[f]{$\AlgoSC$ guided by $\widetilde{\Trace} \Concat \Read$}\label{line:exploration_vsc}\\
    \If(\tcp*[f]{$(X', \GoodWrites')$ has a witness})
    {$\Trace' \neq \bot$}
    {\label{line:exploration_ifwitness}
      $\NegativeAnnotation' \gets \NegativeAnnotation$\label{line:exploration_copycausal}\\
      $\ancestors(\Read) \gets \backtrack$\tcp*[f]{Record ancestor}\label{line:exploration_recancestor}\\
      $\RVFSMC(X', \GoodWrites', \Trace', \NegativeAnnotation')$\label{line:exploration_recursivecall}\\
      $\backtrack \gets \ancestors(\Read)$\tcp*[f]{Retrieve backtrack signal}\label{line:exploration_retrieveback}\\
      delete $\Read$ from $\ancestors$\tcp*[f]{Unrecord ancestor}\label{line:exploration_unrecancestor}\\
    }
  }
  \ForEach(\tcp*[f]{Update causal map $\NegativeAnnotation(\Read)$ for each thread})
  {$\Process \in \Threads$}
  {\label{line:exploration_updatethread}
    $\NegativeAnnotation(\Read)(\Process)\gets |\Events{\widetilde{\Trace}}_{\Process}|$
    \tcp*[f]{Number of events of $\Process$ in $\widetilde{\Trace}$}\label{line:exploration_updatecausal}\\
  }
}
\end{algorithm}

\Paragraph{Algorithm.}
We are now ready to describe our algorithm $\RVFSMC$ in detail,
\cref{algo:exploration} captures the pseudocode of
$\RVFSMC(X, \GoodWrites, \Trace, \NegativeAnnotation)$.
First, in \cref{line:exploration_extendtrace} we extend $\Trace$ to
$\widetilde{\Trace}$ maximally such that no event of the extension is a read.
Then in Lines~\ref{line:exploration_foreachextwrite}--\ref{line:exploration_setbacktrack}
we update the backtrack signals for ancestors of our current
recursion call. After this, in
Lines~\ref{line:exploration_initmutate}--\ref{line:exploration_lin}
we construct a sequence of reads enabled in $\widetilde{\Trace}$.
Finally, we proceed with the main while-loop in \cref{line:exploration_whileback}.
In each while-loop iteration we process an enabled read $\Read$
(\cref{line:exploration_popread}),
and we perform no more while-loop iterations in case we
receive a $\false$ backtrack signal for $\Read$.
When processing $\Read$, first we collect its viable reads-from sources
in \cref{line:exploration_mutwrites}, then we group the sources
by value they write in \cref{line:exploration_mutvalues}, and then
in iterations of the for-loop in \cref{line:exploration_foreachvalue}
we consider each value-group. In
\cref{line:exploration_newevents} we form the event set,
and in \cref{line:exploration_newgw} we form the good-write function
that designates the value-group as the good-writes of $\Read$.
In \cref{line:exploration_vsc} we use $\AlgoSC$ to generate a witness,
and in case it exists, we recursively call $\RVFSMC$
in \cref{line:exploration_recursivecall} with the newly obtained
events, good-write constraint for $\Read$, and witness.

To preserve completeness of $\RVFSMC$, the backtrack-signals technique
can be utilized only for reads $\Read$ with undefined causal map
$\Read \not\in \Domain(\NegativeAnnotation)$
(cf. \cref{lem:lembacktrack}). The order of the enabled reads imposed by
Lines~\ref{line:exploration_initmutate}--\ref{line:exploration_lin}
ensures that subsequently, in iterations of the loop in
\cref{line:exploration_whileback} we first consider all the reads
where we can utilize the backtrack signals.
This is an insightful heuristic that often helps in practice,
though it does not improve the worst-case complexity.

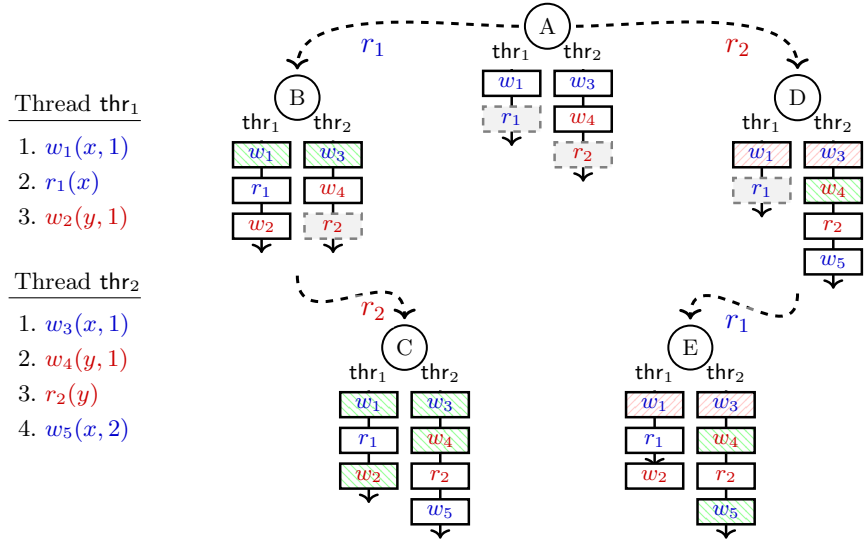
\begin{figure}[h]
  \begin{subfigure}{0.15\textwidth}
    \small
    \vspace{-2mm}
    \begin{align*}
      \text{Th}&\text{read}~\Process_{1}\\
      \hline\\[-1em]
      1.~& \textcolor{\myblue}{\Write_1(x, 1)}\\
      \\[-1.6em]
      2.~& \textcolor{\myblue}{\Read_1(x)}\\
      \\[-1.6em]
      3.~& \textcolor{\myred}{\Write_2(y, 1)}\\
    \end{align*}
    \vspace{-10mm}
    \begin{align*}
      \text{Th}&\text{read}~\Process_{2}\\
      \hline\\[-1em]
      1.~& \textcolor{\myblue}{\Write_3(x, 1)}\\
      \\[-1.6em]
      2.~& \textcolor{\myred}{\Write_4(y, 1)}\\
      \\[-1.6em]
      3.~& \textcolor{\myred}{\Read_2(y)}\\
      \\[-1.6em]
      4.~& \textcolor{\myblue}{\Write_5(x, 2)}\\
    \end{align*}
  \end{subfigure}
  \qquad
  \begin{subfigure}{0.75\textwidth}
    \centering
    \scalebox{0.95}{
    \begin{tikzpicture}[thick,
    pre/.style={<-,shorten >= 2pt, shorten <=2pt, very thick},
    post/.style={->,shorten >= 2pt, shorten <=2pt,  very thick},
    seqtrace/.style={->, line width=1},
    aux_seqtrace/.style={->, line width=1, draw=gray},
    und/.style={very thick, draw=gray},
    event/.style={rectangle, minimum height=3.5mm, draw=black, fill=white, minimum width=8mm,   line width=1pt, inner sep=2, font={\footnotesize}},
    aux_event/.style={event, draw=gray},
    virt/.style={circle,draw=black!50,fill=black!20, opacity=0},
    bad/.style={preaction={fill, white}, pattern color=red!20, pattern=north east lines},
    good/.style={preaction={fill, white}, pattern color=green!40, pattern=north west lines},
    enabled/.style={event, draw=gray, dashed, fill=black!05},
    isLabel/.style={rectangle, fill opacity=0.5, fill=white, text opacity=1},
    recnode/.style={circle, draw=black},
    ]

    \newcommand{\xstep}{2.}
    \newcommand{\ystep}{0.5}
    \newcommand{\yaux}{0.17727}


    \newcommand{\xmove}{0.}
    \newcommand{\ymove}{0.}
    \node[recnode]           (a)        at (\xmove + 0.25*\xstep, \ymove + 0.6*\ystep) {A};
    \node[]                  (a_t1_0)   at (\xmove + 0*\xstep, \ymove + -0.1*\ystep) {\small$\Process_1$};
    \node[]                  (a_t1_end) at (\xmove + 0*\xstep, \ymove + -3*\ystep) {};
    \node[]                  (a_t2_0)   at (\xmove + 0.5*\xstep, \ymove + -0.1*\ystep) {\small$\Process_2$};
    \node[]                  (a_t2_end) at (\xmove + 0.5*\xstep, \ymove + -4*\ystep) {};
    \draw[seqtrace]          (a_t1_0) to (a_t1_end);
    \draw[seqtrace]          (a_t2_0) to (a_t2_end);
    \node[event]             (a_t1_w1)  at (\xmove + 0*\xstep, \ymove + -1*\ystep) {\textcolor{\myblue}{$\Write_1$}};
    \node[event, enabled]    (a_t1_r1)  at (\xmove + 0*\xstep, \ymove + -2*\ystep) {\textcolor{\myblue}{$\Read_1$}};
    \node[event]             (a_t2_w3)  at (\xmove + 0.5*\xstep, \ymove + -1*\ystep) {\textcolor{\myblue}{$\Write_3$}};
    \node[event]             (a_t2_w4)  at (\xmove + 0.5*\xstep, \ymove + -2*\ystep) {\textcolor{\myred}{$\Write_4$}};
    \node[event, enabled]    (a_t2_r2)  at (\xmove + 0.5*\xstep, \ymove + -3*\ystep) {\textcolor{\myred}{$\Read_2$}};

    \renewcommand{\xmove}{-3.5}
    \renewcommand{\ymove}{-1.}
    \node[recnode]           (b)        at (\xmove + 0.25*\xstep, \ymove + 0.6*\ystep) {B};
    \node[]                  (b_t1_0)   at (\xmove + 0*\xstep, \ymove + -0.1*\ystep) {\small$\Process_1$};
    \node[]                  (b_t1_end) at (\xmove + 0*\xstep, \ymove + -4*\ystep) {};
    \node[]                  (b_t2_0)   at (\xmove + 0.5*\xstep, \ymove + -0.1*\ystep) {\small$\Process_2$};
    \node[]                  (b_t2_end) at (\xmove + 0.5*\xstep, \ymove + -4*\ystep) {};
    \node[]                  (b_mid_end) at (\xmove + 0.25*\xstep, \ymove + -4*\ystep) {};
    \draw[seqtrace]          (b_t1_0) to (b_t1_end);
    \draw[seqtrace]          (b_t2_0) to (b_t2_end);
    \node[event, good]       (b_t1_w1)  at (\xmove + 0*\xstep, \ymove + -1*\ystep) {\textcolor{\myblue}{$\Write_1$}};
    \node[event]             (b_t1_r1)  at (\xmove + 0*\xstep, \ymove + -2*\ystep) {\textcolor{\myblue}{$\Read_1$}};
    \node[event]             (b_t1_w2)  at (\xmove + 0*\xstep, \ymove + -3*\ystep) {\textcolor{\myred}{$\Write_2$}};
    \node[event, good]       (b_t2_w3)  at (\xmove + 0.5*\xstep, \ymove + -1*\ystep) {\textcolor{\myblue}{$\Write_3$}};
    \node[event]             (b_t2_w4)  at (\xmove + 0.5*\xstep, \ymove + -2*\ystep) {\textcolor{\myred}{$\Write_4$}};
    \node[event, enabled]    (b_t2_r2)  at (\xmove + 0.5*\xstep, \ymove + -3*\ystep) {\textcolor{\myred}{$\Read_2$}};
    \draw[post, dashed]   (a) to[out=180, in=90, distance=1cm] (b)
                          node [midway, xshift=-55pt, yshift=-0pt, rotate=0, isLabel] {\large {\textcolor{\myblue}{$\Read_1$}}};

    \renewcommand{\xmove}{3.5}
    \renewcommand{\ymove}{-1.}
    \node[recnode]           (d)        at (\xmove + 0.25*\xstep, \ymove + 0.6*\ystep) {D};
    \node[]                  (d_t1_0)   at (\xmove + 0*\xstep, \ymove + -0.1*\ystep) {\small$\Process_1$};
    \node[]                  (d_t1_end) at (\xmove + 0*\xstep, \ymove + -3*\ystep) {};
    \node[]                  (d_t2_0)   at (\xmove + 0.5*\xstep, \ymove + -0.1*\ystep) {\small$\Process_2$};
    \node[]                  (d_t2_end) at (\xmove + 0.5*\xstep, \ymove + -5*\ystep) {};
    \node[]                  (d_mid_end) at (\xmove + 0.25*\xstep, \ymove + -4.5*\ystep) {};
    \draw[seqtrace]          (d_t1_0) to (d_t1_end);
    \draw[seqtrace]          (d_t2_0) to (d_t2_end);
    \node[event, bad]        (d_t1_w1)  at (\xmove + 0*\xstep, \ymove + -1*\ystep) {\textcolor{\myblue}{$\Write_1$}};
    \node[event, enabled]    (d_t1_r1)  at (\xmove + 0*\xstep, \ymove + -2*\ystep) {\textcolor{\myblue}{$\Read_1$}};
    \node[event, bad]        (d_t2_w3)  at (\xmove + 0.5*\xstep, \ymove + -1*\ystep) {\textcolor{\myblue}{$\Write_3$}};
    \node[event, good]       (d_t2_w4)  at (\xmove + 0.5*\xstep, \ymove + -2*\ystep) {\textcolor{\myred}{$\Write_4$}};
    \node[event]             (d_t2_r2)  at (\xmove + 0.5*\xstep, \ymove + -3*\ystep) {\textcolor{\myred}{$\Read_2$}};
    \node[event]             (d_t2_w5)  at (\xmove + 0.5*\xstep, \ymove + -4*\ystep) {\textcolor{\myblue}{$\Write_5$}};
    \draw[post, dashed]   (a) to[out=0, in=90, distance=1cm] (d)
                          node [midway, xshift=90pt, yshift=-0pt, rotate=0, isLabel] {\large {\textcolor{\myred}{$\Read_2$}}};

    \renewcommand{\xmove}{-2.}
    \renewcommand{\ymove}{-4.5}
    \node[recnode]           (c)        at (\xmove + 0.25*\xstep, \ymove + 0.6*\ystep) {C};
    \node[]                  (c_t1_0)   at (\xmove + 0*\xstep, \ymove + -0.1*\ystep) {\small$\Process_1$};
    \node[]                  (c_t1_end) at (\xmove + 0*\xstep, \ymove + -4*\ystep) {};
    \node[]                  (c_t2_0)   at (\xmove + 0.5*\xstep, \ymove + -0.1*\ystep) {\small$\Process_2$};
    \node[]                  (c_t2_end) at (\xmove + 0.5*\xstep, \ymove + -5*\ystep) {};
    \draw[seqtrace]          (c_t1_0) to (c_t1_end);
    \draw[seqtrace]          (c_t2_0) to (c_t2_end);
    \node[event, good]       (c_t1_w1)  at (\xmove + 0*\xstep, \ymove + -1*\ystep) {\textcolor{\myblue}{$\Write_1$}};
    \node[event]             (c_t1_r1)  at (\xmove + 0*\xstep, \ymove + -2*\ystep) {\textcolor{\myblue}{$\Read_1$}};
    \node[event, good]       (c_t1_w2)  at (\xmove + 0*\xstep, \ymove + -3*\ystep) {\textcolor{\myred}{$\Write_2$}};
    \node[event, good]       (c_t2_w3)  at (\xmove + 0.5*\xstep, \ymove + -1*\ystep) {\textcolor{\myblue}{$\Write_3$}};
    \node[event, good]       (c_t2_w4)  at (\xmove + 0.5*\xstep, \ymove + -2*\ystep) {\textcolor{\myred}{$\Write_4$}};
    \node[event]             (c_t2_r2)  at (\xmove + 0.5*\xstep, \ymove + -3*\ystep) {\textcolor{\myred}{$\Read_2$}};
    \node[event]             (c_t2_w5)  at (\xmove + 0.5*\xstep, \ymove + -4*\ystep) {\textcolor{\myblue}{$\Write_5$}};
    \draw[post, dashed]   (b_mid_end) to[out=270, in=90, distance=1cm] (c)
                          node [midway, xshift=-55pt, yshift=-105pt, rotate=0, isLabel] {\large {\textcolor{\myred}{$\Read_2$}}};

    \renewcommand{\xmove}{2.}
    \renewcommand{\ymove}{-4.5}
    \node[recnode]           (e)        at (\xmove + 0.25*\xstep, \ymove + 0.6*\ystep) {E};
    \node[]                  (e_t1_0)   at (\xmove + 0*\xstep, \ymove + -0.1*\ystep) {\small$\Process_1$};
    \node[]                  (e_t1_end) at (\xmove + 0*\xstep, \ymove + -3*\ystep) {};
    \node[]                  (e_t2_0)   at (\xmove + 0.5*\xstep, \ymove + -0.1*\ystep) {\small$\Process_2$};
    \node[]                  (e_t2_end) at (\xmove + 0.5*\xstep, \ymove + -5*\ystep) {};
    \draw[seqtrace]          (e_t1_0) to (e_t1_end);
    \draw[seqtrace]          (e_t2_0) to (e_t2_end);
    \node[event, bad]        (e_t1_w1)  at (\xmove + 0*\xstep, \ymove + -1*\ystep) {\textcolor{\myblue}{$\Write_1$}};
    \node[event]             (e_t1_r1)  at (\xmove + 0*\xstep, \ymove + -2*\ystep) {\textcolor{\myblue}{$\Read_1$}};
    \node[event]             (e_t1_w2)  at (\xmove + 0*\xstep, \ymove + -3*\ystep) {\textcolor{\myred}{$\Write_2$}};
    \node[event, bad]        (e_t2_w3)  at (\xmove + 0.5*\xstep, \ymove + -1*\ystep) {\textcolor{\myblue}{$\Write_3$}};
    \node[event, good]       (e_t2_w4)  at (\xmove + 0.5*\xstep, \ymove + -2*\ystep) {\textcolor{\myred}{$\Write_4$}};
    \node[event]             (e_t2_r2)  at (\xmove + 0.5*\xstep, \ymove + -3*\ystep) {\textcolor{\myred}{$\Read_2$}};
    \node[event, good]       (e_t2_w5)  at (\xmove + 0.5*\xstep, \ymove + -4*\ystep) {\textcolor{\myblue}{$\Write_5$}};
    \draw[post, dashed]   (d_mid_end) to[out=270, in=90, distance=1cm] (e)
                          node [midway, xshift=90pt, yshift=-110pt, rotate=0, isLabel] {\large {\textcolor{\myblue}{$\Read_1$}}};

    \end{tikzpicture}
    }
  \end{subfigure}
  \caption{
    $\RVFSMC$ (\cref{algo:exploration}).
    Circles represent nodes of the recursion tree. Below each circle is its
    corresponding event set $\Events{\widetilde{\Trace}}$
    and the enabled reads (dashed grey).
    Writes with green background are good-writes $(\GoodWrites)$ of its corresponding-variable read.
    Writes with red background are forbidden by $\NegativeAnnotation$ for its corresponding-variable read.
    Dashed arrows represent recursive calls.
  }
  \label{fig:rvfsmc}
\end{figure}

\smallskip\noindent{\em Example.}
\cref{fig:rvfsmc} displays a simple concurrent program on the left,
and its corresponding $\RVFSMC$ (\cref{algo:exploration}) run on the right.
We start with $\RVFSMC(\emptyset, \emptyset, \epsilon, \emptyset)$
(A). By performing the extension (\cref{line:exploration_extendtrace})
we obtain the events and enabled reads as shown below (A).
First we process read
\textcolor{\myblue}{$\Read_1$} (\cref{line:exploration_popread}).
The read can read-from
\textcolor{\myblue}{$\Write_1$} and \textcolor{\myblue}{$\Write_3$},
both write the same value so they are grouped together as good-writes
of \textcolor{\myblue}{$\Read_1$}.
A witness is found and a recursive call to (B) is performed.
In (B), the only enabled event is
\textcolor{\myred}{$\Read_2$}. It can read-from
\textcolor{\myred}{$\Write_2$} and \textcolor{\myred}{$\Write_4$},
both write the same value so they are grouped for
\textcolor{\myred}{$\Read_2$}. A witness is found, a recursive call
to (C) is performed, and (C) concludes with a maximal trace.
Crucially, in (C) the event \textcolor{\myblue}{$\Write_5$} is
discovered, and since it is a potential new reads-from source for
\textcolor{\myblue}{$\Read_1$}, a backtrack signal is sent to (A).
Hence after $\RVFSMC$ backtracks to (A), in (A) it needs to perform
another iteration of \cref{line:exploration_whileback} while-loop.
In (A), first the causal map $\NegativeAnnotation$ is updated to forbid
\textcolor{\myblue}{$\Write_1$} and \textcolor{\myblue}{$\Write_3$}
for $\Read_1$. Then, read \textcolor{\myred}{$\Read_2$}
is processed from (A), creating (D). In (D),
\textcolor{\myblue}{$\Read_1$} is the only enabled event, and
\textcolor{\myblue}{$\Write_5$} is its only
$\NegativeAnnotation$-allowed write. This results in (E) which
reports a maximal trace. The algorithm backtracks
and concludes, reporting two maximal traces in total.

\begin{restatable}{theorem}{themexploration}\label{them:exploration}
Consider a concurrent program $\System$ of
$k$ threads and $d$ variables, with $n$
the length of the
longest trace in $\System$.
$\RVFSMC$ is a sound and complete algorithm for local safety properties in $\System$.
The time complexity of $\RVFSMC$ is
$k^{d} \cdot n^{O(k)} \cdot \beta$,
where
$\beta$ is the size of the $\RVF$ trace partitioning of $\System$.
\end{restatable}

\Paragraph{Novelties of the exploration.}
Here we highlight some key aspects of $\RVFSMC$.
First, we note that $\RVFSMC$ constructs the traces incrementally with each recursion step, as opposed to other approaches such as~\cite{Abdulla14,Abdulla19} that always work with maximal traces.
The reason of incremental traces is technical and has to do with the value-based treatment of the $\RVF$ partitioning.
We note that the other two value-based approaches \cite{HUANG15,Chatterjee19} also operate with incremental traces.
However, $\RVFSMC$ brings certain novelties compared to these two methods.
First, the exploration algorithm of \cite{HUANG15} can visit the same class of the partitioning (and even the same trace) an exponential number of times by different recursion branches, leading to significant performance degradation.
The exploration algorithm of~\cite{Chatterjee19} alleviates this issue using the causal map data structure, similar to our algorithm.
The causal map data structure can provably limit the number of revisits to polynomial (for a fixed number of threads),
and although it offers an improvement over the exponential revisits, it can still affect performance.
To further improve performance in this work,
our algorithm combines causal maps with a new technique, which is the backtrack signals.
Causal maps and backtrack signals together are very effective in avoiding having different branches of the recursion visit the same $\RVF$ class.

\Paragraph{Beyond $\RVF$ partitioning.}
While $\RVFSMC$ explores the $\RVF$ partitioning in the
worst case, in practice it often operates on a partitioning coarser
than the one induced by the $\RVF$ equivalence. Specifically,
$\RVFSMC$ may treat two traces $\Trace_1$ and $\Trace_2$ with
same events ($\Events{\Trace_1} = \Events{\Trace_2}$) and
value function ($\Value_{\Trace_1} = \Value_{\Trace_2}$) as
equivalent even when they differ in some causal orderings
($\CHB{}{\Trace_1}{} \Project \SysReads \neq \CHB{}{\Trace_2}{} \Project \SysReads$).
To see an example of this, consider the program and the $\RVFSMC$ run
in~\cref{fig:rvfsmc}. The recursion node (C) spans all traces where
(i)
\textcolor{\myblue}{$\Read_1$} reads-from either
\textcolor{\myblue}{$\Write_1$} or \textcolor{\myblue}{$\Write_3$},
and (ii)
\textcolor{\myred}{$\Read_2$} reads-from either
\textcolor{\myred}{$\Write_2$} or \textcolor{\myred}{$\Write_4$}.
Consider two such traces $\Trace_1$ and $\Trace_2$, with
$\RF{\Trace_1}(\textcolor{\myred}{\Read_2}) =
 \textcolor{\myred}{\Write_2}$ and
$\RF{\Trace_2}(\textcolor{\myred}{\Read_2}) =
 \textcolor{\myred}{\Write_4}$.
We have $\CHB{\textcolor{\myblue}{\Read_1}}{\Trace_1}{\textcolor{\myred}{\Read_2}}$
and $\NCHB{\textcolor{\myblue}{\Read_1}}{\Trace_2}{\textcolor{\myred}{\Read_2}}$,
and yet $\Trace_1$ and $\Trace_2$ are (soundly) considered equivalent
by $\RVFSMC$.
Hence the $\RVF$ partitioning is used to upper-bound the time complexity of $\RVFSMC$.
We remark that the algorithm is always sound, i.e.,
it is guaranteed to discover all thread states even when it does not explore the $\RVF$ partitioning in full.

\section{Experiments}\label{sec:experiments}

In this section we describe the
experimental evaluation of our SMC approach $\RVFSMC$.
We have implemented $\RVFSMC$
as an extension in Nidhugg~\cite{Abdulla2015},
a state-of-the-art stateless model checker for multithreaded
C/C++ programs that operates on LLVM
Intermediate Representation.
First we assess the advantages of utilizing
the $\RVF$ equivalence in SMC as compared to other
trace equivalences.
Then we perform ablation studies to demonstrate the impact of
the backtrack signals technique (cf. \cref{sec:smc}) and
the $\AlgoSC$ heuristics (cf. \cref{sec:verification_efficiency}).

In our experiments we compare $\RVFSMC$ with several
state-of-the-art SMC tools utilizing different trace equivalences.
First we consider $\VCDPOR$~\cite{Chatterjee19}, the SMC approach
operating on the value-centric equivalence.
Then we consider $\ReadsFrom$~\cite{Abdulla19}, the SMC algorithm
utilizing the reads-from equivalence.
Further we consider $\DCDPOR$~\cite{Chalupa17} that operates on
the data-centric equivalence, and finally
we compare with $\Source$~\cite{Abdulla14} utilizing the
Mazurkiewicz equivalence.
\footnote{
The MCR algorithm \cite{HUANG15} is beyond the experimental
scope of this work, as that tool handles Java programs
and uses heavyweight SMT solvers that require fine-tuning.}
The works of~\cite{Abdulla19} and \cite{LangS20}
in turn compare the $\ReadsFrom$ algorithm with additional
SMC tools, namely
$\mathsf{GenMC}$~\cite{Kokologiannakis19}
(with reads-from equivalence),
$\mathsf{RCMC}$~\cite{Kokologiannakis17}
(with Mazurkiewicz equivalence), and
$\mathsf{CDSChecker}$~\cite{NorrisD16}
(with Mazurkiewicz equivalence),
and thus we omit those tools from our evaluation.

There are two main objectives to our evaluation.
First, from~\cref{sec:equiv} we know that the $\RVF$
equivalence can be up to exponentially coarser than the
other equivalences, and we want to discover how often this
happens in practice.
Second, in cases where $\RVF$ does provide reduction in
the trace-partitioning size, we aim to see
whether this reduction is accompanied by the reduction
in the runtime of $\RVFSMC$ operating on $\RVF$ equivalence.

\Paragraph{Setup.}
We consider 119 benchmarks in total in our evaluation.
Each benchmark comes with a scaling parameter, called
the \emph{unroll} bound. The parameter controls the bound
on the number of iterations in all loops of the benchmark.
For each benchmark and unroll bound, we capture the number of
explored maximal traces,
and the total running time,
subject to a timeout of one hour.
In \cref{sec:app_experiments} we provide further details on our setup.

\begin{figure}[h]
  \raggedright
  \begin{minipage}{0.50\textwidth}
    \includegraphics[width=6.2cm]{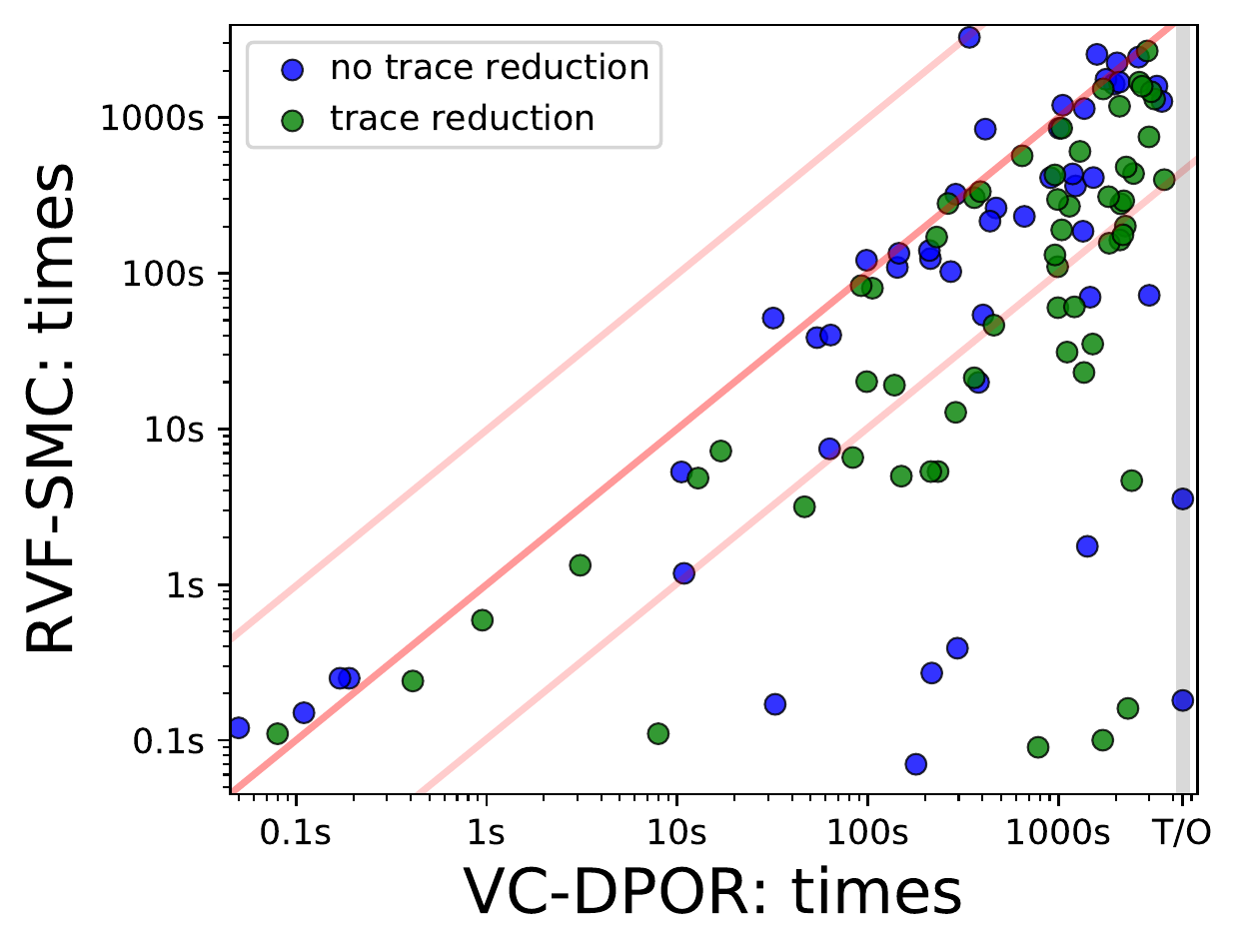}
  \end{minipage}
  \begin{minipage}{0.47\textwidth}
    \includegraphics[width=6.2cm]{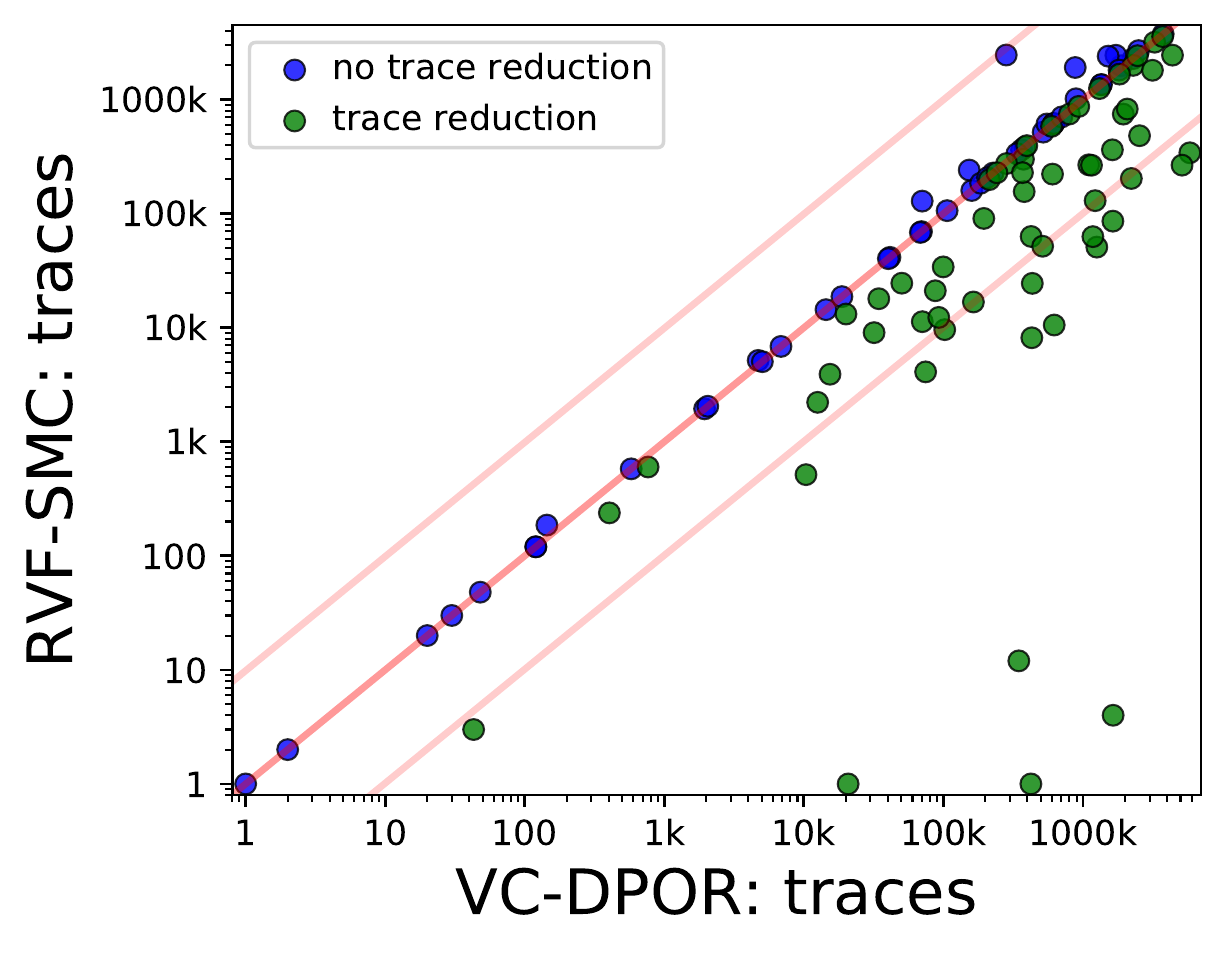}
  \end{minipage}
  \caption{
    Runtime and traces comparison of $\RVFSMC$ with $\VCDPOR$.
  }
  \label{fig:exp_vs_vcdpor}
\end{figure}

\begin{figure}[h]
  \raggedright
  \begin{minipage}{0.50\textwidth}
    \includegraphics[width=6.2cm]{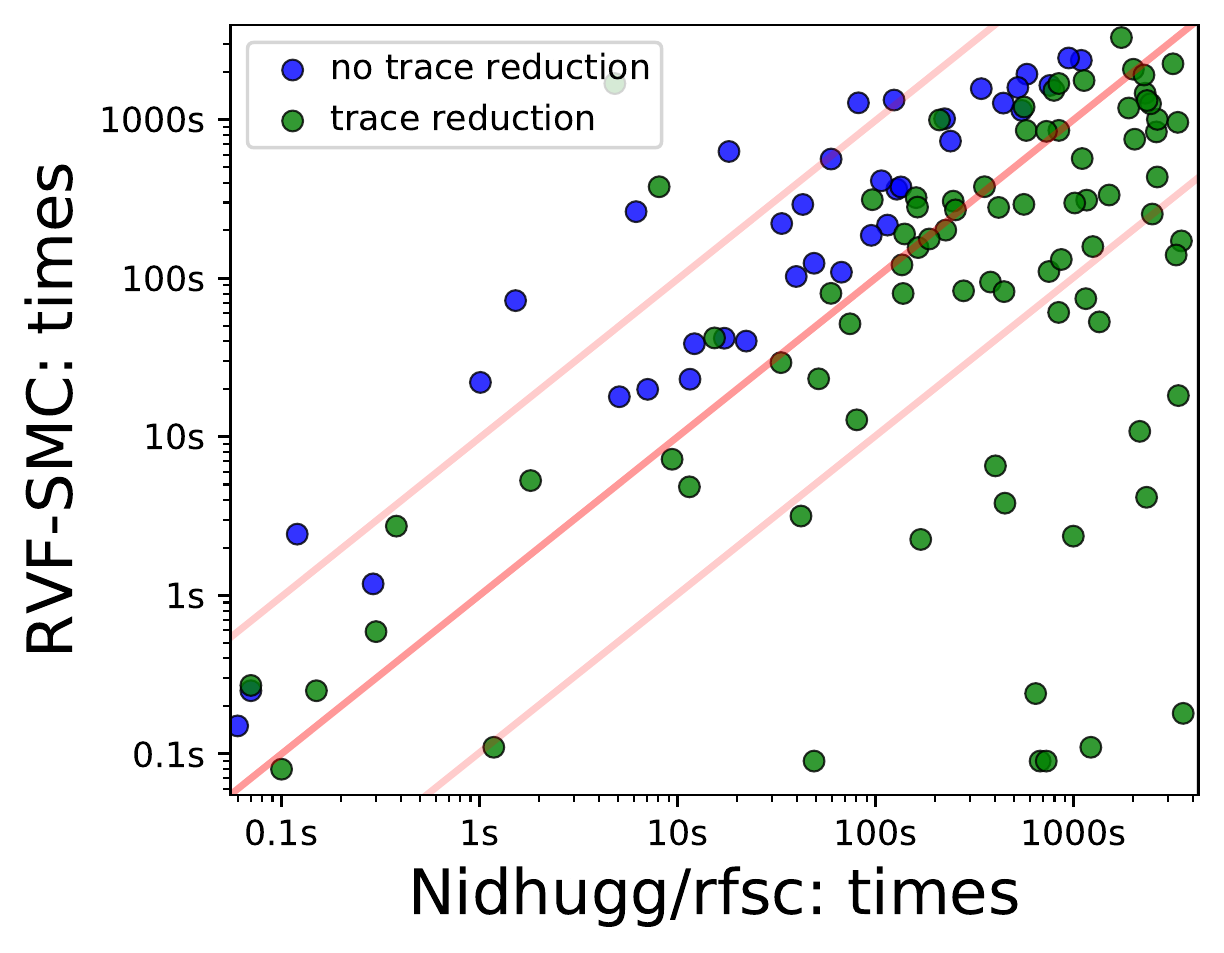}
  \end{minipage}
  \begin{minipage}{0.47\textwidth}
    \includegraphics[width=6.2cm]{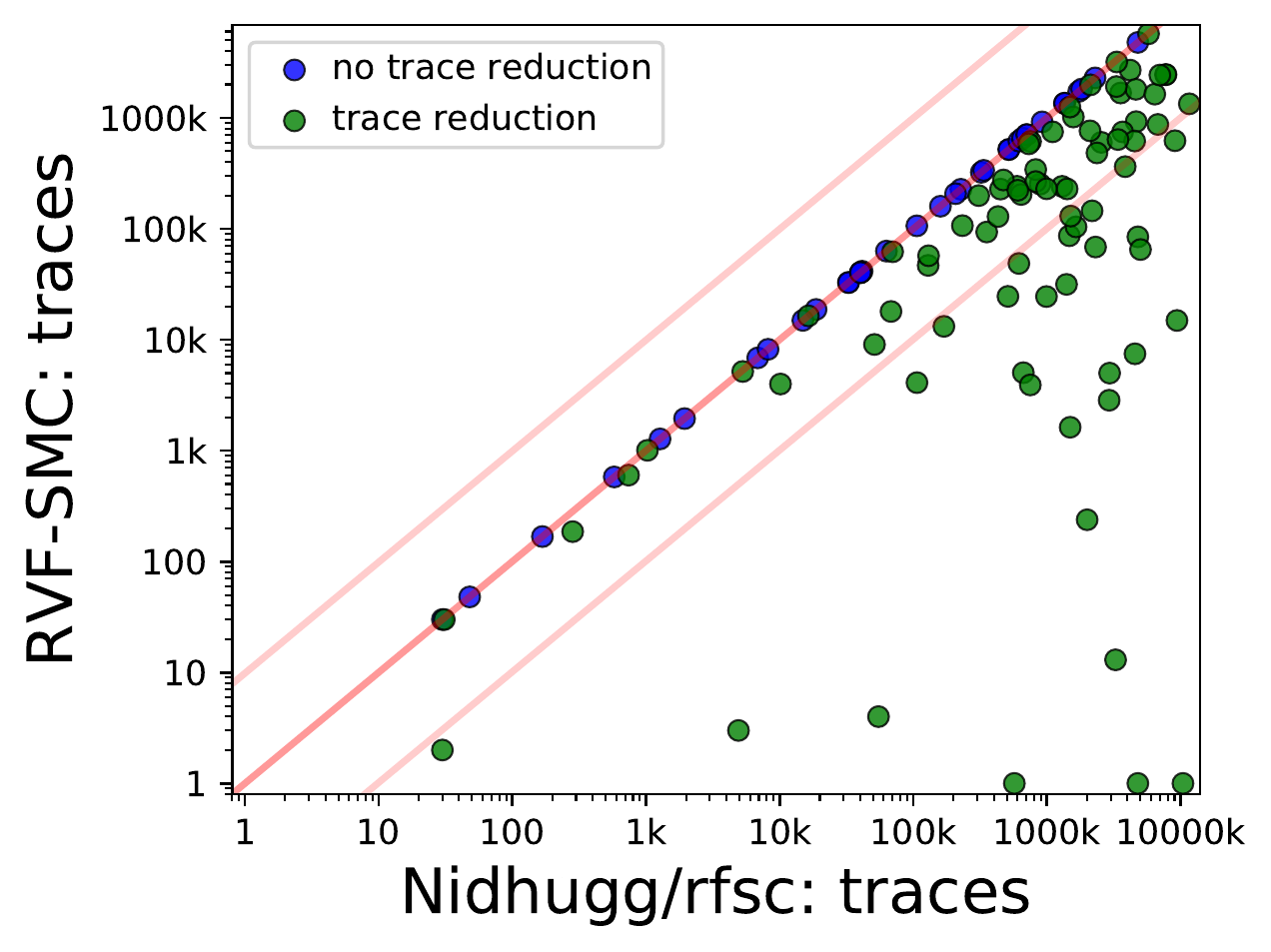}
  \end{minipage}
  \caption{
    Runtime and traces comparison of $\RVFSMC$ with $\mathsf{Nidhugg/rfsc}$.
  }
  \label{fig:exp_vs_rf}
\end{figure}

\begin{figure}[h]
  \raggedright
  \begin{minipage}{0.50\textwidth}
    \includegraphics[width=6.2cm]{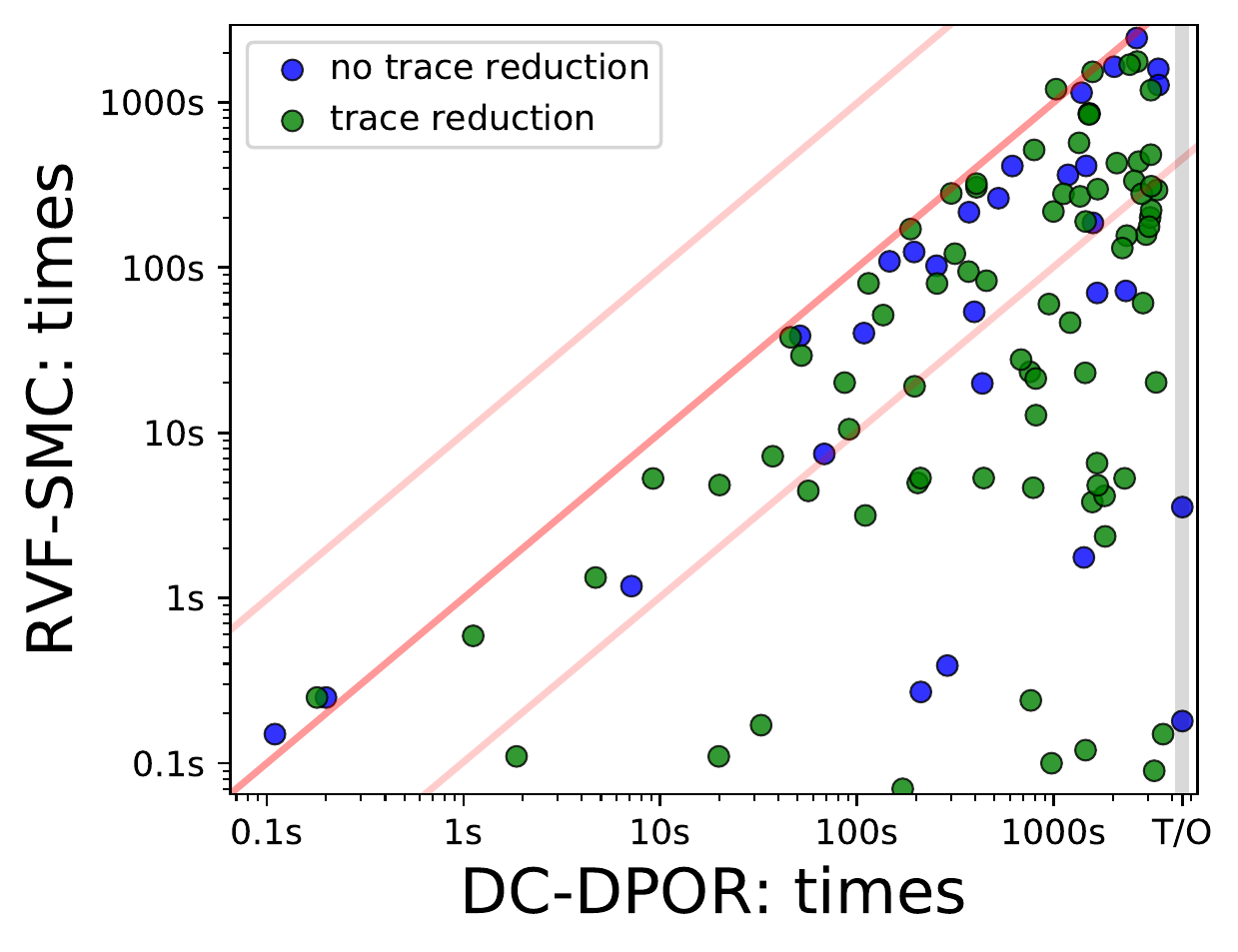}
  \end{minipage}
  \begin{minipage}{0.47\textwidth}
    \includegraphics[width=6.2cm]{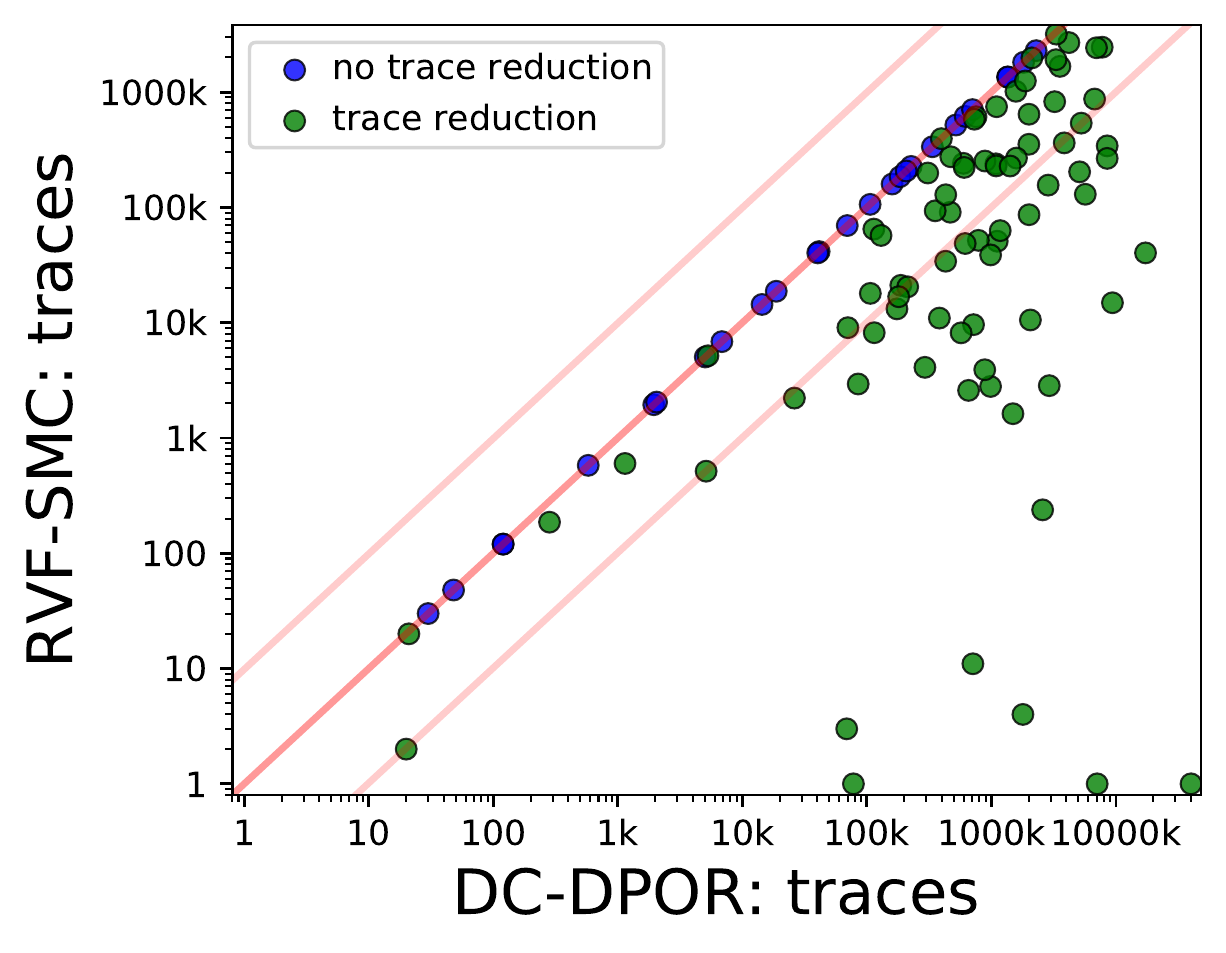}
  \end{minipage}
  \caption{
    Runtime and traces comparison of $\RVFSMC$ with $\DCDPOR$.
  }
  \label{fig:exp_vs_dcdpor}
\end{figure}

\begin{figure}[h]
  \raggedright
  \begin{minipage}{0.50\textwidth}
    \includegraphics[width=6.2cm]{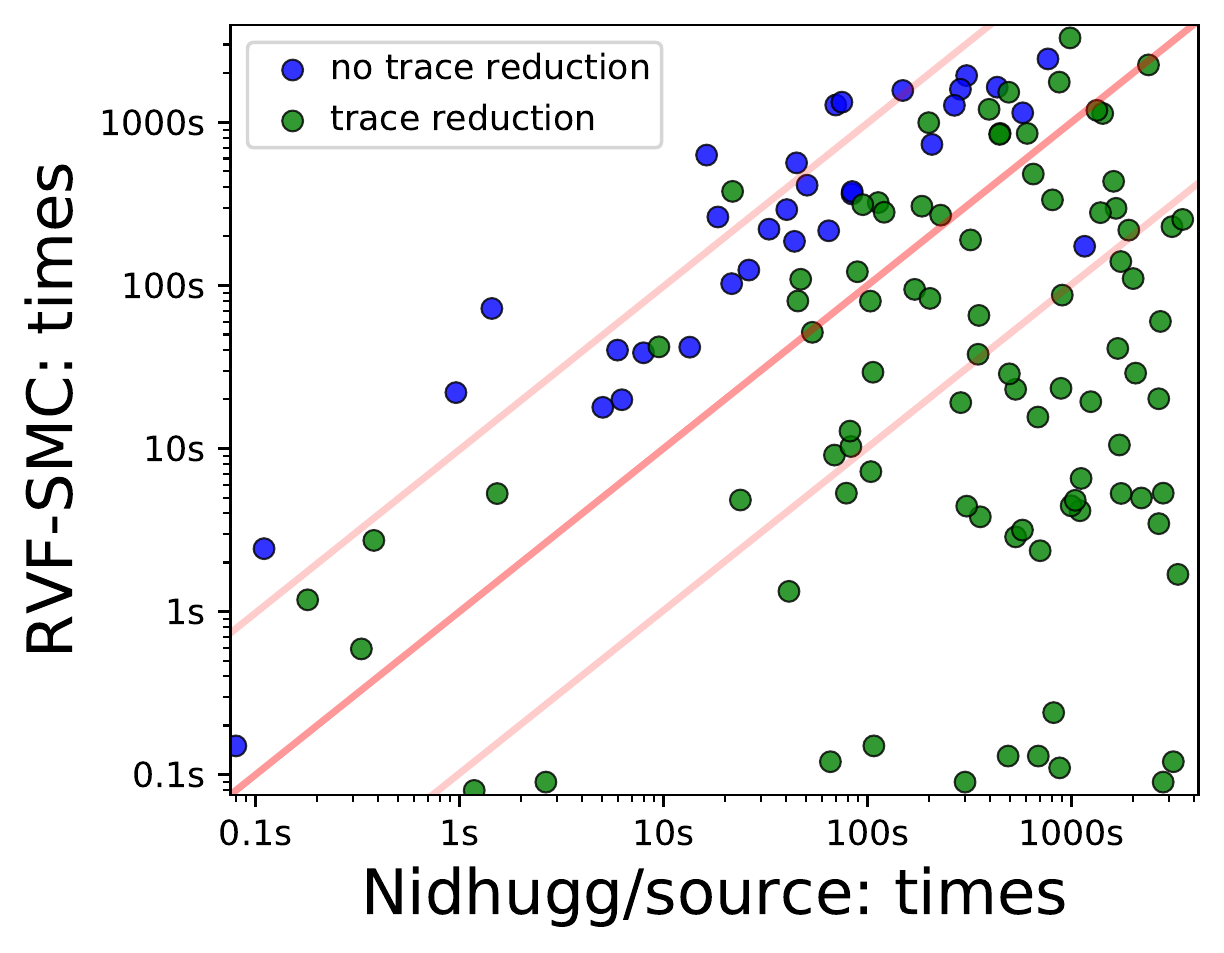}
  \end{minipage}
  \begin{minipage}{0.47\textwidth}
    \includegraphics[width=6.2cm]{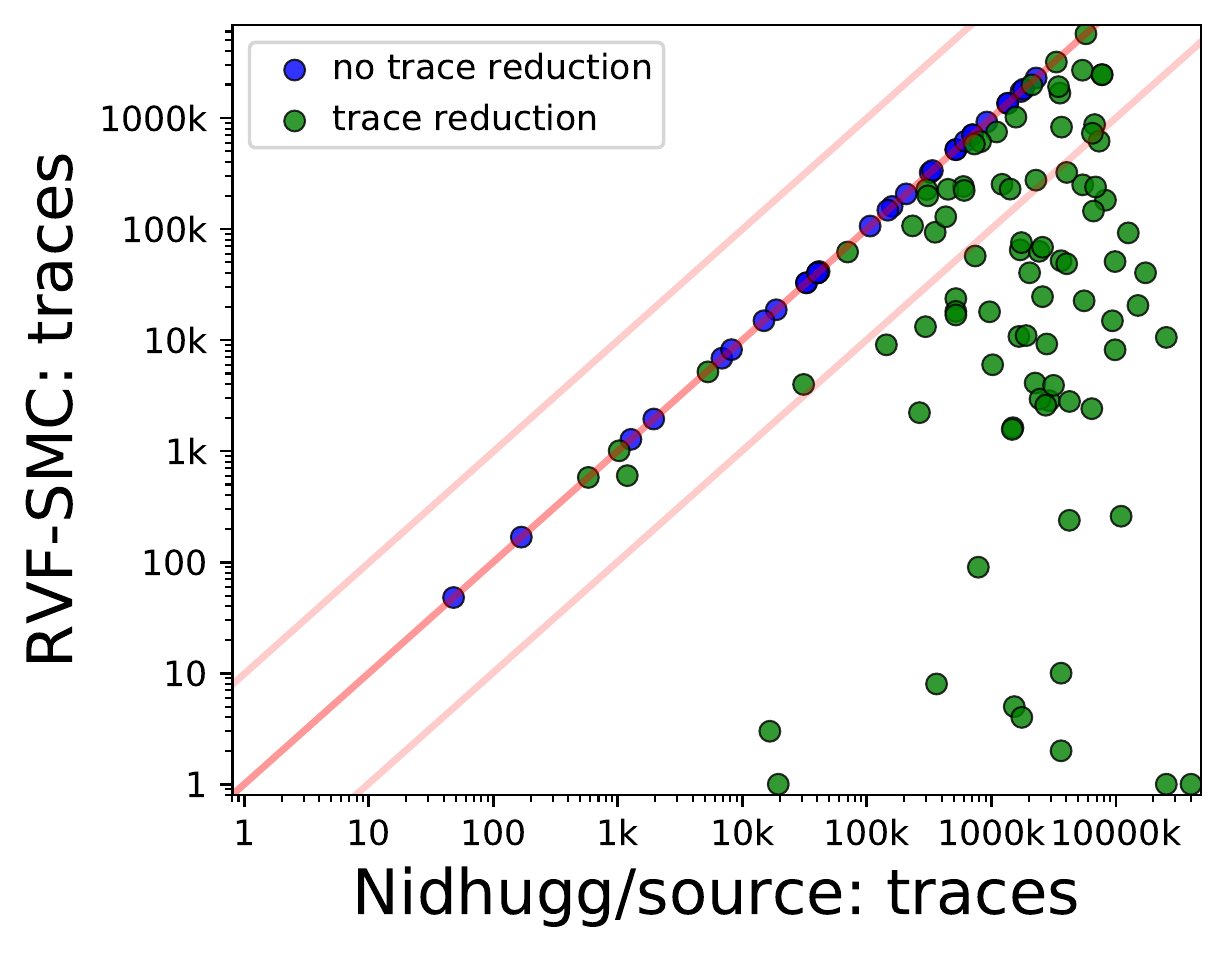}
  \end{minipage}
  \caption{
    Runtime and traces comparison of $\RVFSMC$ with $\mathsf{Nidhugg/source}$.
  }
  \label{fig:exp_vs_source}
\end{figure}

\Paragraph{Results.}
We provide a number of scatter plots summarizing the comparison
of $\RVFSMC$ with other state-of-the-art tools.
In \cref{fig:exp_vs_vcdpor},
\cref{fig:exp_vs_rf}, \cref{fig:exp_vs_dcdpor} and
\cref{fig:exp_vs_source}
we provide comparison both in runtimes and explored traces,
for $\VCDPOR$, $\ReadsFrom$, $\DCDPOR$, and $\Source$,
respectively.
In each scatter plot, both its axes are log-scaled,
the opaque red line represents equality, and the two semi-transparent
lines represent an order-of-magnitude difference.
The points are colored green when $\RVFSMC$ achieves trace reduction
in the underlying benchmark, and blue otherwise.

\Paragraph{Discussion: Significant trace reduction.}
In \cref{tab:better} we provide the results
for several benchmarks where $\RVF$ achieves significant reduction
in the trace-partitioning size. This is typically accompanied
by significant runtime reduction, allowing is to scale the benchmarks
to unroll bounds that other tools cannot handle.
Examples of this are \textsf{27\_Boop4} and
\textsf{scull\_loop}, two toy Linux kernel drivers.

In several benchmarks the number of explored traces
remains the same for $\RVFSMC$ even when scaling up the unroll bound,
see \textsf{45\_monabsex1}, \textsf{reorder\_5} and
\textsf{singleton} in \cref{tab:better}. The
\textsf{singleton} example is further interesting, in that
while $\VCDPOR$ and $\DCDPOR$ also explore few traces, they still
suffer in runtime due to additional redundant exploration,
as described in Sections~\ref{sec:intro}~and~\ref{sec:smc}.

\Paragraph{Discussion: Little-to-no trace reduction.}
\cref{tab:notbetter} presents several benchmarks
where the $\RVF$ partitioning achieves little-to-no reduction.
In these cases the well-engineered $\ReadsFrom$ and
$\Source$ dominate the runtime.

\begin{figure}[h]
  \raggedright
  \begin{minipage}{0.50\textwidth}
    \includegraphics[width=6.2cm]{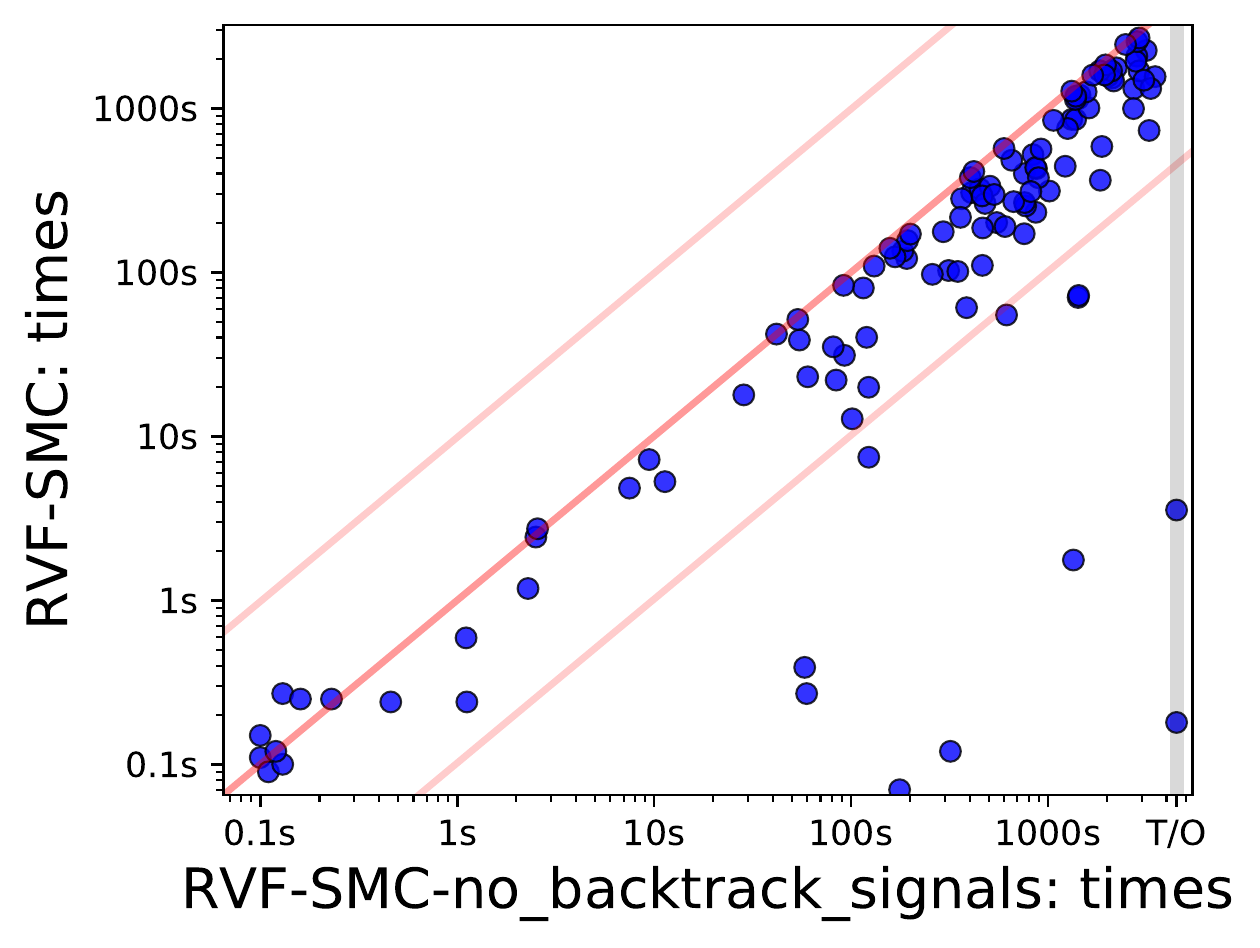}
  \end{minipage}
  \begin{minipage}{0.47\textwidth}
    \includegraphics[width=6.2cm]{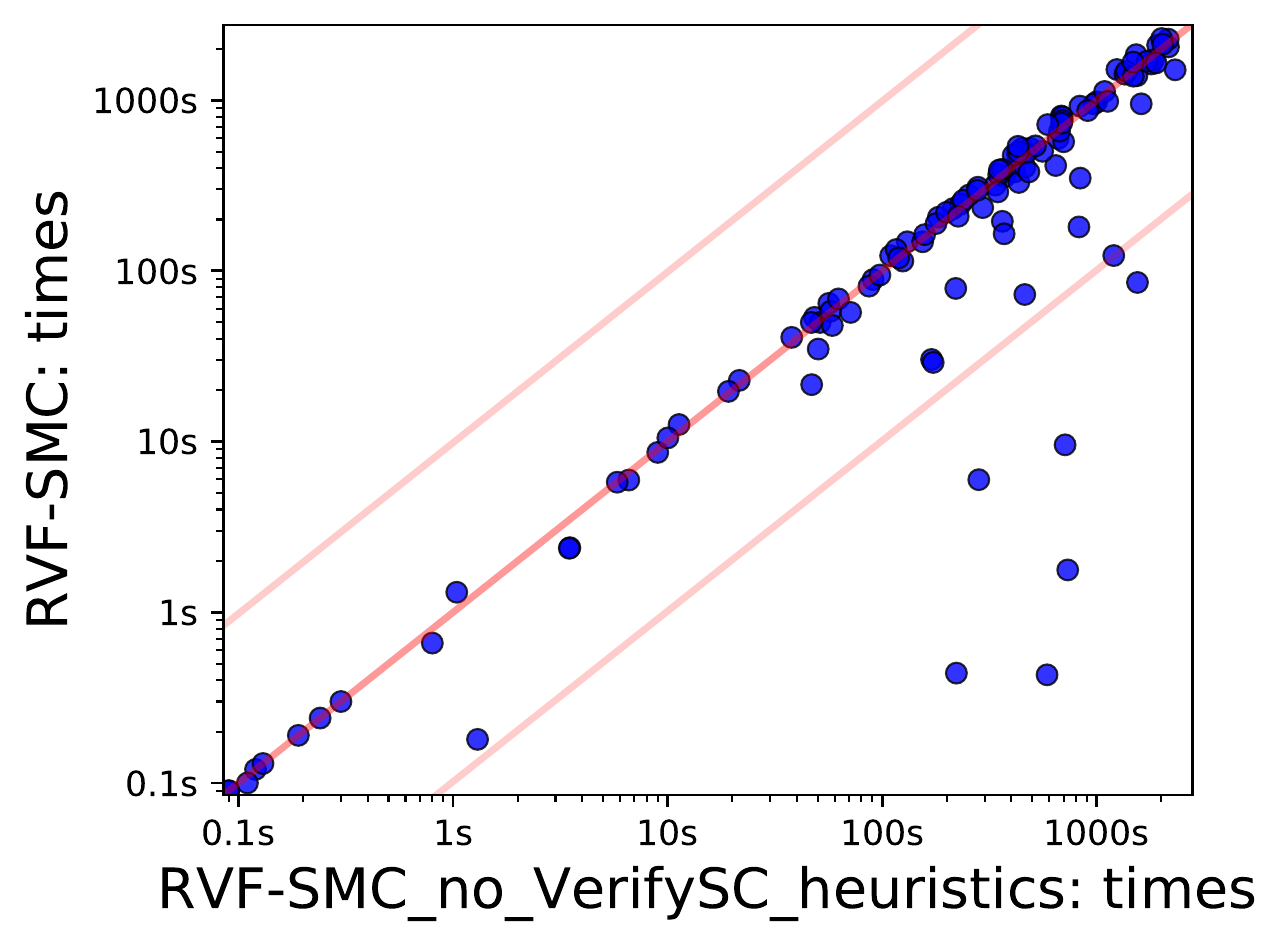}
  \end{minipage}
  \caption{
    Ablation studies of $\RVFSMC$. The left plot compares $\RVFSMC$
    with and without backtrack signals. The right plots compares
    $\RVFSMC$ with and without the closure and auxiliary-trace
    heuristics of \cref{sec:verification_efficiency}.
  }
  \label{fig:exp_time_ablations}
\end{figure}

\Paragraph{$\RVFSMC$ ablation studies.}
Here we demonstrate the effect that follows from our $\RVFSMC$
algorithm utilizing the approach of backtrack signals
(see~\cref{sec:smc}) and the heuristics of $\AlgoSC$
(see~\cref{sec:verification_efficiency}).
These techniques have no effect on the number of the explored traces,
thus we focus on the runtime.
The left plot of \cref{fig:exp_time_ablations} compares $\RVFSMC$
as is with a $\RVFSMC$ version that does not utilize the backtrack
signals (achieved by simply keeping the $\backtrack$ flag
in \cref{algo:exploration} always $\true$). The right plot
of \cref{fig:exp_time_ablations} compares $\RVFSMC$ as is with
a $\RVFSMC$ version that employs $\AlgoSC$ without the closure and
auxiliary-trace heuristics.
We can see that the techniques almost always result in improved
runtime. The improvement is mostly within an order of magnitude,
and in a few cases there is several-orders-of-magnitude improvement.

\begin{table}[t]
\scriptsize
\centering
\newcolumntype{?}{!{\vrule width 1.5pt}}
\setlength{\extrarowheight}{.00em}
\begin{tabular}{? l | c | c ? c c c c c ?}
\specialrule{.15em}{0em}{0em}
                                                             \multicolumn{2}{?c|}{\textbf{Benchmark}} &                                                U &                                        $\RVFSMC$ &                                        $\VCDPOR$ &                             $\mathsf{Nidh/rfsc}$ &                                        $\DCDPOR$ &                           $\mathsf{Nidh/source}$ \\
\specialrule{.1em}{0em}{0em}
\multirow{4}{*}{\begin{tabular}{l}
\textbf{27\_Boop4}\\
threads: 4
\end{tabular}}
&
\multirow{2}{*}{Traces}
                                                                                                     &                                               10 &                                 \textbf{1337215} &                                          1574287 &                                         11610040 &                                                - &                                                - \\
                                                   &                                                  &                                               12 &                                 \textbf{2893039} &                                                - &                                                - &                                                - &                                                - \\
\cline{2-8}
&
\multirow{2}{*}{Times}
                                                                                                     &                                               10 &                                    \textbf{837s} &                                            1946s &                                            2616s &                                                - &                                                - \\
                                                   &                                                  &                                               12 &                                   \textbf{2017s} &                                                - &                                                - &                                                - &                                                - \\
\specialrule{.1em}{0em}{0em}
\multirow{4}{*}{\begin{tabular}{l}
\textbf{45\_monabsex1}\\
threads: U
\end{tabular}}
&
\multirow{2}{*}{Traces}
                                                                                                     &                                                7 &                                       \textbf{1} &                                           423360 &                                           262144 &                                          7073803 &                                         25401600 \\
                                                   &                                                  &                                                8 &                                       \textbf{1} &                                                - &                                          4782969 &                                                - &                                                - \\
\cline{2-8}
&
\multirow{2}{*}{Times}
                                                                                                     &                                                7 &                                   \textbf{0.09s} &                                             784s &                                              33s &                                            3239s &                                            2819s \\
                                                   &                                                  &                                                8 &                                   \textbf{0.09s} &                                                - &                                             677s &                                                - &                                                - \\
\specialrule{.1em}{0em}{0em}
\multirow{4}{*}{\begin{tabular}{l}
\textbf{reorder\_5}\\
threads: U+1
\end{tabular}}
&
\multirow{2}{*}{Traces}
                                                                                                     &                                                9 &                                       \textbf{4} &                                          1644716 &                                             1540 &                                          1792290 &                                                - \\
                                                   &                                                  &                                               30 &                                       \textbf{4} &                                                - &                                            54901 &                                                - &                                                - \\
\cline{2-8}
&
\multirow{2}{*}{Times}
                                                                                                     &                                                9 &                                   \textbf{0.10s} &                                            1711s &                                            0.44s &                                             974s &                                                - \\
                                                   &                                                  &                                               30 &                                   \textbf{0.09s} &                                                - &                                              49s &                                                - &                                                - \\
\specialrule{.1em}{0em}{0em}
\multirow{4}{*}{\begin{tabular}{l}
\textbf{scull\_loop}\\
threads: 3
\end{tabular}}
&
\multirow{2}{*}{Traces}
                                                                                                     &                                                2 &                                    \textbf{3908} &                                            15394 &                                           749811 &                                           884443 &                                          3157281 \\
                                                   &                                                  &                                                3 &                                  \textbf{115032} &                                                - &                                                - &                                                - &                                                - \\
\cline{2-8}
&
\multirow{2}{*}{Times}
                                                                                                     &                                                2 &                                   \textbf{6.55s} &                                              83s &                                             403s &                                            1659s &                                            1116s \\
                                                   &                                                  &                                                3 &                                    \textbf{266s} &                                                - &                                                - &                                                - &                                                - \\
\specialrule{.1em}{0em}{0em}
\multirow{4}{*}{\begin{tabular}{l}
\textbf{singleton}\\
threads: U+1
\end{tabular}}
&
\multirow{2}{*}{Traces}
                                                                                                     &                                               20 &                                       \textbf{2} &                                       \textbf{2} &                                               20 &                                               20 &                                                - \\
                                                   &                                                  &                                               30 &                                       \textbf{2} &                                                - &                                               30 &                                                - &                                                - \\
\cline{2-8}
&
\multirow{2}{*}{Times}
                                                                                                     &                                               20 &                                   \textbf{0.07s} &                                             179s &                                            0.08s &                                             171s &                                                - \\
                                                   &                                                  &                                               30 &                                   \textbf{0.08s} &                                                - &                                            0.10s &                                                - &                                                - \\
\specialrule{.1em}{0em}{0em}
\end{tabular}
\caption{
Benchmarks with trace reduction achieved by $\RVFSMC$.
The unroll bound is shown in the column \textbf{U}.
Symbol ``-'' indicates one-hour timeout.
Bold-font entries indicate the smallest numbers
for respective benchmark and unroll.
}
\label{tab:better}
\end{table}

\begin{table}[h]
\scriptsize
\centering
\newcolumntype{?}{!{\vrule width 1.5pt}}
\setlength{\extrarowheight}{.00em}
\begin{tabular}{? l | c | c ? c c c c c ?}
\specialrule{.15em}{0em}{0em}
                                                             \multicolumn{2}{?c|}{\textbf{Benchmark}} &                                                U &                                        $\RVFSMC$ &                                        $\VCDPOR$ &                             $\mathsf{Nidh/rfsc}$ &                                        $\DCDPOR$ &                           $\mathsf{Nidh/source}$ \\
\specialrule{.1em}{0em}{0em}
\multirow{4}{*}{\begin{tabular}{l}
\textbf{13\_unverif}\\
threads: U
\end{tabular}}
&
\multirow{2}{*}{Traces}
                                                                                                     &                                                5 &                                   \textbf{14400} &                                   \textbf{14400} &                                   \textbf{14400} &                                   \textbf{14400} &                                   \textbf{14400} \\
                                                   &                                                  &                                                6 &                                  \textbf{518400} &                                                - &                                  \textbf{518400} &                                                - &                                  \textbf{518400} \\
\cline{2-8}
&
\multirow{2}{*}{Times}
                                                                                                     &                                                5 &                                            7.45s &                                              63s &                                            3.33s &                                              68s &                                   \textbf{2.72s} \\
                                                   &                                                  &                                                6 &                                             376s &                                                - &                                             134s &                                                - &                                     \textbf{84s} \\
\specialrule{.1em}{0em}{0em}
\multirow{4}{*}{\begin{tabular}{l}
\textbf{approxds\_append}\\
threads: U
\end{tabular}}
&
\multirow{2}{*}{Traces}
                                                                                                     &                                                6 &                                   \textbf{50897} &                                          1256381 &                                           198936 &                                          1114746 &                                          9847080 \\
                                                   &                                                  &                                                7 &                                  \textbf{923526} &                                                - &                                          4645207 &                                                - &                                                - \\
\cline{2-8}
&
\multirow{2}{*}{Times}
                                                                                                     &                                                6 &                                     \textbf{60s} &                                             995s &                                              67s &                                             944s &                                            2733s \\
                                                   &                                                  &                                                7 &                                            2078s &                                                - &                                   \textbf{2003s} &                                                - &                                                - \\
\specialrule{.1em}{0em}{0em}
\multirow{4}{*}{\begin{tabular}{l}
\textbf{chase-lev-dq}\\
threads: 3
\end{tabular}}
&
\multirow{2}{*}{Traces}
                                                                                                     &                                                4 &                                   \textbf{87807} &                         $\dagger$ &                                           175331 &                         $\dagger$ &                                           175331 \\
                                                   &                                                  &                                                5 &                                  \textbf{227654} &                         $\dagger$ &                                           448905 &                         $\dagger$ &                                           448905 \\
\cline{2-8}
&
\multirow{2}{*}{Times}
                                                                                                     &                                                4 &                                             289s &                         $\dagger$ &                                     \textbf{71s} &                         $\dagger$ &                                     \textbf{71s} \\
                                                   &                                                  &                                                5 &                                             995s &                         $\dagger$ &                                             210s &                         $\dagger$ &                                    \textbf{200s} \\
\specialrule{.1em}{0em}{0em}
\multirow{4}{*}{\begin{tabular}{l}
\textbf{linuxrwlocks}\\
threads: U+1
\end{tabular}}
&
\multirow{2}{*}{Traces}
                                                                                                      &                                                1 &                                      \textbf{56} &                         $\dagger$ &                                               59 &                         $\dagger$ &                                               59 \\
                                                   &                                                  &                                                2 &                                   \textbf{62018} &                         $\dagger$ &                                            70026 &                         $\dagger$ &                                            70026 \\
\cline{2-8}
&
\multirow{2}{*}{Times}
                                                                                                      &                                                1 &                                            0.12s &                         $\dagger$ &                                   \textbf{0.09s} &                         $\dagger$ &                                            0.13s \\
                                                   &                                                  &                                                2 &                                              42s &                         $\dagger$ &                                              15s &                         $\dagger$ &                                   \textbf{9.50s} \\
\specialrule{.1em}{0em}{0em}
\multirow{4}{*}{\begin{tabular}{l}
\textbf{pgsql}\\
threads: 2
\end{tabular}}
&
\multirow{2}{*}{Traces}
                                                                                                     &                                                3 &                                    \textbf{3906} &                                    \textbf{3906} &                                    \textbf{3906} &                                    \textbf{3906} &                                    \textbf{3906} \\
                                                   &                                                  &                                                4 &                                  \textbf{335923} &                                  \textbf{335923} &                                  \textbf{335923} &                                  \textbf{335923} &                                  \textbf{335923} \\
\cline{2-8}
&
\multirow{2}{*}{Times}
                                                                                                     &                                                3 &                                            3.30s &                                            5.98s &                                            1.01s &                                            4.00s &                                   \textbf{0.51s} \\
                                                   &                                                  &                                                4 &                                             412s &                                             911s &                                             107s &                                             616s &                                     \textbf{51s} \\
\specialrule{.1em}{0em}{0em}
\end{tabular}
\caption{
Benchmarks with little-to-no trace reduction by $\RVFSMC$.
Symbol $\dagger$ indicates that a particular benchmark operation
is not handled by the tool.
}
\label{tab:notbetter}
\end{table}

\begin{figure}[h]
  \raggedright
  \begin{minipage}{0.70\textwidth}
    \includegraphics[width=8.4cm]{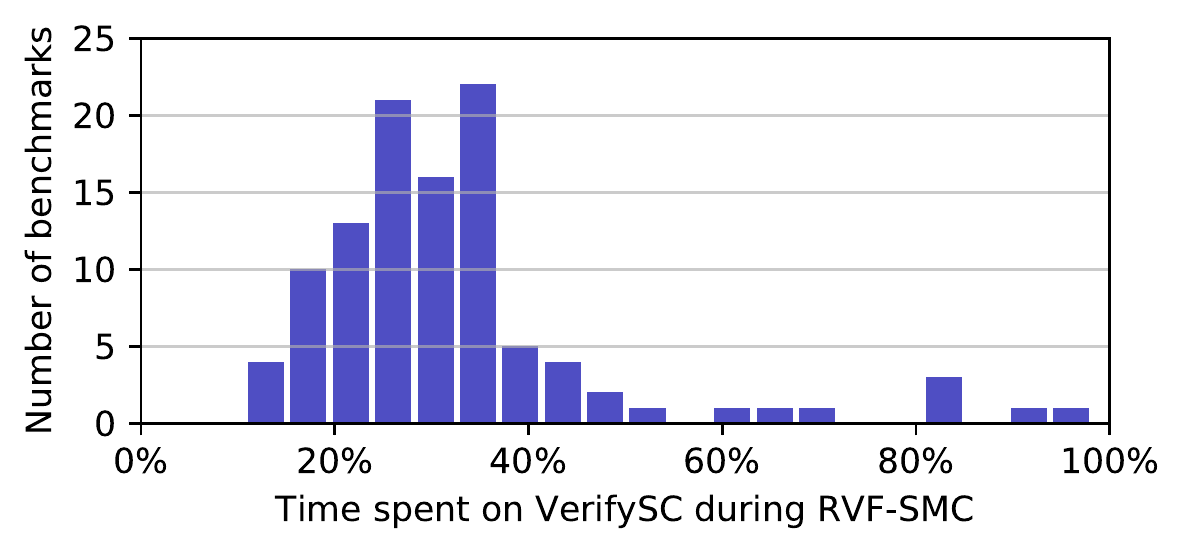}
  \end{minipage}
  \begin{minipage}{0.27\textwidth}
    \vspace{-3mm}
    \caption{
      Histogram that illustrates the percentage
      of time spent solving $\VSC$ instances during $\RVFSMC$.
    }
    \label{fig:exp_time_doing_lin}
  \end{minipage}
\end{figure}

Finally, in \cref{fig:exp_time_doing_lin} we illustrate
how much time during $\RVFSMC$ is typically spent on $\AlgoSC$
(i.e., on solving $\VSC$ instances generated during $\RVFSMC$).

\section{Conclusions}\label{sec:conclusions}

In this work we developed $\RVFSMC$, a new SMC algorithm for the verification of concurrent programs using a novel equivalence called \emph{reads-value-from (RVF)}.
On our way to  $\RVFSMC$, we have revisited the famous $\VSC$ problem~\cite{Gibbons97}.
Despite its NP-hardness, we  have shown that the problem is parameterizable in $k+d$ (for $k$ threads and $d$ variables),
and becomes even fixed-parameter tractable in $d$ when $k$ is constant.
Moreover we have developed practical heuristics that solve the problem efficiently in many practical settings.

Our $\RVFSMC$ algorithm couples our solution for $\VSC$  to a novel exploration of the underlying RVF partitioning,
and is able to model check many concurrent programs where previous approaches time-out.
Our experimental evaluation reveals that RVF is very often the most effective equivalence, as the underlying partitioning is exponentially coarser than other approaches.
Moreover, $\RVFSMC$ generates representatives very efficiently, as
the reduction in the partitioning is often met with significant speed-ups in the model checking task.
Interesting future work includes further improvements over the $\VSC$, as well as extensions of $\RVFSMC$ to relaxed memory models.

\Paragraph{Acknowledgments.}
The research was partially funded by the ERC CoG 863818 (ForM-SMArt)
and the Vienna Science and Technology Fund (WWTF) through project ICT15-003.


\pagebreak
\bibliographystyle{unsrt}
\bibliography{bibliography}

\newpage
\appendix

\section{Extensions of the concurrent model}\label{sec:app_model}

For presentation clarity, in our exposition we considered a simple
concurrent model with a static set of threads, and with only read
and write events.
Here we describe how our approach handles the following extensions
of the concurrent model:
\begin{enumerate}[noitemsep,topsep=0pt,partopsep=0px]
\item Read-modify-write and compare-and-swap events.
\item Mutex events lock-acquire and lock-release.
\item Spawn and join events for dynamic thread creation.
\end{enumerate}

\Paragraph{Read-modify-write and compare-and-swap events.}
We model a read-modify-write atomic operation on a variable $x$ as
a pair of two events $\RMW_{\Read}$ and $\RMW_{\Write}$,
where $\RMW_{\Read}$ is a read event of $x$,
$\RMW_{\Write}$ is a write event of $x$, and for each trace $\Trace$
either the events are both not present in $\Trace$, or they are
both present and appearing together in $\Trace$
($\RMW_{\Read}$ immediately followed by $\RMW_{\Write}$ in $\Trace$).
We model a compare-and-swap atomic operation similarly,
obtaining a pair of events $\CAS_{\Read}$ and $\CAS_{\Write}$.
In addition we consider a local event happening immediately
after the read event $\CAS_{\Read}$, evaluating the ``compare'' condition
of the compare-and-swap instruction.
Thus, in traces $\Trace$ that contain $\CAS_{\Read}$ and
the ``compare'' condition evaluates to
$\true$, we have that $\CAS_{\Read}$ is immediately followed by
$\CAS_{\Write}$ in $\Trace$.
In traces $\Trace'$ that contain $\CAS_{\Read}$ and
the ``compare'' condition evaluates to
$\false$, we have that $\CAS_{\Write}$ is not present in $\Trace'$.

We now discuss our extension of $\AlgoSC$ to handle the $\VSC(X, \GoodWrites)$ problem (\cref{sec:verification})
in presence of read-modify-write and compare-and-swap events.
First, observe that as the event set $X$ and the good-writes function
$\GoodWrites$ are fixed, we possess the information on whether each
compare-and-swap instruction satisfies its ``compare'' condition or not.
Then, in case we have in our event set a read-modify-write event
pair $\Event_1 = \RMW_{\Read}$ and $\Event_2 = \RMW_{\Write}$
(resp. a compare-and-swap event pair $\Event_1 = \CAS_{\Read}$ and $\Event_2 = \CAS_{\Write}$),
we proceed as follows.
When the first of the two events $\Event_1$ becomes executable in~\cref{algo:vsc_ifextend}
of~\cref{algo:vsc} for $\Seq$, we proceed only in case
$\Event_2$ is also executable in $\Seq \Concat \Event_1$, and in such a case
in \cref{algo:vsc_execute_event} we consider straight away a sequence
$\Seq \Concat \Event_1 \Concat \Event_2$.
This ensures that in all sequences we consider,
the event pair of the read-modify-write (resp. compare-and-swap)
appears as one event immediately followed by the other event.

In the presence of read-modify-write and compare-and-swap events,
the SMC approach $\RVFSMC$ can be utilized as presented in
\cref{sec:smc}, after an additional corner case is handled
for backtrack signals.
Specifically, when processing the extension events in
\cref{line:exploration_foreachextwrite} of \cref{algo:exploration},
we additionally process in the same fashion reads $\CAS_{\Read}$ enabled
in $\widetilde{\Trace}$ that are part of a compare-and-swap instruction.
These reads $\CAS_{\Read}$ are then treated as potential novel
reads-from sources for ancestor mutations
$\CAS^{*}_{\Read} \in \Domain(\ancestors)$ (\cref{line:exploration_ifnewsource})
where $\CAS^{*}_{\Read}$ is also a read-part of a compare-and-swap instruction.

\Paragraph{Mutex events.}
Mutex events $\Aquire$ and $\Release$
are naturally handled by our approach as follows. We consider each
lock-release event $\Release$ as a write event and
each lock-acquire event $\Aquire$ as a read event,
the corresponding unique mutex they access is considered
a global variable of $\Globals$.

In SMC, we enumerate good-writes functions whose domain
also includes the lock-acquire events.
Further, a good-writes set of each lock-acquire admits only
a single conflicting lock-release event, thus obtaining
constraints of the form $\GoodWrites(\Aquire) = \{ \Release \}$.
During closure (\cref{sec:verification_efficiency}),
given $\GoodWrites(\Aquire) = \{ \Release \}$, we consider
the following condition:
$\Proc{\Aquire} \neq \Proc{\Release}$ implies $\Release <_P \Aquire$.
Thus $P$ totally orders the critical sections of each mutex,
and therefore $\AlgoSC$ does not need to take additional care
for mutexes.
Indeed, respecting $P$ trivially solves all $\GoodWrites$ constraints
of lock-acquire events, and further
preserves the property that no thread tries to acquire
an already acquired (and so-far unreleased) mutex.
No modifications to the $\RVFSMC$ algorithm are needed to
incorporate mutex events.

\Paragraph{Dynamic thread creation.}
For simplicity of presentation,
we assumed a static set of threads for a given concurrent program.
However, our approach straightforwardly handles dynamic thread creating,
by including in the
program order $\TO$ the orderings naturally induced by spawn and join events.
In our experiments, all our considered benchmarks spawn threads dynamically.

\section{Details of {\cref{sec:equiv}}}\label{sec:app_equiv}

\begin{figure}[h]
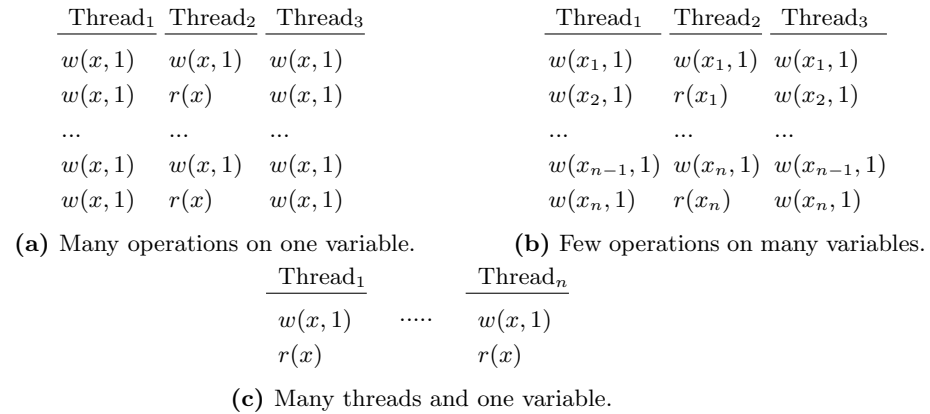

  \begin{subfigure}{0.45\textwidth}
    \centering
    \small
    \begin{minipage}{0.10\textwidth}
      \centering
      \begin{align*}
        &\;\text{Thread}_1\\
        \hline\\[-1em]
        \,& \Write(x, 1)\\
        \\[-1.6em]
        \,& \Write(x, 1)\\
        \\[-1.6em]
        \,& ...\\
        \\[-1.6em]
        \,& \Write(x, 1)\\
        \\[-1.6em]
        \,& \Write(x, 1)\\
      \end{align*}
    \end{minipage}
    \begin{minipage}{0.10\textwidth}
      \centering
      \begin{align*}
        &\text{Thread}_2\\
        \hline\\[-1em]
        \,& \Write(x, 1)\\
        \\[-1.6em]
        \,& \Read(x)\\
        \\[-1.6em]
        \,& ...\\
        \\[-1.6em]
        \,& \Write(x, 1)\\
        \\[-1.6em]
        \,& \Read(x)\\
      \end{align*}
    \end{minipage}
    \begin{minipage}{0.10\textwidth}
      \centering
      \begin{align*}
        &\;\text{Thread}_3\\
        \hline\\[-1em]
        \,& \Write(x, 1)\\
        \\[-1.6em]
        \,& \Write(x, 1)\\
        \\[-1.6em]
        \,& ...\\
        \\[-1.6em]
        \,& \Write(x, 1)\\
        \\[-1.6em]
        \,& \Write(x, 1)\\
      \end{align*}
    \end{minipage}
    \vspace{-5mm}
    \caption{
      Many operations on one variable.
    }
  \end{subfigure}
  \qquad
  \begin{subfigure}{0.45\textwidth}
    \centering
    \small
    \begin{minipage}{0.10\textwidth}
      \centering
      \begin{align*}
        &\;\text{Thread}_1\\
        \hline\\[-1em]
        \,& \Write(x_1, 1)\\
        \\[-1.6em]
        \,& \Write(x_2, 1)\\
        \\[-1.6em]
        \,& ...\\
        \\[-1.6em]
        \,& \Write(x_{n-1}, 1)\\
        \\[-1.6em]
        \,& \Write(x_n, 1)\\
      \end{align*}
    \end{minipage}
    \begin{minipage}{0.10\textwidth}
      \centering
      \begin{align*}
        &\text{Thread}_2\\
        \hline\\[-1em]
        \,& \Write(x_1, 1)\\
        \\[-1.6em]
        \,& \Read(x_1)\\
        \\[-1.6em]
        \,& ...\\
        \\[-1.6em]
        \,& \Write(x_n, 1)\\
        \\[-1.6em]
        \,& \Read(x_n)\\
      \end{align*}
    \end{minipage}
    \begin{minipage}{0.10\textwidth}
      \centering
      \begin{align*}
        &\;\text{Thread}_3\\
        \hline\\[-1em]
        \,& \Write(x_1, 1)\\
        \\[-1.6em]
        \,& \Write(x_2, 1)\\
        \\[-1.6em]
        \,& ...\\
        \\[-1.6em]
        \,& \Write(x_{n-1}, 1)\\
        \\[-1.6em]
        \,& \Write(x_n, 1)\\
      \end{align*}
    \end{minipage}
    \vspace{-5mm}
    \caption{
      Few operations on many variables.
    }
  \end{subfigure}
  \begin{subfigure}{0.9\textwidth}
    \vspace{-2mm}
    \small
    \centering
    \vspace{-2mm}
    \begin{minipage}{0.08\textwidth}
      \centering
      \begin{align*}
        &\text{Thread}_1\\
        \hline\\[-1em]
        \,~& \Write(x, 1)\\
        \\[-1.6em]
        \,~& \Read(x)\\
      \end{align*}
    \end{minipage}
    \begin{minipage}{0.10\textwidth}
      \centering
      .....
    \end{minipage}
    \begin{minipage}{0.08\textwidth}
      \centering
      \begin{align*}
        &\text{Thread}_n\\
        \hline\\[-1em]
        \,~& \Write(x, 1)\\
        \\[-1.6em]
        \,~& \Read(x)\\
      \end{align*}
    \end{minipage}
    \vspace{-5mm}
    \caption{
      Many threads and one variable.
    }
  \end{subfigure}
  \caption{
    Programs with one $\RVF$-equivalence class, and
    $\Omega(2^n)$ equivalence classes for the reads-from, value-centric,
    data-centric, and Mazurkiewicz equivalence.
  }
  \label{fig:coarseprograms}
\end{figure}

Consider the simple programs of~\cref{fig:coarseprograms}.
In each program, all traces of the program are
pairwise $\RVF$-equivalent, while the other equivalences induce
exponentially many inequivalent traces.

\section{Details of {\cref{sec:verification}}}\label{sec:app_verification}

Here we present the proof of \cref{them:vsc}.

\themvsc*

\smallskip
\begin{proof}

We argue separately about soundness, completeness, and complexity
of $\AlgoSC$ (\cref{algo:vsc_no_witness}).

\Paragraph{Soundness.}
We prove by induction that each sequence in the worklist $\Worklist$
is a witness prefix. The base case with an empty sequence trivially
holds. For an inductive case, observe that extending a witness prefix
$\Seq$ with an event $\Event$ executable in $\Seq$ yields
a witness prefix. Indeed, if $\Event$ is a read, it has an active
good-write in $\Seq$, thus its good-writes condition is satisfied in
$\Seq \Concat \Event$. If $\Event$ is a write, new reads $\Read$
may start holding the variable $\Location{\Event}$ in $\Seq \Concat \Event$,
but for all these reads $\Event$ is its good-write, and it shall
be active in $\Seq \Concat \Event$.
Hence the soundness follows.

\Paragraph{Completeness.}
First notice that for each witness $\Seq$ of $\VSC(X, \GoodWrites)$,
each prefix of $\Seq$ is a witness prefix.
What remains to prove is that given two witness prefixes
$\Seq_1$ and $\Seq_2$ with equal induced witness state, if a suffix
exists to extend $\Seq_1$ to a witness of $\VSC(X, \GoodWrites)$,
then such suffix also exists for $\Seq_2$. Note that since
$\Seq_1$ and $\Seq_2$ have an equal witness state, their length equals
too (since $\Events{\Seq_1} = \Events{\Seq_2}$). We thus prove the
argument by induction with respect to $|X \setminus \Events{\Seq_1}|$,
i.e., the number of events remaining to add to $\Seq_1$ resp. $\Seq_2$.
The base case with $|X \setminus \Events{\Seq_1}| = 0$ is trivially
satisfied. For the inductive case,
let there be an arbitrary suffix $\Seq^*$ such that
$\Seq_1 \Concat \Seq^*$ is a witness of $\VSC(X, \GoodWrites)$.
Let $\Event$ be the first event of $\Seq^*$, we have that
$\Seq_1 \Concat \Event$ is a witness prefix.
Note that $\Seq_2 \Concat \Event$ is also a witness prefix.
Indeed, if $\Event$ is a read, the equality of the memory maps
$\MemoryMap_{\Seq_1}$ and $\MemoryMap_{\Seq_2}$ implies that
since $\Event$ reads a good-write in $\Seq_1 \Concat \Event$,
it also reads the same good-write in $\Seq_2 \Concat \Event$.
If $\Event$ is a write, since $\Events{\Seq_1} = \Events{\Seq_2}$,
each read either holds its variable in both $\Seq_1$ and $\Seq_2$
or it does not hold its variable in either of $\Seq_1$ and $\Seq_2$.
Finally observe that
$\MemoryMap_{\Seq_1 \Concat \Event} = \MemoryMap_{\Seq_2 \Concat \Event}$.
We have $\MemoryMap_{\Seq_1} = \MemoryMap_{\Seq_2}$, if $\Event$ is a read
both memory maps do not change, and if $\Event$ is a write the only change
of the memory maps as compared to $\MemoryMap_{\Seq_1}$ and $\MemoryMap_{\Seq_2}$
is that
$\MemoryMap_{\Seq_1 \Concat \Event}(\Location{\Event}) = \MemoryMap_{\Seq_2 \Concat \Event}(\Location{\Event}) = \Process(\Event)$.
Hence we have that $\Seq_1 \Concat \Event$ and
$\Seq_2 \Concat \Event$ are both witness prefixes with the same
induced witness state, and we can apply our induction hypothesis.

\Paragraph{Complexity.}
There are at most $n^k \cdot k^d$ pairwise distinct witness states,
since the number of different lower sets of $(X,\TO)$ is bounded by
$n^k$, and the number of different memory maps is bounded by
$k^d$. Hence we have a bound $n^k \cdot k^d$ on the number of iterations
of the main while-loop in \cref{algo:vsc_main_while}.
Further, each iteration of the main while-loop spends
$O(n \cdot k)$ time. Indeed, there are at most $k$ iterations of the
for-loop in \cref{algo:vsc_ifextend}, in each iteration it takes
$O(n)$ time to check whether the event is executable, and the other
items take constant time (manipulating $\DoneSet$ in
\cref{algo:vsc_if_new} and \cref{algo:vsc_insert_worklist} takes
amortized constant time with hash sets).

\qed
\end{proof}

\section{Details of {\cref{sec:smc}}}\label{sec:app_smc}

In this section we present the proofs of \cref{lem:lembacktrack}
and \cref{them:exploration}. We first prove
\cref{lem:lembacktrack}, and then we refer to it when
proving \cref{them:exploration}.

\lembacktrack*

\smallskip
\begin{proof}
We prove the statement by a sequence of reasoning steps.

\begin{enumerate}[noitemsep,topsep=0pt,partopsep=0px]
\item
Let $S$ be the set of writes $\Write^* \not\in \Writes{\widetilde{\Trace}}$
such that there exists a trace $\Trace^*$ that
(i) satisfies $\GoodWrites$ and $\NegativeAnnotation$, and
(ii) contains $\Read$ with $\RF{\Trace^*}(\Read) = \Write^*$.
Observe that $\RF{\Trace'}(\Read) \in S$, hence $S$ is nonempty.
\item
Let $Y$ be the set of events containing $\Read$ and the causal future
of $\Read$ in $\Trace'$, formally
$Y = \{ \Event \in \Events{\Trace'} \;|\;  \CHB{\Read}{\Trace}{\Event}  \} \cup \{ \Read \}$.
Observe that $Y \cap \Events{\widetilde{\Trace}} = \emptyset$.
Consider a subsequence $\Trace'' = \Trace' \Project (\Events{\Trace'} \setminus Y)$
(i.e., subsequence of all events except $Y$).
$\Trace''$ is a valid trace, and from $Y \cap \Events{\widetilde{\Trace}} = \emptyset$
we get that $\Trace''$ satisfies $\GoodWrites$ and $\NegativeAnnotation$.
\item
Let $T_1$ be the set of all partial traces that
(i) contain all of $\Events{\widetilde{\Trace}}$,
(ii) satisfy $\GoodWrites$ and $\NegativeAnnotation$,
(iii) do not contain r, and
(iv) contain some $\Write^* \in S$.
$T_1$ is nonempty due to (2).
\item
Let $T_2 \subseteq T_1$ be the traces $\Trace^*$ of $T_1$ where
for each $\Write^* \in S \cap \Events{\Trace^*}$, we have that
$\Write^*$ is the last event of its thread in $\Trace^*$.
The set $T_2$ is nonempty: since $\Write^* \not\in \Writes{\widetilde{\Trace}}$,
the events of its causal future in $\Trace^*$ are also not in
$\Events{\widetilde{\Trace}}$,
and thus they are not good-writes to any read in $\GoodWrites$.
\item
Let $T_3 \subseteq T_2$ be the traces of $T_2$ with the least
amount of read events in total. Trivially $T_3$ is nonempty.
Further note that in each trace $\Trace^* \in T_3$, no read reads-from
any write $\Write^* \in S \cap \Events{\Trace^*}$.
Indeed, such write can only be read-from by reads $\Read^*$ out of
$\Events{\widetilde{\Trace}}$ (traces of $T_3$ satisfy $\GoodWrites$).
Further, events of the causal future of such reads $\Read^*$ are not good-writes
to any read in $\GoodWrites$ (they are all out of $\Events{\widetilde{\Trace}}$).
Thus the presence of $\Read^*$ violates the property of having the least
amount of read events in total.
\item
Let $\Trace_1$ be an arbitrary partial trace from $T_3$.
Let $S_1 = S \cap \Events{\Trace_1}$, by (3) we have that $S_1$
is nonempty. Let $\Trace_2 = \Trace_1 \Project (\Events{\Trace_1} \setminus S_1)$.
Note that $\Trace_2$ is a valid trace, as for each $\Write^* \in S_1$,
by (4) it is the last event of its thread, and
by (5) it is not read-from by any read in $\Trace_1$.
\item
Since $\Read \in \Enabled(\widetilde{\Trace})$ and
$\Events{\widetilde{\Trace}} \subseteq \Events{\Trace_2}$ and
$\Read \not\in \Events{\Trace_2}$, we have that
$\Read \in \Enabled(\Trace_2)$.
Let $\Write^* \in S_1$ arbitrary, by the previous step we have
$\Write^* \in \Enabled(\Trace_2)$.
Now consider $\bad{\Trace} = \Trace_2 \Concat \Read \Concat \Write^*$.
Notice that
(i) $\bad{\Trace}$ satisfies $\GoodWrites$ and $\NegativeAnnotation$,
(ii) $\RF{\bad{\Trace}}(\Read) \in \Writes{\widetilde{\Trace}}$
(there is no write out of $\Writes{\widetilde{\Trace}}$ present in
$\Trace_2$), and
(iii) for $\Write^* \in \Events{\bad{\Trace}}$, since $\Write^* \in S$
we have $\Confl{\Read}{\Write^*}$ and $\Process(\Read) \neq \Process(\Write^*)$.
\end{enumerate}
\qed
\end{proof}

Now we are ready to prove \cref{them:exploration}.

\themexploration*

\smallskip
\begin{proof}

We argue separately about soundness, completeness, and complexity.

\Paragraph{Soundness.}
The soundness of $\RVFSMC$ follows from the soundness
of $\AlgoSC$ used as a subroutine to generate traces that
$\RVFSMC$ considers.

\Paragraph{Completeness.}
Let $nd = \RVFSMC(X, \GoodWrites, \Trace, \NegativeAnnotation)$
be an arbitrary recursion node of $\RVFSMC$.
Let $\Trace'$ be an arbitrary valid full program trace
satisfying $\GoodWrites$ and $\NegativeAnnotation$.
The goal is to prove that
the exploration rooted at $a$ explores a good-writes function
$\GoodWrites'\colon \Reads{\Trace'} \to 2^{\Writes{\Trace'}}$
such that for each $\Read \in \Reads{\Trace'}$
we have $\RF{\Trace'}(\Read) \in \GoodWrites'(\Read)$.

We prove the statement by induction in the length of
maximal possible extension, i.e., the largest possible number of reads
not defined in $\GoodWrites$ that a valid full program trace
satisfying $\GoodWrites$ and $\NegativeAnnotation$ can have.
As a reminder, given $nd = \RVFSMC(X, \GoodWrites, \Trace, \NegativeAnnotation)$
we first consider a trace
$\widetilde{\Trace} = \Trace \Concat \widehat{\Trace}$
where $\widehat{\Trace}$ is a maximal extension such that no event
of $\widehat{\Trace}$ is a read.

\smallskip\noindent{\em Base case: 1.}
There is exactly one enabled read $\Read \in \Enabled(\widetilde{\Trace})$.
All other threads have no enabled event, i.e., they are fully extended
in $\widetilde{\Trace}$. Because of this, our algorithm considers
every possible source $\Read$ can read-from in traces satisfying
$\GoodWrites$ and $\NegativeAnnotation$. Completeness of $\AlgoSC$
then implies completeness of this base case.

\smallskip\noindent{\em Inductive case.}
Let ${\sf MAXEXT}$ be the length of maximal possible extension of
$nd = \RVFSMC(X, \GoodWrites, \Trace, \NegativeAnnotation)$.
By induction hypothesis, $\RVFSMC$ is complete when rooted at any node
with maximal possible extension length ${\sf <MAXEXT}$.
The rest of the proof is to prove completeness when rooted at $nd$, and
the desired result then follows.

\smallskip\noindent{\em Inductive case: $\RVFSMC$ without backtrack signals.}
We first consider a simpler version of $\RVFSMC$, where the boolean
signal $\backtrack$ is always set to $\true$
(i.e., \cref{algo:exploration} without \cref{line:exploration_backfalse}).
After we prove the inductive case of this version, we use it
to prove the inductive case of the full version of $\RVFSMC$.

Let $\Read_1,...,\Read_k$ be the enabled events in $\widetilde{\Trace}$,
$nd$ proceeds with recursive calls in that order
(i.e., first with $\Read_1$, then with $\Read_2$, ..., last with
$\Read_k$).
Let $\Trace'$ be an arbitrary valid full program trace
satisfying $\GoodWrites$ and $\NegativeAnnotation$.
Trivially, $\Trace'$ contains all of $\Read_1,...,\Read_k$.
Consider their reads-from sources
$\RF{\Trace'}(\Read_1),...,\RF{\Trace'}(\Read_k)$. Now consider two cases:
\begin{enumerate}[noitemsep,topsep=0pt,partopsep=0px]
\item There exists $1\leq i \leq k$ such that $\RF{\Trace'}(\Read_i) \in \Events{\widetilde{\Trace}} \cup \{\initev\}$.
\item There exists no such $i$.
\end{enumerate}
Let us prove that (2) is impossible. By contradiction, consider it possible,
let $\Read_j$ be the first read out of $\Read_1,...,\Read_k$ in the
order as appearing in $\Trace'$. Consider the thread of
$\RF{\Trace'}(\Read_j)$. It has to be one of $\Process{\Read_1},...,\Process{\Read_k}$,
as other threads have no enabled event in $\widetilde{\Trace}$, thus they
are fully extended in $\widetilde{\Trace}$.
It cannot be $\Process{\Read_j}$, because all thread-predecessors of
$\Read_j$ are in $\Events{\widetilde{\Trace}}$.
Thus let it be a thread $1 \leq m \leq k, m \neq j$.
Since $\RF{\Trace'}(\Read_j) \not \in \Events{\widetilde{\Trace}}$,
$\RF{\Trace'}(\Read_j)$ comes after $\Read_m$ in $\Trace'$.
This gives us $\Read_m <_{\Trace'} \RF{\Trace'}(\Read_j) <_{\Trace'} \Read_j$,
which is a contradiction with $\Read_j$ being the first out of
$\Read_1,...,\Read_k$ in $\Trace'$. Hence we know that above case
(1) is the only possibility.

Let $1 \leq j \leq k$ be the smallest with
$\RF{\Trace'}(\Read_j) \in \Events{\widetilde{\Trace}} \cup \{\initev\}$.
Since $\Trace'$ satisfies $\GoodWrites$ and $\NegativeAnnotation$,
we have $\RF{\Trace'}(\Read_j) \not\in \NegativeAnnotation(\Read_j)$.
Consider $nd$ performing a recursive call with $\Read_j$.
Since $\RF{\Trace'}(\Read_j) \in \Events{\widetilde{\Trace}} \cup \{\initev\}$
and $\RF{\Trace'}(\Read_j) \not\in \NegativeAnnotation(\Read_j)$,
$nd$ considers for $\Read_j$ (among others) a good-writes set
$\GoodWrites_j$ that contains $\RF{\Trace'}(\Read_j)$,
and by completeness of $\VSC$, we correctly classify
$\GoodWrites \cup \{ (\Read_j, \GoodWrites_j) \}$ as realizable.
This creates a recursive call with the following
$\bad{nd} = \RVFSMC(\bad{X}, \bad{\GoodWrites}, \bad{\Trace}, \bad{\NegativeAnnotation)}$:
\begin{enumerate}[noitemsep,topsep=0pt,partopsep=0px]
\item $\bad{X} = \Events{\widetilde{\Trace}} \cup \{ \Read_j \}$.
\item $\bad{\GoodWrites} = \GoodWrites \cup \{ (\Read_j, \GoodWrites_j) \}$.
\item $\bad{\Trace}$ is a witness trace, i.e., valid program trace satisfying $\bad{\GoodWrites}$.
\item $\bad{\NegativeAnnotation)} = \NegativeAnnotation \cup \{  (\Read_i, \NegativeAnnotation_i) \;|\: 1\leq i < j \}$
where $\NegativeAnnotation_i$ are the writes of $\Location{\Read_i}$ in $\Events{\widetilde{\Trace}} \cup \{\initev\}$.
\end{enumerate}
Clearly $\Trace'$ satisfies $\bad{\GoodWrites}$. Note that
$\Trace'$ also satisfies $\bad{\NegativeAnnotation)}$, as it satisfies
$\NegativeAnnotation$, and for each $1 \leq i < j$,
we have $\RF{\Trace'}(\Read_i) \not\in \NegativeAnnotation_i$, as
$\RF{\Trace'}(\Read_i) \not\in \Events{\widetilde{\Trace}} \cup \{\initev\}$.
Hence we can apply our inductive hypothesis for $\bad{nd}$, and we're done.

\smallskip\noindent{\em Inductive case: $\RVFSMC$.}
Let $\Read_1,...,\Read_m$ be the enabled events (reads) in $\widetilde{\Trace}$
not defined in $\NegativeAnnotation$, and let
$\Read_{m+1},...,\Read_k$ be the enabled events defined in $\NegativeAnnotation$.
The node $nd$ proceeds with the recursive calls as follows:
\begin{enumerate}[noitemsep,topsep=0pt,partopsep=0px]
\item $nd$ processes calls with $\Read_1$, stops if $\backtrack = \false$ (\cref{line:exploration_whileback}), else:
\item $nd$ processes calls with $\Read_2$, stops if $\backtrack = \false$, else .....
\item $nd$ processes calls with $\Read_m$, stops if $\backtrack = \false$, else
\item $nd$ processes calls with $\Read_m$, then $\Read_{m+1}$, ..., finally $\Read_k$.
\end{enumerate}

Let $\Trace'$ be an arbitrary valid full program trace satisfying
$\GoodWrites$ and $\NegativeAnnotation$.
Trivially, $\Trace'$ contains all of $\Read_1,...,\Read_k$.
Consider their reads-from sources
$\RF{\Trace'}(\Read_1),...,\RF{\Trace'}(\Read_k)$.
Let $1\leq j\leq k$ be the smallest with
$\RF{\Trace'}(\Read_j) \in \Events{\widetilde{\Trace}} \cup \{\initev\}$.
From the above paragraph, we have that such $j$ exists.
From the above paragraph we also have that $nd$ processing calls with
$\Read_j$ explores a good-writes function
$\GoodWrites'\colon \Reads{\Trace'} \to 2^{\Writes{\Trace'}}$
such that for each $\Read \in \Reads{\Trace'}$
we have $\RF{\Trace'}(\Read) \in \GoodWrites'(\Read)$.
What remains to prove is that $nd$ will reach the point of processing
calls with $\Read_j$.
That amounts to proving that for each $1\leq x \leq \min(m, j-1)$,
$nd$ receives $\backtrack = \true$ when processing calls with $\Read_x$.
For each such $x$, $\RF{\Trace'}(\Read_x) \not\in \Events{\widetilde{\Trace}} \cup \{\initev\}$.
We construct a trace that matches the antecedent of
\cref{lem:lembacktrack}.

First denote $\Trace' = \Trace_1 \Concat \RF{\Trace'}(\Read_x) \Concat \Trace_2$.
Let $\Trace_3$ be the subsequence of $\Trace_2$, containing only
events of $\Events{\widetilde{\Trace}} \cap \Events{\Trace_2}$.
Now consider $\Trace_4 = \Trace_1 \Concat \Trace_3 \Concat \RF{\Trace'}(\Read_x) \Concat \Read_x$.
Note that $\Trace_4$ is a valid trace because:
\begin{enumerate}[noitemsep,topsep=0pt,partopsep=0px]
\item Each event of $\Events{\Trace_2} \setminus \Events{\widetilde{\Trace}}$ appears in its thread
only after all events of that thread in $\Events{\widetilde{\Trace}}$.
\item Each read of $\Trace_3$ has $\GoodWrites$ defined, and $\Trace_3$ does not contain
$\RF{\Trace'}(\Read_x)$ nor any events of $\Events{\Trace_2} \setminus \Events{\widetilde{\Trace}}$.
\item $\RF{\Trace'}(\Read_x)$ is enabled in $\Trace_1 \Concat \Trace_3$,
since it is already enabled in $\Trace_1$.
\item $\Read_1$ is not in $\Trace_1$ as those events appear before
$\RF{\Trace'}(\Read_x)$ in $\Trace'$, also $\Read_1$ is not in
$\Trace_3$ because $\Read_1 \not\in \Events{\widetilde{\Trace}}$.
$\Read_1$ is enabled in $\Trace_1 \Concat \Trace_3 \Concat \RF{\Trace'}(\Read_x)$
as it contains all events of $\Events{\widetilde{\Trace}}$.
\end{enumerate}
Hence $\Trace_4$ is a valid trace containing all events $\Events{\widetilde{\Trace}}$.
Further $\Trace_4$ satisfies $\GoodWrites$ and $\NegativeAnnotation$,
because $\Trace'$ satisfies $\GoodWrites$ and $\NegativeAnnotation$ and
$\Trace_4$ contains the same subsequence of the events $\Events{\widetilde{\Trace}}$.
Finally, $\RF{\Trace_4}(\Read_x) = \RF{\Trace'}(\Read_x) \not\in \Writes{\widetilde{\Trace}}$.
The inductive case, and hence the completeness result, follows.

\Paragraph{Complexity.}
Each recursive call of $\RVFSMC$ (\cref{algo:exploration}) trivially
spends $n^{O(k)}$ time in total except the $\AlgoSC$ subroutine
of \cref{line:exploration_vsc}. For the $\AlgoSC$ subroutine we utilize
the complexity bound $O(n^{k + 1} \cdot k^{d + 1})$ from
\cref{them:vsc}, thus the total time spent in each call of $\RVFSMC$
is $n^{O(k)} \cdot O(k^d)$.

Next we argue that no two leaves of the recursion tree of $\RVFSMC$
correspond to the same class of the $\RVF$ trace partitioning.
For the sake of reaching contradiction, consider two such distinct leaves
$l_1$ and $l_2$. Let $a$ be their last (starting from the root recursion
node) common ancestor. Let
$c_1$ and $c_2$ be the child of $a$ on the way to $l_1$ and $l_2$
respectively. We have $c_1 \neq c_2$ since $a$ is the last common ancestor
of $l_1$ and $l_2$. The recursion proceeds from $a$ to $c_1$ (resp. $c_2$)
by issuing a good-writes set to some read $\Read_1$ (resp. $\Read_2$).
If $\Read_1 = \Read_2$, then the two good-writes set issued to
$\Read_1 = \Read_2$ in $a$ differ in the value that the writes of the
two sets write (see \cref{line:exploration_mutvalues}
of~\cref{algo:exploration}). Hence $l_1$ and $l_2$ cannot represent
the same $\RVF$ partitioning class, as representative traces of the two
classes shall differ in the value that $\Read_1 = \Read_2$ reads.
Hence the only remaining possibility is $\Read_1 \neq \Read_2$.
In iterations of \cref{line:exploration_whileback} in $a$,
wlog assume that $\Read_1$ is processed before $\Read_2$.
For any pair of traces $\Trace_1$ and $\Trace_2$ that are class
representatives of $l_1$ and $l_2$ respectively, we have
that $\RF{\Trace_1}(\Read_1) \neq \RF{\Trace_2}(\Read_1)$.
This follows from the update of the causal map $\NegativeAnnotation$
in \cref{line:exploration_updatethread} of the
\cref{line:exploration_whileback}-iteration of $a$
processing $\Read_1$.
Further, we have that $\RF{\Trace_2}(\Read_1)$ is a thread-successor
of a read $\bad{\Read} \neq \Read_1$ that was among the enabled reads of
$\mutate$ in $a$. From this we have $\CHB{\bad{\Read}}{\Trace_2}{\Read_1}$
and $\NCHB{\bad{\Read}}{\Trace_1}{\Read_1}$.
Thus the traces $\Trace_1$ and $\Trace_2$ differ in the causal orderings
of the read events, contradicting that $l_1$ and $l_2$
correspond to the same class of the $\RVF$ trace partitioning.

Finally we argue that for each class of the $\RVF$ trace partitioning,
represented by the $(X, \GoodWrites)$ of its $\RVFSMC$ recursion leaf,
at most $n^k$ calls of $\RVFSMC$ can be performed where its
$X'$ and $\GoodWrites'$ are subsets of $X$ and $\GoodWrites$,
respectively. This follows from two observations. First, in each call
of $\RVFSMC$, the event set is extended maximally by enabled writes,
and further by one read, while the good-writes function is extended
by defining one further read. Second,
the amount of lower sets of the partial order $(\Reads{X}, \TO)$
is bounded by $n^k$.

The desired complexity result follows.
\qed
\end{proof}

\section{Details of {\cref{sec:experiments}}}\label{sec:app_experiments}

Here we present additional details on our experimental setup.

\Paragraph{Handling assertion violations.}
Some of the benchmarks in our experiments contain assertion violations,
which are successfully detected by all algorithms we consider in our
experimental evaluation. After performing this sanity check,
we have disabled all assertions, in order to not have the measured
parameters be affected by how fast a violation is discovered, as the
latter is arbitrary. Our primary experimental goal is to characterize
the size of the underlying partitionings, and the time
it takes to explore these partitionings.

\Paragraph{Identifying events.}
As mentioned in~\cref{sec:prel}, an event is uniquely identified
by its predecessors in $\TO$, and by the values its $\TO$-predecessors
have read.
In our implementation, we rely on the
interpreter built inside Nidhugg to identify events. An event $\Event$ is
defined by a pair ($a_{\Event}, b_{\Event}$), where $a_{\Event}$
is the thread identifier of $\Event$ and $b_{\Event}$ is the
sequential number of the last LLVM instruction
(of the corresponding thread) that is part of $\Event$
(the $\Event$ corresponds to zero or several LLVM instructions not
accessing shared variables, and exactly one LLVM instruction
accessing a shared variable).
It can happen
that there exist two traces $\Trace_1$ and $\Trace_2$, and two different events
$\Event_1 \in \Trace_1$, $\Event_2 \in \Trace_2$, such that their identifiers are
equal, i.e., $a_{\Event_1}=a_{\Event_2}$ and $b_{\Event_2}=b_{\Event_2}$.
However, this means that the control-flow leading
to each event is different. In this case, $\Trace_1$ and $\Trace_2$ differ in
the value read by a common event that is ordered by the program
order $\TO$ both before $\Event_1$ and before $\Event_2$,
hence $\Event_1$ and $\Event_2$ are treated as inequivalent.

\Paragraph{Technical details.}
For our experiments we have used a Linux machine with Intel(R) Xeon(R)
CPU E5-1650 v3 @ 3.50GHz (12 CPUs) and 128GB of RAM.
We have run Nidhugg with Clang and LLVM version 8.

\Paragraph{Scatter plots setup.}
Each scatter plot compares our algorithm $\RVFSMC$ with some other
algorithm $X$. In a fixed plot, each benchmark provides a single data
point, obtained as follows. For the benchmark, we consider the
highest unroll bound where neither of the algorithms $\RVFSMC$ and $X$
timed out.\footnote{
In case one of the algorithms timed out on all attempted unroll bounds,
we do not consider this benchmark when reporting on explored traces,
and when reporting on execution times we consider the results on the
lowest unroll bound, reporting the time-out accordingly.
}
Then we plot the times resp. traces obtained on that benchmark and
unroll bound by the two algorithms $\RVFSMC$ and $X$.

\end{document}